\documentclass[12pt]{article}
\usepackage[margin=1.5cm]{geometry}
\usepackage{amsthm}
\usepackage{amssymb}
\usepackage[utf8]{inputenc}
\usepackage[english]{babel}
\newtheorem{lemma}{Lemma}

\usepackage{graphicx}
\usepackage{newtxtext}
\usepackage{newtxmath}
\usepackage{natbib}
\usepackage{hyperref}
\newcommand{\RomanNumeralCaps}[1]
\linenumbers
\usepackage[font=small,labelfont=bf]{caption}
\usepackage{xcolor}
\usepackage{color}
\usepackage{amsmath}
\usepackage{comment}
\usepackage{epic}
\usepackage{wrapfig}

\usepackage{bm}
\newsavebox{\tthirdbox}
\sbox{\tthirdbox}{$\textstyle\frac{1}{3}$}

\newsavebox{\tsixthbox}
\sbox{\tsixthbox}{$\textstyle\frac{1}{6}$}

\newsavebox{\ttwelvethbox}
\sbox{\ttwelvethbox}{$\textstyle\frac{1}{12}$}

\def\R {\mathbb{R}}

\def\det {\text{\rm det}\,}

\def\tanh {\text{\rm tanh}\,}

\def\sgn {\text{\rm sgn}\,}

\newcommand{\fpar}[2]{\frac{\partial #1}{\partial #2}}

\newcommand*{\mydprime}{^{\prime\prime}\mkern-1.2mu}

\newcommand{\beq}{\begin{equation}}
	\newcommand{\eeq}{\end{equation}}
\newcommand{\beqa}{\begin{eqnarray}}
	
	\newcommand{\eeqa}{\end{eqnarray}}
\newcommand{\labeleq}[1]{\label{#1}} 
\newcommand{\req}[1]{(\ref{#1})}

\newcommand{\secta}[1]{\section*{#1}\setcounter{section}{1}
	\setcounter{equation}{0}
	
	\def\theequation{\Alph{section}.\arabic{equation}}}

\newcommand{\sectb}[1]{\section*{#1}\addtocounter{section}{1}
	\setcounter{equation}{0}
	\def\theequation{\Alph{section}.\arabic{equation}}}

\newcommand{\sectc}[1]{\section*{#1}\addtocounter{section}{1}
	\setcounter{equation}{0}                                                                       
	\def\theequation{\Alph{section}.\arabic{equation}}}

\newcommand{\sectd}[1]{\section*{#1}\addtocounter{section}{1}
	\setcounter{equation}{0}
	\def\theequation{\Alph{section}.\arabic{equation}}}

\newlength{\defbaselineskip}
\newcommand{\setlinespacing}[1]
{\setlength{\baselineskip}{#1 \defbaselineskip}}
\setlinespacing{1.5}

\newcommand\lrr[1]{\ensuremath{\left( #1 \right)}}

\usepackage{overpic}

\title{A classification of mode-1 internal solitary waves in a three-layer fluid}

\author
{R.~Barros$^1$
	\thanks{Email address for correspondence: r.barros@lboro.ac.uk},
	A. Doak$^2$,
	W.~Choi$^{3}$,
	and P.~A.~Milewski$^{2,4}$}
\date{%
	\small {$^1$1. Department of Mathematical Sciences, Loughborough University , Loughborough LE11 3TU, UK}\\ %
	$^2$Department of Mathematical Sciences, University of Bath, Bath, BA2 7AY, UK \\ 
	$^3$Department of Mathematical Sciences,  New Jersey Institute of Technology, Newark, NJ 07102-1982, USA \\ 
	$^4$Department of Mathematics, The Pennsylvania State University, State College, PA 16802, USA
	\today}

\begin{document}

	\maketitle
	
	\abstract{We explore the bifurcation structure of mode-1 solitary waves in a three-layer fluid confined between two rigid boundaries. A recent study (\citeauthor{lamb_2023}, \textit{J. Fluid Mech.}  2023, 962, A17) proposed a method to predict the coexistence of solitary waves with opposite polarity in a continuously stratified fluid with a double pycnocline by examining the conjugate states for the Euler equations. 
		We extend this line of inquiry to a piecewise-constant three-layer stratification, taking advantage of the fact that the conjugate states for the Euler equations are exactly preserved by the strongly nonlinear model that we will refer to as the three-layer Miyata-Maltseva-Choi-Camassa (MMCC3) equations. 
		In this reduced setting, solitary waves are governed by a Hamiltonian system with two degrees of freedom, whose critical points are used to explain the bifurcation structure.
		Through this analysis, we also discover families of solutions that have not been previously reported. 
		Using the shared conjugate state structure between the MMCC3 model and the full Euler equations, we propose criteria for distinguishing the full range of solution behaviours. This alignment between the reduced and full models provides strong evidence that partitioning the parameter space into regions associated with distinct solution types is valid within both theories. This classification is further substantiated by numerical solutions to both models, which show excellent agreement. 
	}

	\section{Introduction}
	
	Internal solitary waves (ISWs) in density stratified fluids have been the focus of extensive theoretical, numerical, and experimental research. Their common occurrence in coastal oceans has prompted numerous field studies aimed at understanding their influence on mass transport and mixing within the global ocean (\citeauthor{shroyer_et_al} \citeyear{shroyer_et_al}; \citeauthor{moum_et_al} \citeyear{moum_et_al}). Oceanic ISWs can be modelled as solutions to the Euler equations for a continuous stratification that, in steady
	form, are known as the Dubreil-Jacotin-Long (DJL) equation (\citeauthor{dubreil_jacotin} \citeyear{dubreil_jacotin}; \citeauthor{long} \citeyear{long}). Although there are infinitely many modes of propagation in the ocean, as predicted by linear theory, most previous literature, both observational and theoretical, concerns waves of the first baroclinic mode (mode-1), which have the simplest vertical structure (see \citeauthor{helfrich_melville} \citeyear{helfrich_melville} and references therein). Mode-1 solutions to the DJL have been computed for different stratifications by using various methods (\citeauthor{tung_et_al} \citeyear{tung_et_al}; \citeauthor{turkington_et_al} \citeyear{turkington_et_al}; \citeauthor{stastna_lamb} \citeyear{stastna_lamb}; \citeauthor{marek_book} \citeyear{marek_book}). These studies suggest that a wide range of solution behaviours can be expected. In particular, while certain stratification profiles lead to wave broadening as amplitude increases, others result in wave narrowing with growing amplitude \citep{lamb_wan}. Additional phenomena, such as the coexistence of solutions with opposite polarity, were recently examined by \cite{lamb_2023}  in the context of a double-pycnocline stratification. This configuration features two hyperbolic tangent functions that represent a smooth density profile with two sharp transitions. Lamb found that wave broadening occurred consistently across all physical parameters considered.

	The double-pycnocline stratification profile can be idealised as a three-layer fluid system as depicted in figure~\ref{setup}, in which the densities of the fluids in each layer are constant. This simplified configuration facilitates analysis and has been extensively studied
	using weakly nonlinear theory, including the Korteweg-de Vries equation \citep{yang_et_al} and the Gardner equation \citep{kurkina_et_al}. Weakly nonlinear theory has revealed a variety of solution behaviours: mode-1 ISWs may appear as waves of elevation, waves of depression, or as two distinct branches with opposite polarity. More importantly, it enables the delineation of regions in the parameter space where different types of solutions are present. However, concerns have been raised regarding the validity of these models when compared to the full Euler theory \citep{lamb_yan}.
	
	As an effort to understand the strongly nonlinear effects on ISWs in this system, \cite{barros_choi_milewski} and \cite{doak_et_al} have adopted the strongly nonlinear theory, based on a long wave model that extends the well-known (two-layer) Miyata-Maltseva-Choi-Camassa (MMCC) model to three layers. While previous studies have commonly referred to this model as the Miyata–Choi–Camassa (MCC) model (\citeauthor{miyata} \citeyear{miyata}; \citeauthor{choi_camassa_99} \citeyear{choi_camassa_99}), recent recognition of the contributions by \citet{maltseva_1989} warrants this rectification.
	Accordingly, the model used in this work will be referred to as the three-layer Miyata–Maltseva–Choi–Camassa (MMCC3) model. Although our previous investigations focused primarily on mode-2 solutions, we use it here to provide a classification of mode-1 solutions. 
	
	Despite being simpler than Euler, the MMCC3 model captures exactly the conjugate states in the sense of \cite{benjamin_71}, which limit the amplitude of mode-1  solutions. This is a promising sign for the accuracy of the MMCC3 model in describing waves well beyond the weakly nonlinear regime, as established for the two-layer system by \cite{camassa_et_al}. Furthermore, as pointed out by \cite{doak_et_al}, the conjugate states are related to the critical points of the dynamical system of the MMCC3 theory, a Hamiltonian system with two degrees of freedom. This offers a dramatic reduction in complexity compared to the Euler equations, which we exploit to shine light on the bifurcation structure of mode-1 solutions. We are able to provide a complete classification of mode-1 solitary waves for the MMCC3 system, covering all of parameter space under the Boussinesq approximation. 
	The analysis of the conjugate states is paramount to delimit regions in the parameter space where different types of solutions can be found. In particular, the criterion for the coexistence of solutions of opposite polarity established by \cite{lamb_2023} is recovered. Other criteria are proposed for other types of solutions, including previously unseen bifurcation behaviour, in which multiple branches with the same polarity are observed. The fact that conjugate states are common to both the MMCC3 and Euler systems provides strong evidence that the results are valid within both theoretical frameworks. This conclusion is further substantiated by numerical tests.
	
	The paper is structured as follows.
	In section \S 2, we introduce the models examined in the paper.
	In section \S 3, we provide an overview of the main findings. 
	In section \S 4, we examine the properties of the critical points of the MMCC3 system, with particular emphasis on their collision.
	In section \S 5 we use these findings to predict different bifurcation structures, 
	and provide a complete description of the solution space, under Boussinesq approximation. 
	In section \S 6, we present and compare numerical solutions from both the MMCC3 and full Euler models. Concluding remarks are given in \S 7.

	\section{Preliminaries}
	
\begin{figure}
		\captionsetup{width=7cm, justification=justified }
		\begin{center}
			\includegraphics*[width=300pt]{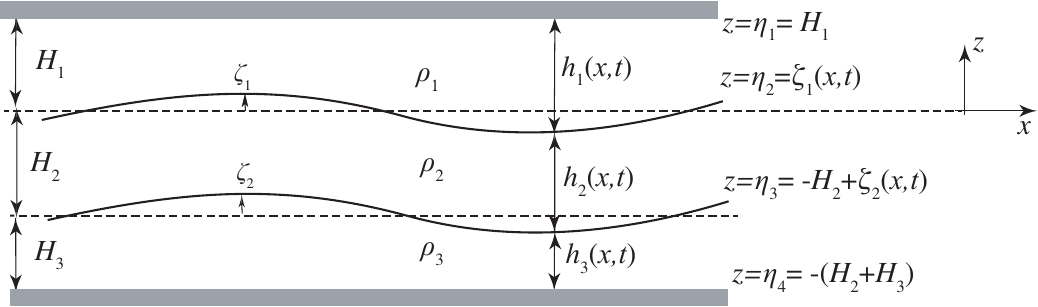}
		\end{center}
		\caption{A three-fluid system. \label{setup}}
	\end{figure}

	We consider long waves in a physical system composed of three fluid layers, bounded above and below by a rigid flat surface, as depicted in figure \ref{setup}. 
	The stratification is assumed gravitationally stable, {\it i.e.,} the constant densities $\rho_i$ for $i=1,2,3$ (from top to bottom) satisfy $\rho_1< \rho_2 < \rho_3$. The location of the upper and lower interfaces are defined by $z=\eta_2 \equiv \zeta_1 (x,t)$ and $z=\eta_3 \equiv -H_2 + \zeta_2 (x,t)$, respectively. The two rigid walls are located at $z=\eta_1 \equiv H_1$ and $z=\eta_4 \equiv -(H_2+H_3)$. The thickness of the $i$-th layer $h_i$ is then given by $h_i=\eta_i-\eta_{i+1}$. More precisely:
	\begin{equation}\label{thickness_def}
		h_1= H_1 - \zeta_1, \quad h_2 = \zeta_1 - \zeta_2 + H_2, \quad h_3 = H_3 + \zeta_2,  
	\end{equation}
	where $H_i$ denotes the undisturbed thickness of the $i$-th layer. The total depth of the water column will be denoted by $H$ and defined by $H=H_1+H_2+H_3$. 

	Assuming the fluid in each layer is inviscid and incompressible the velocity vector $\bm{u_i}(x,z,t)=(u_i,w_i)$ in each layer is governed by the Euler equations
	\begin{align}\label{eq:euler1}
		\rho_i \frac{D_i\bm{u_i}}{Dt} &= -\nabla p_i - \rho_i g \bm{e_z}, && i=1,2,3,\\
		\nabla \cdot \bm{u_i} &=0, && i=1,2,3,
	\end{align}
	where $p_i$ is the pressure in layer $i$, the operator $D_i/Dt(=\partial_t + \bm{u_i}\cdot \nabla)$ is the material derivative with respect to velocity field in layer $i$, $g$ is the acceleration due to gravity, and $\bm{e_z}$ is the unit vector in the $z$ direction. 
	On the top and bottom walls, we have no normal flow, that is
	\begin{align}
		\bm{u_3}\cdot \bm{e_z} &=0,&&\text{at } z=\eta_4, \\
		\bm{u_1}\cdot \bm{e_z}&=0,&&\text{at } z=\eta_1 .
	\end{align}
	On the a priori unknown interfaces, we have both kinematic boundary conditions, given by
	\begin{align}
		\frac{D_{i-1}(\eta_i-z)}{Dt} &=0, && \text{at } z=\eta_i, \,\, i=2,3, \\
		\frac{D_{i}(\eta_i-z)}{Dt} &=0, && \text{at } z=\eta_i, \,\, i=2,3,
	\end{align}
	and continuity of pressure, i.e.
	\begin{align}\label{eq:cont_p}
		p_i&=p_{i-1} && \text{at } z=\eta_i, \,\, i=2,3.
	\end{align}
	We impose that $\bm{u_i}$ and $\eta_i$ decay to zero as $|x|\rightarrow\infty$. The fully nonlinear system \eqref{eq:euler1}-\eqref{eq:cont_p} is referred to as the \emph{Euler system} throughout this paper. 
	
	To examine large amplitude long waves in this system, we also adopt the strongly nonlinear multi-layer model proposed by \citet{choi}, which, starting from the Euler system above, under weak horizontal vorticity and long wave ($\epsilon= H_i/\lambda \ll 1$) assumptions, is 
	formulated in terms of the thicknesses of each layer, denoted by $h_i$, and depth-averaged horizontal velocities $\overline{u}_i$. Namely, consisting of the mass conservation laws for each layer, 
	\begin{equation}\label{mass_conservation_laws}
		h_{i,t} + (h_i \,\overline{u}_i)_x =0, 
	\end{equation}
	and the momentum equations,
	\begin{equation}\label{momentum_eq0}
		{\overline u}_{i,t}
		+{\overline u}_i{\overline u}_{i,x}+g \eta_{i,x}+{P_{i,x}\over \rho_i} ={1\over h_i}
		\left (\frac{1}{3} h_i^3 a_i+\frac{1}{2} h_i^2 b_i \right )_x
		+\left (\frac{1}{2} h_i a_i + b_i \right )\,{\eta_{i+1,x}}\,,
	\end{equation}
	where $a_i(x,t)=-(D_i^2 h_i)/h_i$ and $b_i(x,t) = - D_i^2 \eta_{i+1}$ with $D_i=\partial/\partial t+  {\overline u_i}\, \partial/\partial x $.
	The pressure at the location $z=\eta_i (x,t)$ denoted by $P_i$ satisfies the recursion formula:
	\begin{equation}\label{peq1}
		P_{i+1} = P_i + \rho_i \left(g h_i -\frac{1}{2} a_i h_i^2 - b_i h_i \right).
	\end{equation}
	The system of equations composed of \eqref{mass_conservation_laws} and \eqref{momentum_eq0} for $i=1,2,3$ and \eqref{peq1} for $i=1,2$, along with a geometric constraint given by $h_1+h_2+h_3=H_1+H_2+H_3=H$ will be referred to as the MMCC3 model.

	According to linear theory, there exist two propagation modes in each direction, known as mode-1 (faster) and mode-2 (slower) according to the magnitude of the wave speed. Linear long wave speeds $c_0^\pm$ ($0<c_0^-<c_0^+$) are obtained as the roots of the equation
	\begin{multline}\label{lin_lw_speed_ast}
		\left[ (\rho_1 H_2 + \rho_2 H_1)c^2 - g (\rho_2-\rho_1) H_1 H_2 \right] \left[ (\rho_2 H_3 + \rho_3 H_2) c^2 - g (\rho_3-\rho_2) H_2 H_3 \right] 
		-\rho_2^2 H_1 H_3 \,c^4  =0,
	\end{multline}
	and, for linear waves, the ratio between the two interface displacements can be found as:
	\begin{equation}\label{ratio_displacements}
		\zeta_2/\zeta_1 = \gamma^\pm,
	\end{equation}
	where $\pm$ refers to the propagation mode considered, and $\gamma^\pm$ is obtained from either equality:
	\begin{equation}\label{def_gamma}
		\gamma^\pm =  1+\frac{\rho_1 H_2}{\rho_2 H_1}  - g \left( \frac{\rho_2-\rho_1}{\rho_2}\right) \frac{H_2}{{c_0^\pm}^2}
		= \left[1+\frac{\rho_3 H_2}{\rho_2 H_3}  - g \left( \frac{\rho_3-\rho_2}{\rho_2}\right) \frac{H_2}{{c_0^\pm}^2} \right]^{-1}.
	\end{equation}
	An alternate form of the equation for the linear long wave speed \eqref{lin_lw_speed_ast} can be written:
	\begin{equation}\label{lin_lw_speeds_gamma}
		\left[  1+\frac{\rho_1 H_2}{\rho_2 H_1}  - g \left( \frac{\rho_2-\rho_1}{\rho_2}\right) \frac{H_2}{c^2} \right] \left[  1+\frac{\rho_3 H_2}{\rho_2 H_3}  - g \left( \frac{\rho_3-\rho_2}{\rho_2}\right) \frac{H_2}{c^2} \right] =1.
	\end{equation}
	In addition, we have $\gamma^+>0$ and $\gamma^-<0$ (see {\it e.g.} \citeauthor{barros_choi_milewski} \citeyear{barros_choi_milewski}).  
	
	It will be assumed, without loss of generality, that the stratification is given as 
	\begin{equation}\label{stratification}
		\rho_1 = \rho_0 (1-\Delta_1), \quad \rho_2=\rho_0, \quad \rho_3 = \rho_0 (1+ \Delta_2),
	\end{equation}
	with $0<\Delta_1<1$, $\Delta_2>0$. Furthermore, we introduce the parameter $\delta$ as the ratio between the two density increments:  
	\begin{equation}\label{delta_def}
		\delta=\Delta_1/\Delta_2.
	\end{equation}

	\subsection{Boussinesq approximation and upside-down symmetry}
	
	The Boussinesq approximation is commonly adopted in the study of stratified fluids 
	when the density variation is small.
	For the physical system under consideration, 
	under the Boussinesq approximation, both $\Delta_1 = (\rho_2-\rho_1)/\rho_2$ and $\Delta_2 = (\rho_3-\rho_2)/\rho_2$ are assumed small ($\ll 1$) and of the same order of magnitude, and can be neglected except where they appear in terms multiplied by $g$. Under this approximation, the linear long wave speeds $c_0$ are roots of: 
	\begin{equation}\label{c0_Boussinesq}
		\left[ (H_1+H_2 )c^2 - g_1^\prime H_1 H_2 \right] \left[ ( H_2 + H_3) c^2 - g_2^\prime H_2 H_3 \right] - H_1 H_3 \,c^4  =0,
	\end{equation}
	and $\gamma$ is defined accordingly as 
	\begin{equation}\label{gamma_def_Boussinesq}
		\gamma^\pm = 1+ \frac{H_2}{H_1} - \frac{g_1^\prime H_2}{{c_0^\pm}^2}
		= \left[ 1+\frac{H_2}{H_3} - \frac{g_2^\prime H_2}{{c_0^\pm}^2} \right]^{-1}.
	\end{equation}
	Here, $g_1^\prime = g \Delta_1$ and $g_2^\prime = g \Delta_2$ denote reduced gravities (The notation $g^\prime$ will be used unambiguously in the case when $\Delta_1=\Delta_2$).
	
	We note that, under the Boussinesq approximation, both the Euler and MMCC3 systems are invariant under the so-called 
	``upside-down" symmetry (see \citeauthor{guan} \citeyear{guan} and \citeauthor{barros_choi_milewski} \citeyear{barros_choi_milewski}, respectively). For every steady solution pair $(\zeta_1,\zeta_2)$ of a given configuration with densities $\rho_0 (1-\Delta_1)$, $\rho_0$, $\rho_0 (1+\Delta_2)$ and undisturbed thickness of layers $H_1$, $H_2$, $H_3$ (from top to bottom), $(-\zeta_2, -\zeta_1)$ is a solution of the modified physical configuration with densities $\rho_0 (1-\Delta_2)$, $\rho_0$, $\rho_0 (1+\Delta_1)$, where the undisturbed thickness of layers are $H_3$, $H_2$, $H_1$ (from top to bottom).

	In the remaining part of this section, we discuss the characterisation of solitary waves provided by weakly nonlinear theory, with particular emphasis on mode-1, also known as the first baroclinic mode. 
	
	\subsection{Korteweg-de Vries theory}
	
	If a weakly nonlinear theory is adopted, by considering uni-directional waves in the limit of $\alpha = a/H_i=O(H_i^2/\lambda^2)=O(\epsilon^2)$  ($\epsilon \ll 1$), with $a$ and $\lambda$ typical values of amplitude and wavelength, a Korteweg-de Vries (KdV) equation can be derived for the upper interface $z=\zeta_1 (x,t)$: 
	\begin{equation}\label{KdV_compact_form}
		\zeta_{1,t} + c_0 \,\zeta_{1,x} + c_1 \,\zeta_1 \,\zeta_{1,x} + c_2 \,\zeta_{1,xxx}=0,
	\end{equation}
	with coefficients $c_1$ and $c_2$  given by:
	\begin{equation}\label{coef_c1}
		c_1 = \frac{3}{2} c_0 \frac{\frac{\rho_3}{H_3^2} \gamma^3 + \frac{\rho_2}{H_2^2} (1-\gamma)^3 -\frac{\rho_1}{H_1^2}}{\frac{\rho_3}{H_3} \gamma^2 + \frac{\rho_2}{H_2} (1-\gamma)^2+\frac{\rho_1}{H_1}},
	\end{equation}
	\begin{equation}\label{coef_c2}
		c_2= \frac{1}{6} c_0  \frac{\rho_3 H_3 \gamma^2 + \rho_2 H_2 (1+\gamma+\gamma^2)+\rho_1 H_1}{\frac{\rho_3}{H_3} \gamma^2 + \frac{\rho_2}{H_2} (1-\gamma)^2+\frac{\rho_1}{H_1}},
	\end{equation}
	where $\gamma$ is defined in \eqref{def_gamma}. Within the KdV theory, the propagation modes are fully decoupled, and solutions for mode-1 (mode-2) are obtained by replacing $c_0$ by $c_0^+$ ($c_0^-$) in \eqref{KdV_compact_form} and \eqref{def_gamma}. For each mode, internal solitary wave (ISW) solutions exist for \eqref{KdV_compact_form} as long as $c>c_0$ and their polarity depends strictly on the sign of the quadratic nonlinearity coefficient $c_1$ ($c_2>0$ regardless of the physical parameters considered): the ISW is of elevation (depression) for $\zeta_1$ if $c_1> 0$ ($<0$). On the other hand, according to the KdV theory, no ISWs exist in the case when the coefficient $c_1$ vanishes, {\it i.e.}, when:
	\begin{equation}\label{criticality_condition}
		\frac{\rho_3}{H_3^2} \gamma^3 + \frac{\rho_2}{H_2^2} (1-\gamma)^3 - \frac{\rho_1}{H_1^2} =0,
	\end{equation}
	often referred to as the {\it criticality condition}. Typically, and regardless of the propagation mode considered, given densities $\rho_i$ ($i=1,2,3$) and a fixed ratio $H_1/H_3$, criticality is attained at one single point in the parameter space on the $(H_1/H, H_2/H)$-plane. An exception arises when the stratification satisfies
	$\rho_2/\rho_1= \rho_3/\rho_2$. In this unique case, if the mode-1 is considered, criticality is maintained along the entire line $H_2/H = -\left( 1+(\rho_3/\rho_1)^{1/2}\right) H_1/H + 1$
	(see Appendix B in \citeauthor{barros_choi_milewski} \citeyear{barros_choi_milewski}).  
	
	We remark that, alternatively, a KdV equation could have been written for the lower interface displacement $\zeta_2(x,t)$: 
	\begin{equation}\label{KdV_zeta2}
		\zeta_{2,t} + c_0 \,\zeta_{2,x} + \tilde{c}_1 \,\zeta_2 \,\zeta_{2,x} + c_2 \,\zeta_{2,xxx}=0.
	\end{equation}
	Here $\tilde{c}_1 = c_1/\gamma$ and the equation can be easily recovered from \eqref{KdV_compact_form} by setting $\zeta_1 = \zeta_2/\gamma$.

	\begin{figure}
		\begin{minipage}{.5\textwidth}
			\centering
			\begin{overpic}[width=160pt]{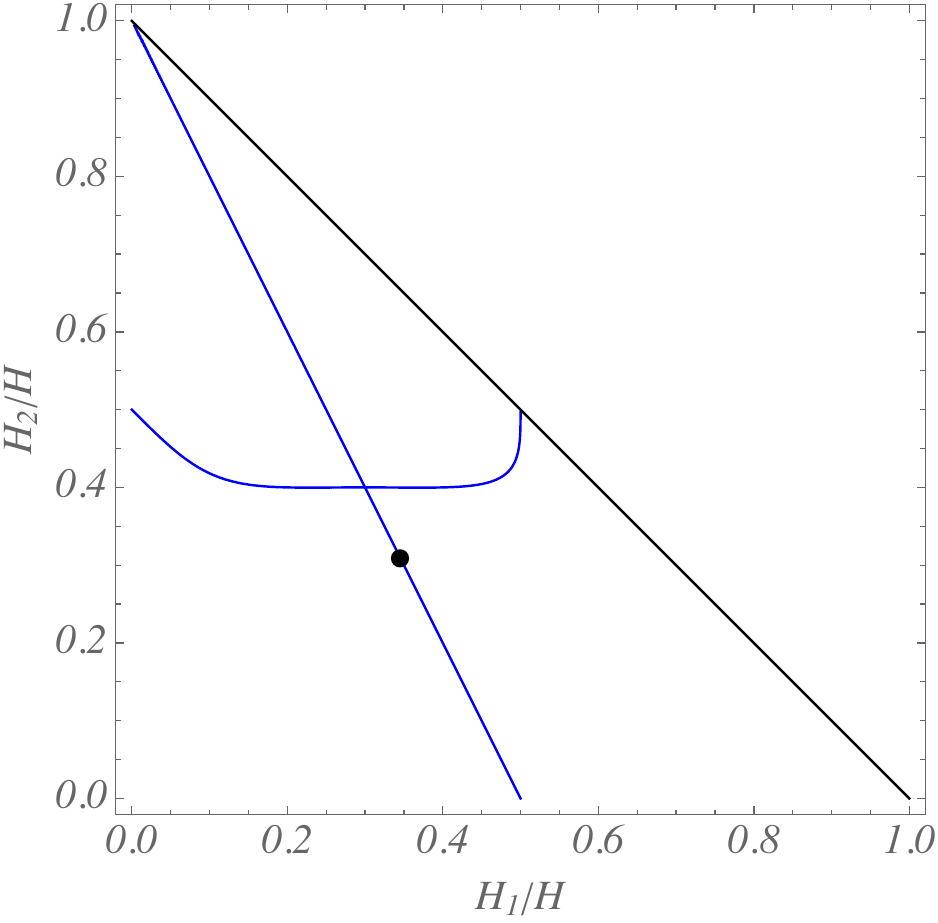}
				\put(-4,95){$(a)$}
			\end{overpic}
		\end{minipage}%
		\begin{minipage}{0.5\textwidth}
			\begin{overpic}[width=160pt]{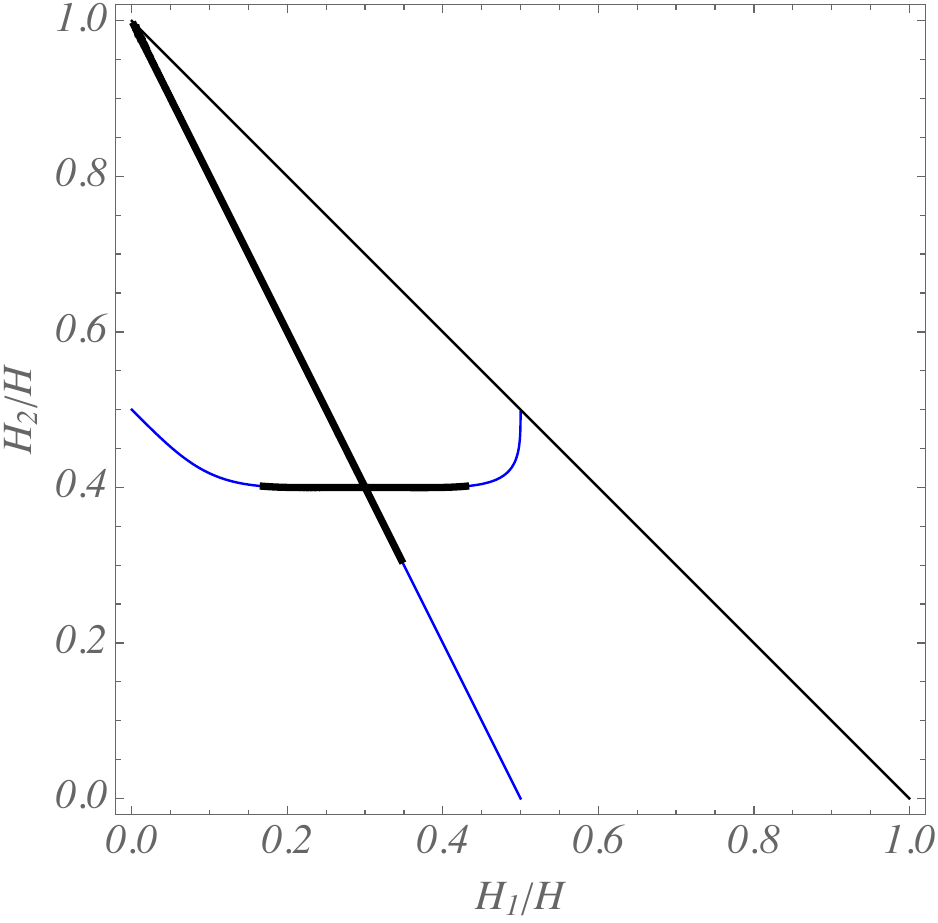}
				\put(-4,95){$(b)$}
			\end{overpic}
		\end{minipage}
		\caption{
			(a) Shows the locus in the parameter space where the criticality condition \eqref{criticality_condition} is satisfied, under the Boussinesq approximation with $\Delta_1=\Delta_2$. The criticality curve (in blue), along which no ISWs exist according to the KdV theory, splits up into two components on the $(H_1/H, H_2/H)$-plane, one of which is the line $H_2/H=-2 H_1/H + 1$, corresponding to a fixed ratio $H_1/H_3=1$. Also marked on this line is the point with coordinates $(9/26,4/13)$ at which a transition from non-existence to the existence of solitary waves to the mKdV equation \eqref{mKdV_eq_sym_case} occurs. This is made clearer in $(b)$ where the mKdV theory is adopted by relaxing the constraint $H_1=H_3$, for which the locus (in thick black) for the coexistence of ISWs of opposite polarity is revealed.
		}\label{locus_criticality_multiple_polarities_delta_1}
	\end{figure}
	
	To obtain the expression of the KdV coefficients under the Boussinesq approximation, as well as the corresponding criticality condition, it suffices to set $\rho_1 = \rho_2 = \rho_3 = \rho_0$, along with $c_0$ defined implicitly by \eqref{c0_Boussinesq}, and consider $\gamma$ defined by \eqref{gamma_def_Boussinesq} instead of \eqref{def_gamma}. By doing so, we recover the KdV equation presented by \cite{yang_et_al}, based on the work by \cite{benney} (see also \citeauthor{benjamin_66} \citeyear{benjamin_66}; \citeauthor{grimshaw_1981} \citeyear{grimshaw_1981}). 
	
	It is well known that when Boussinesq approximation is adopted, with equal density increments ($\Delta_1=\Delta_2$) and same undisturbed depths for the top and bottom layers ($H_1=H_3$), then mode-1 solitary waves cannot exist according to KdV theory. The reason being that, for these parameters, we have $\gamma^+=1$, which trivially satisfies the criticality condition. 
	On the other hand, 
	if the modified Korteweg-de Vries (mKdV) equation
	\begin{equation}\label{mKdV_eq_sym_case}
		\zeta_{1,t} + c_0^+ \,\zeta_{1,x} + \frac{1}{12} c_0^+ H_1(2H_1 + 3H_2) \,\zeta_{1,xxx} - \frac{1}{8 H_1^3} c_0^+ (8H_1 -9H_2) \left(\zeta_1^3\right)_x =0,
	\end{equation}
	is used for the first baroclinic mode (with $c_0^+= \sqrt{g^\prime H_1}$) different predictions can be made (see \citeauthor{talipova_et_al}  \citeyear{talipova_et_al}). When $H_1>(9/8) \,H_2$ (or equivalently $H_2/H<4/13$ along the line $H_2/H = -2 H_1/H+1$, as in figure~\ref{locus_criticality_multiple_polarities_delta_1}$(a)$), no mode-1 ISWs exist according to the mKdV theory, in agreement with the KdV theory. In contrast, when $H_1 <(9/8) \,H_2$,
	two branches of solutions with opposite polarity can be predicted by the mKdV theory (see \citeauthor{grimshaw_et_al} \citeyear{grimshaw_et_al}). We note that for the same density stratification, {\it i.e.,} $\delta=1$, a mKdV equation can be written by relaxing the constraint $H_1=H_3$. Such equation is defined solely along the criticality curve. Hence, the locus for the coexistence of ISWs of opposite polarity, according to such mKdV equation, is a subset of the criticality curve. In figure~\ref{locus_criticality_multiple_polarities_delta_1}$(b)$ such set is depicted in black. In the remaining subset (in blue) of the criticality curve, no ISWs can exist according to the theory. 
	
	We remark that, under the Boussinesq approximation, the KdV theory carries over the upside-down symmetry of the Euler equations. This results from \eqref{KdV_compact_form} and \eqref{KdV_zeta2} and the conditions:  
	\begin{align}
		c_0(H_3,H_2,H_1,1/\delta) &= c_0(H_1,H_2,H_3,\delta), \label{cond1}\\
		c_1(H_3,H_2,H_1,1/\delta) &= -\tilde{c}_1(H_1,H_2,H_3,\delta), \label{cond2} \\ c_2(H_3,H_2,H_1,1/\delta) &= c_2(H_1,H_2,H_3,\delta), \label{cond3}   
	\end{align}
	which are all satisfied throughout the entire parameter space. 
	
	\subsection{Gardner theory}
	Following \cite{choi_camassa_99}, we can extend the KdV equation \eqref{KdV_compact_form} to various orders in an attempt to better describe larger amplitude waves. Namely, the Gardner equation:
	\begin{equation}\label{Gardner_3layer_dim}
		\zeta_{1,t} +c_0 \,\zeta_{1,x} + c_1 \zeta_1 \zeta_{1,x} +  c_2 \,\zeta_{1,xxx} + c_3 \left(\zeta_1^3\right)_x = 0.
	\end{equation}

	\begin{figure}
		\begin{minipage}{.47\textwidth}
			\centering
			\begin{overpic}[width=.9\textwidth]{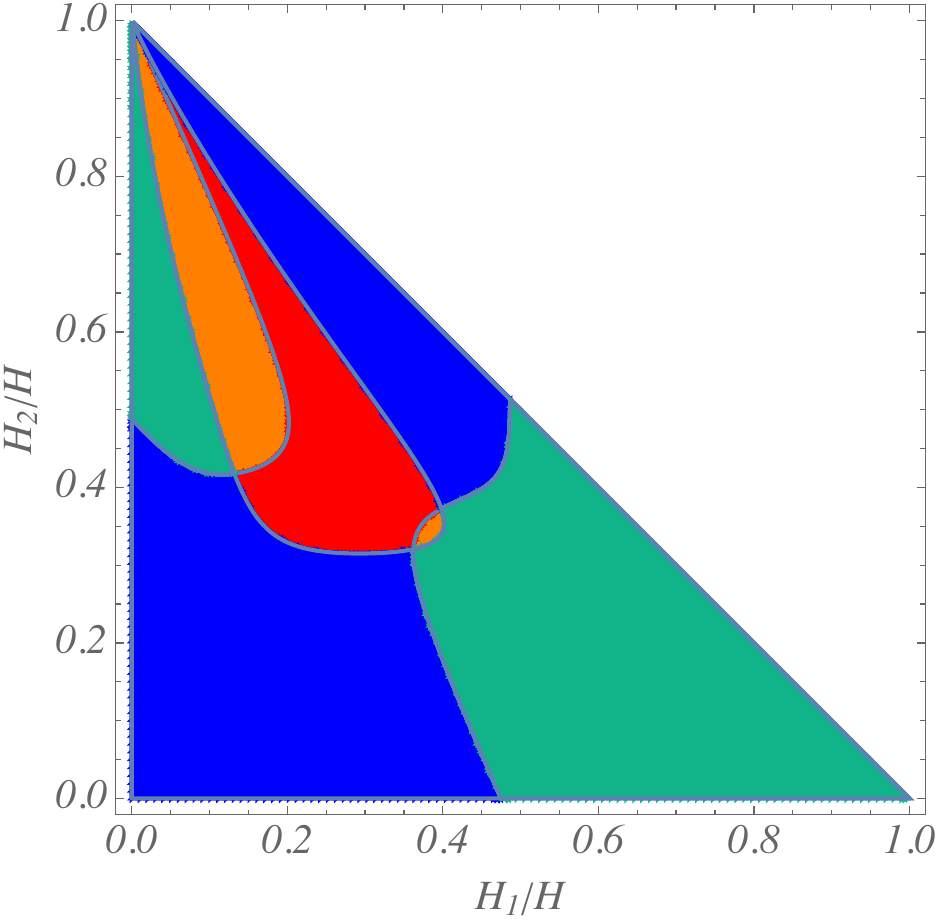}
				\put(50,90){$\delta=1$}
				\put(-3,95){$(a)$}
			\end{overpic}
		\end{minipage}
		\begin{minipage}{.05\textwidth}
		\end{minipage}
		\begin{minipage}{.47\textwidth}
			\centering
			\begin{overpic}[width=0.8\textwidth]{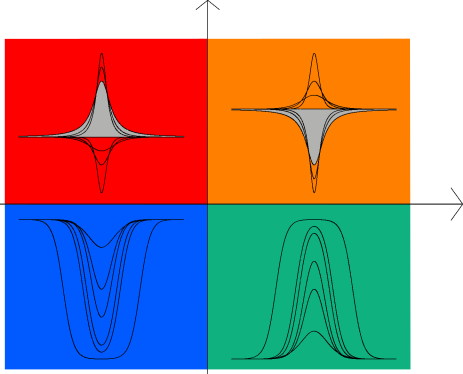}
				\put(45,83){$c_3$}
				\put(95,42){$c_1$}
				\put(-5,90){$(b)$}
			\end{overpic}
		\end{minipage}
		\caption{Behaviour of the quadratic and cubic nonlinearity coefficients ($c_1$, $c_3$, respectively) of the Gardner equation \eqref{Gardner_3layer_dim} for a three-layer fluid, when restricted to mode-1 waves. The stratification is given by \eqref{stratification} with $\Delta_1=\Delta_2=0.1$. It is shown how the signs for the nonlinearity coefficients may vary across the whole parameter space. The colour scheme set in panel $(a)$ is the same as used by Kurkina {\it et al.} (2015) {\it cf.} panel $(b)$, and allows to easily distinguish the various solution behaviours. Shaded regions in $(b)$ represent a minimum amplitude along a solution branch. 
			\label{Gardner_predictions}}
	\end{figure}
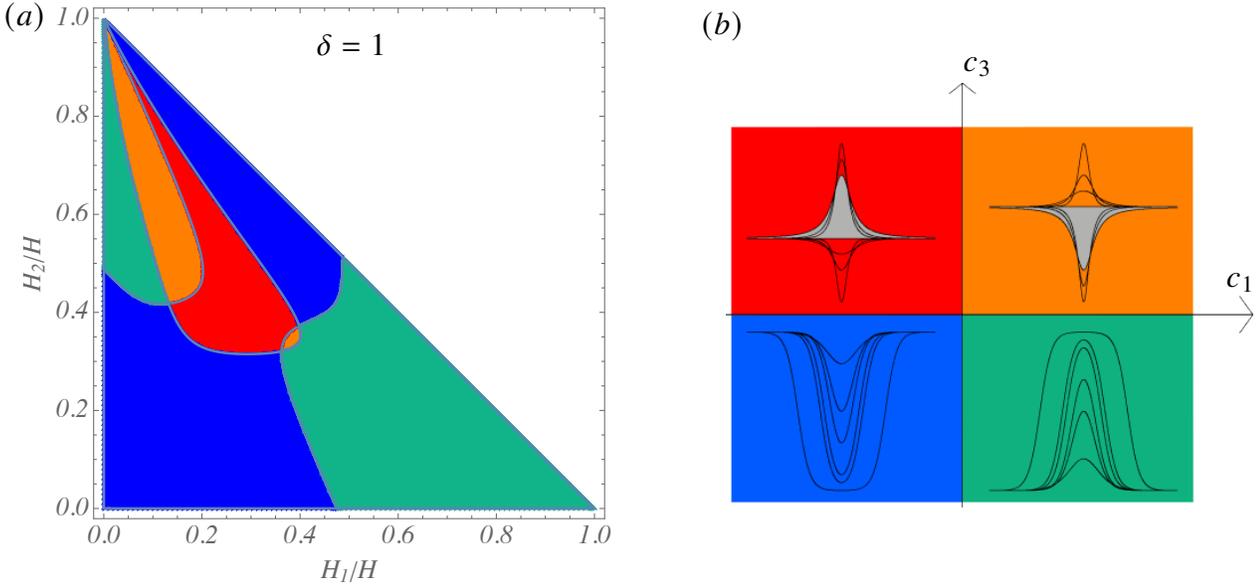

	As expected, the coefficients $c_1$, $c_2$ are the same as for the KdV equation, given by \eqref{coef_c1} and \eqref{coef_c2}, respectively. The cubic nonlinearity coefficient $c_3$ is found (see Appendix A) as: 
	\begin{equation}\label{cubic_non_coef}
		c_3= \frac{T_1+T_2+T_3}{\frac{\rho_3}{H_3}\gamma^2+\frac{\rho_2}{H_2}(1-\gamma)^2+\frac{\rho_1}{H_1}},
	\end{equation}
	with $T_1$, $T_2$, $T_3$ given by:
	\begin{multline}\label{def_T1}
		T_1=-\frac{1}{4} c_0 \Bigg[ 9 \frac{H_2}{\rho_2}\left( \frac{\rho_3}{H_3^2}\gamma^2-\frac{\rho_2}{H_2^2}(1-\gamma)^2\right)\left(\frac{\rho_2}{H_2^2}(1-\gamma)^2-\frac{\rho_1}{H_1^2}\right)+\\
		+4\left( \frac{\rho_3}{H_3^3}\gamma^4+\frac{\rho_2}{H_2^3}(1-\gamma)^4+\frac{\rho_1}{H_1^3}\right)\Bigg],
	\end{multline}
	
	\begin{multline}
		T_2=-\frac{1}{6} \frac{c_1}{\rho_2} \Bigg[ 10 \frac{\rho_2^2}{H_2^2}(1-\gamma)^3+\frac{3\rho_1}{H_1} \frac{\rho_3}{H_3} \left( \frac{2 H_2}{H_1}-3\frac{H_2}{H_3}\gamma \right) \,\gamma + 3\frac{\rho_2}{H_2}(1-\gamma)^2 \left( 3 \frac{\rho_1}{H_1}-2\frac{\rho_3}{H_3}\gamma \right) + \\
		+\frac{\rho_1}{H_1}\frac{\rho_2}{H_1} (-1+6\gamma)+ \frac{\rho_2}{H_3} \frac{\rho_3}{H_3} \gamma^2 \,(-9+4\gamma)\Bigg],
	\end{multline}
	\begin{equation}
		T_3=-\frac{1}{6}\frac{c_1^2}{c_0} \Bigg[ -3\left( \frac{\rho_3}{H_3}\gamma^2+\frac{\rho_2}{H_2}(1-\gamma)^2+\frac{\rho_1}{H_1}\right) + \frac{4 \gamma}{\rho_2 H_1 H_3} (\rho_1 \rho_2 H_3+\rho_1 \rho_3 H_2 +\rho_2 \rho_3 H_1)\Bigg].
	\end{equation}
	As before, to recover the expression of the Gardner coefficients under the Boussinesq approximation (see \citeauthor{kurkina_et_al} \citeyear{kurkina_et_al}), it suffices to set $\rho_1 = \rho_2 = \rho_3 = \rho_0$, along with $c_0$ defined implicitly by \eqref{c0_Boussinesq}, and consider $\gamma$ defined by \eqref{gamma_def_Boussinesq}. Here, 
	the cubic nonlinearity coefficient $c_3$ can change sign depending on the physical parameters considered. Given that the coefficient $c_2$ of the linear dispersive term $\zeta_{1,xxx}$ is always positive, it is well known that when $c_3>0$ waves of either polarity exist with no maximum amplitude. Minimum wave amplitudes exist for waves of elevation (depression) for $c_1<0$ ($>0$). Specifically, these have a minimum amplitude of $-2c_1/(3 c_3)$.
	On the other hand, when $c_3<0$ a single branch of solutions exist with the same polarity as the one predicted by the KdV equation, with the important difference that such solutions now broaden with increasing wave speeds (or amplitude), until a front is reached (see {\it e.g.} \citeauthor{grimshaw_et_al_99} \citeyear{grimshaw_et_al_99}). A schematic illustration of all solution behaviours is given in figure~\ref{Gardner_predictions}$(b)$ following \cite{grimshaw_et_al_99}.
	
\begin{figure}
		\begin{center}
			\begin{overpic}[width=150pt]{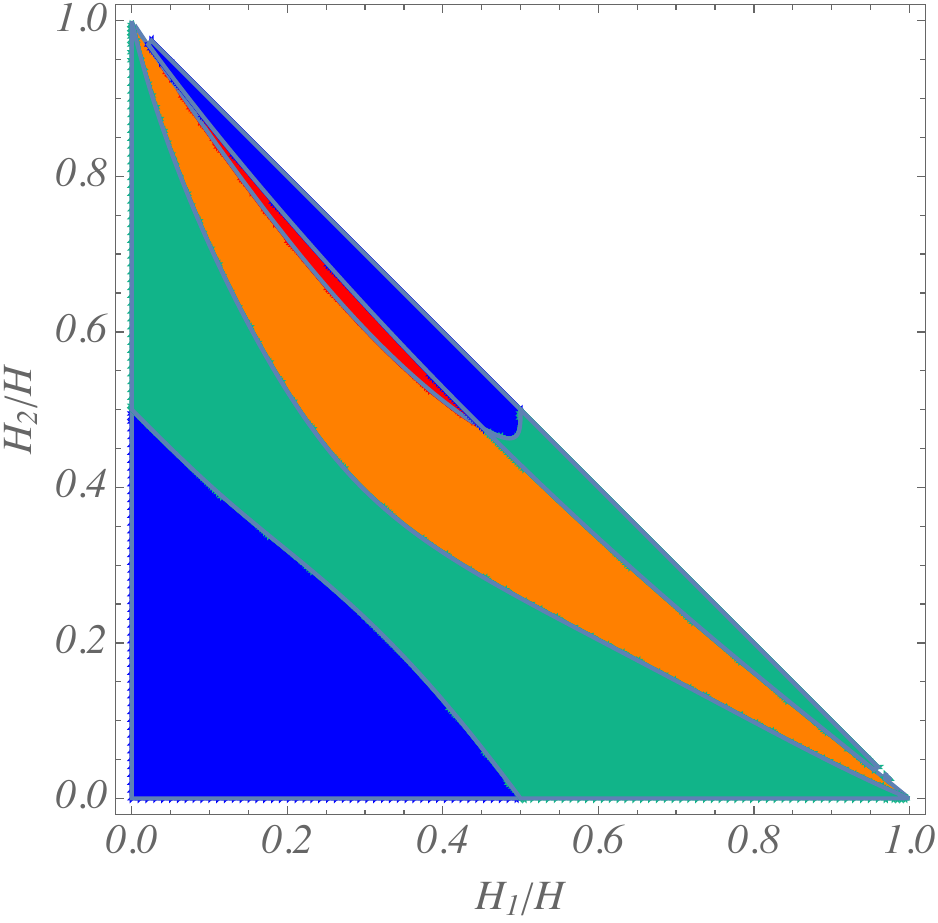}
				\put(45,90){$\delta=0.5$}
				\put(-5,93){$(a)$}
			\end{overpic}
			\,\, 
			\begin{overpic}[width=150pt]{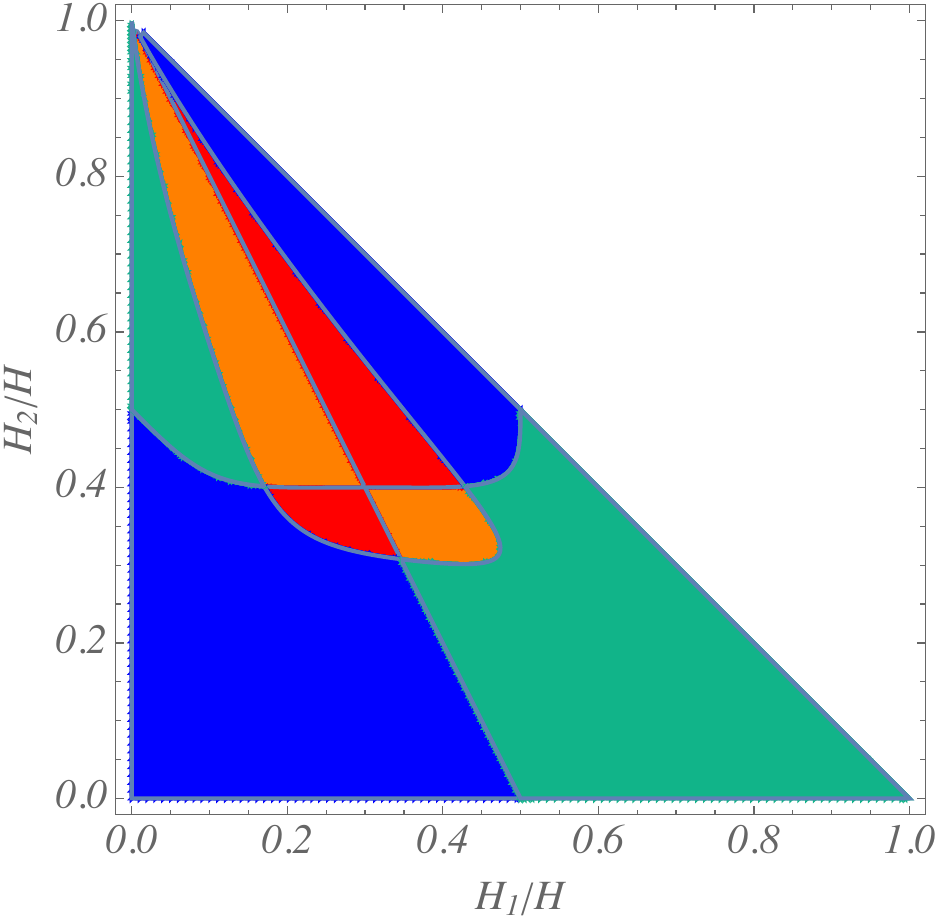}
				\put(45,90){$\delta=1$}
				\put(-5,93){$(b)$}
			\end{overpic}
			\,\, 
			\begin{overpic}[width=150pt]{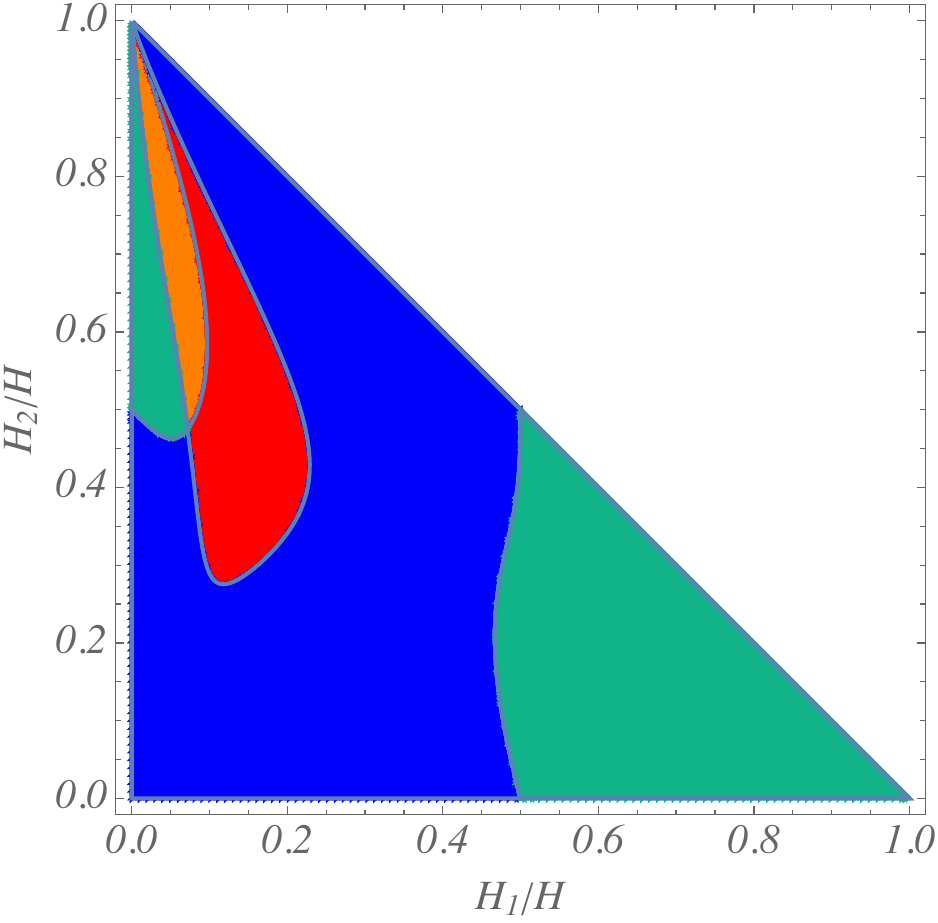}
				\put(45,90){$\delta=2$}
				\put(-5,93){$(c)$}
			\end{overpic}
		\end{center}
		\caption{Behaviour of the quadratic and cubic nonlinearity coefficients ($c_1$, $c_3$, respectively) of the Gardner equation \eqref{Gardner_3layer_dim} for a three-layer fluid, when restricted to mode-1 waves under the Boussinesq approximation. 
			Different values of $\delta$ are considered, and the same colour scheme as for figure~\ref{Gardner_predictions} is adopted. 
			\label{Gardner_predictions_Boussinesq}
		}
	\end{figure}

	\cite{kurkina_et_al} have examined in great detail how the quadratic and cubic nonlinearity coefficients can vary with the different physical parameters considered in a three-layer fluid, under the Boussinesq approximation. In particular, it was highlighted through extensive numerical tests that the coefficient $c_3$ could never change sign for mode-2 waves, and remained negative. As a consequence, Gardner theory predicts that the coexistence of ISWs of opposite polarity can only occur for mode-1 waves. As in \cite{kurkina_et_al}, when graphically presenting the parameter space for different solutions, we may fix the density increments $\Delta_i$ and show the result of varying the undisturbed depths $H_i$. The domain of physically permissible parameters is then a triangle with vertices $(0,0)$, $(0,1)$, and $(1,0)$ in the $(H_1/H,H_2/H)$-space. We show in figure \ref{Gardner_predictions} how the sign for the quadratic and cubic nonlinearity coefficients changes with $\Delta_1=\Delta_2=0.1$. If we consider $\Delta_1=\delta \Delta_2$, with $\delta\neq 1$ and $\Delta_2=0.1$, and draw similar pictures, we observe that, despite the somewhat large density contrasts considered, the results obtained are in good agreement with those obtained by \cite{kurkina_et_al}, reproduced here in figure~\ref{Gardner_predictions_Boussinesq}.
	Notice that when the Boussinesq approximation is considered, sweeping the parameter space becomes easier, since it is defined only by the three parameters $\delta$, $H_1/H$, and $H_2/H$.
	We can infer from the figures that given fixed densities, coexistence of solutions of opposite polarity can be predicted within a limited range of values of $H_2/H$, or none at all, depending on the ratio between the undisturbed thicknesses of the top and bottom layer (see figure~\ref{Gardner_along_lines}).

	\begin{figure}
		\begin{center}
			\begin{overpic}[width=150pt]{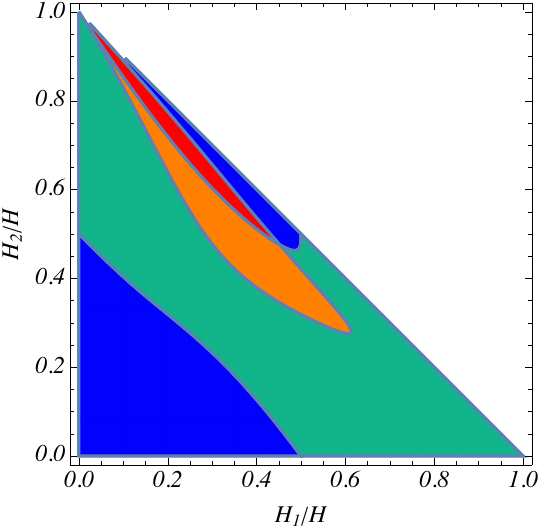}
				\put(45,90){$\delta=0.5$}
				\put(-5,93){$(a)$}
			\end{overpic}
			\,\, 
			\begin{overpic}[width=150pt]{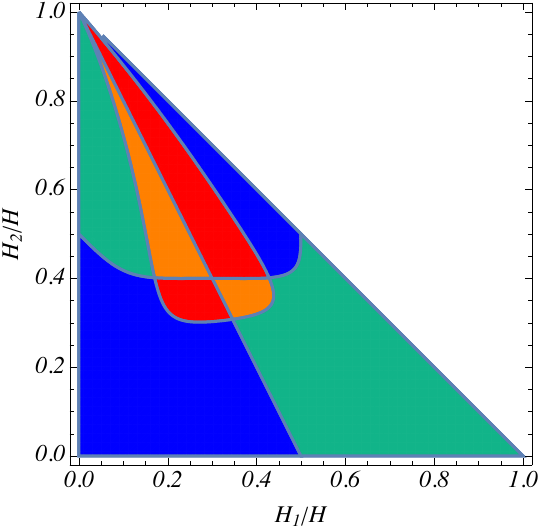}
				\put(45,90){$\delta=1$}
				\put(-5,93){$(b)$}
			\end{overpic}
			\,\, 
			\begin{overpic}[width=150pt]{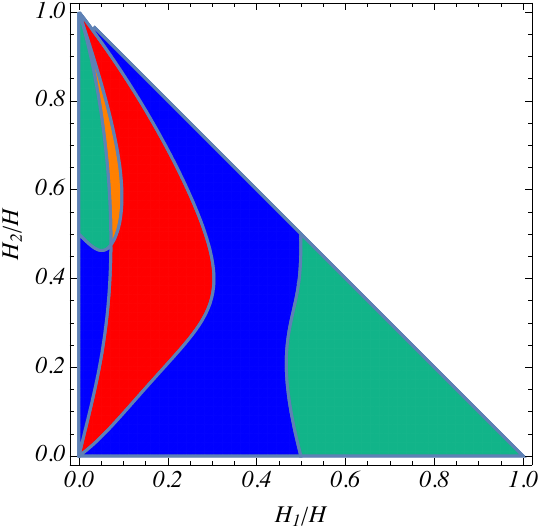}
				\put(45,90){$\delta=2$}
				\put(-5,93){$(c)$}
			\end{overpic}
		\end{center}
		\caption{Behaviour of the quadratic and cubic nonlinearity coefficients ($\tilde{c}_1$, $\tilde{c}_3$, respectively) of the Gardner equation \eqref{Gardner_zeta2} for the lower interface displacement $\zeta_2(x,t)$, when restricted to mode-1 waves under the Boussinesq approximation. 
			Different values of $\delta$ are considered, and the colour scheme remains consistent with that of figure~\ref{Gardner_predictions}; however, the coordinate axes have been redefined.
			\label{Gardner_predictions_Boussinesq_zeta2}
		}
	\end{figure}
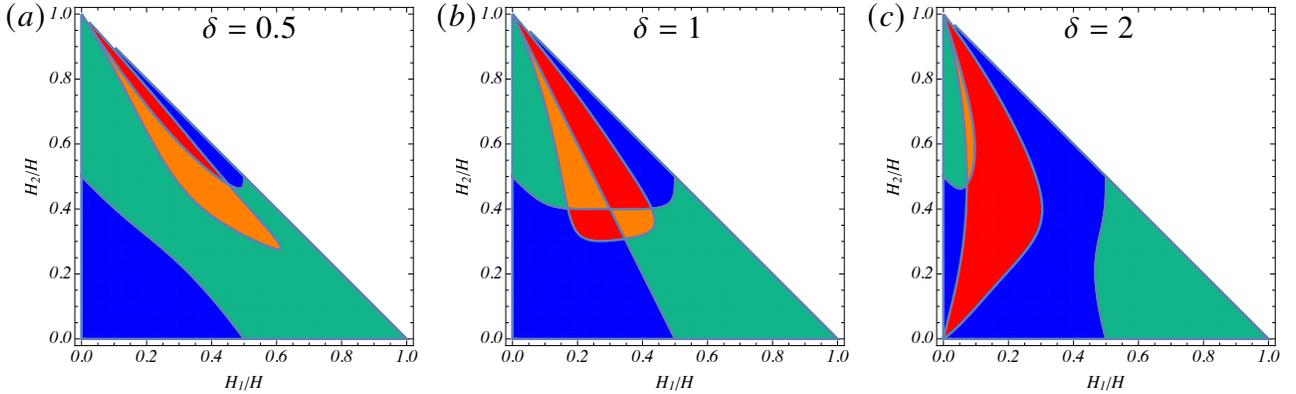

	We remark that, alternatively, a Gardner equation could have been written for the lower interface displacement $\zeta_2(x,t)$: 
	\begin{equation}\label{Gardner_zeta2}
		\zeta_{2,t} + c_0 \,\zeta_{2,x} + \tilde{c}_1 \,\zeta_2 \,\zeta_{2,x} + c_2 \,\zeta_{2,xxx} + \tilde{c}_3 (\zeta_2^3)_x=0.
	\end{equation}
	As shown in \cite{kurkina_et_al}, under the Boussinesq approximation, the following relationship holds:
	\begin{equation}
		c_3(H_3,H_2,H_1,1/\delta) = \tilde{c}_3(H_1,H_2,H_3,\delta),
	\end{equation}
	which shows that the Gardner theory does preserve the upside-down symmetry of the Euler equations. Consequently, figure~\ref{Gardner_predictions_Boussinesq}$(c)$ can be transformed into figure~\ref{Gardner_predictions_Boussinesq_zeta2}$(a)$ using the mapping
	$(x,y)\mapsto(1-x-y,y)$, with the caveat of a reversed colour scheme (green replacing blue, and orange replacing red), as we would be describing the behaviour of $-\zeta_2(x,t)$. What is intriguing is that figure~\ref{Gardner_predictions_Boussinesq_zeta2}$(a)$ and figure~\ref{Gardner_predictions_Boussinesq}$(a)$ exhibit clear discrepancies, suggesting that the predicted solution behaviour may depend on which form of the Gardner equation, \eqref{Gardner_3layer_dim} or \eqref{Gardner_zeta2}, is used {\it cf.} \cite{kurkina_et_al}
	Similar considerations apply to panels $(a)$ and $(b)$ of figure~\ref{Gardner_predictions_Boussinesq}. 
	
	\begin{figure}
		\begin{center}
			\includegraphics*[width=8cm]{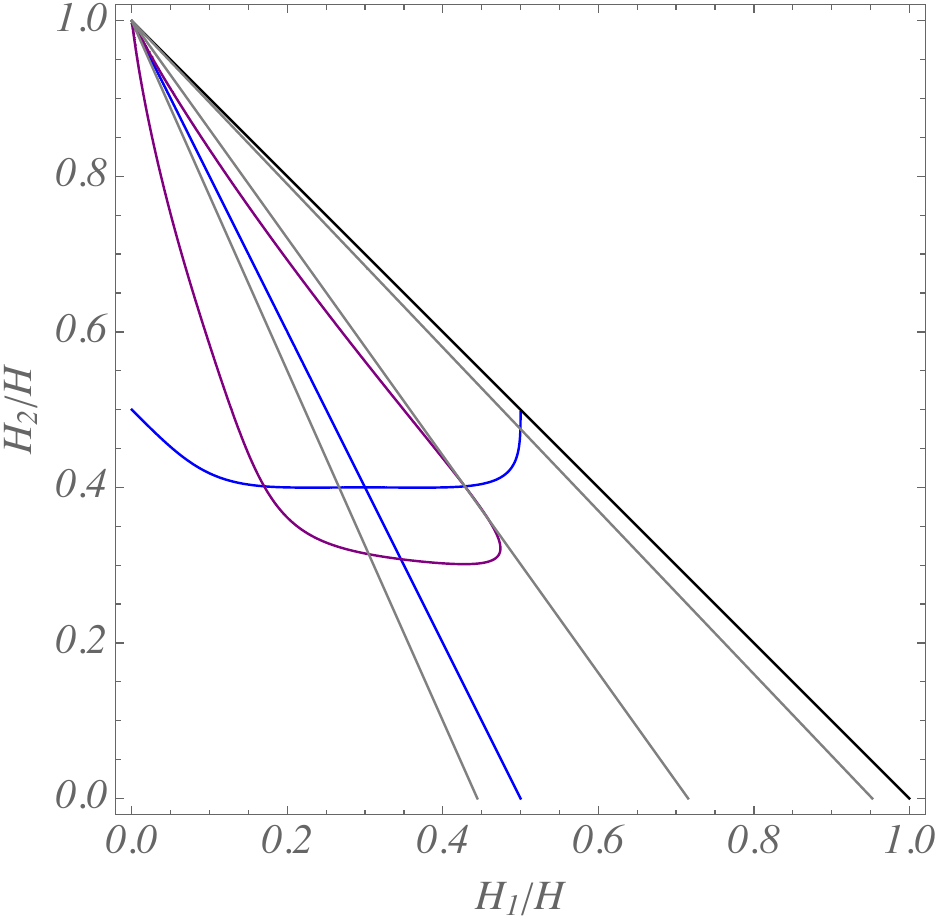}
		\end{center}
		\caption{Vanishing of the quadratic (in blue) and cubic (in purple) nonlinearity coefficients ($c_1$, $c_3$, respectively) of the Gardner equation \eqref{Gardner_3layer_dim} for a three-layer fluid, when restricted to mode-1 waves under Boussinesq approximation. The parameter $\delta$ is set to 1, as in figure~\ref{Gardner_predictions_Boussinesq}($b$). The grey lines from the top vertex to the base of the triangle correspond to different ratios $H_1/H_3$. Note that a fixed ratio $H_1/H_3=m$ determines the line with equation $H_2/H = -(1+m^{-1} ) H_1/H +1$ in the parameter space. From left to right we have: $H_1/H_3=0.8, \, \approx 2.5155, \,20$. For $H_1/H_3>2.5155$, according to the Gardner theory no ISWs of opposite polarity can coexist. This will be shown to be in disagreement with the predictions by the MMCC3 and Euler theories (e.g. see figure~\ref{Regions_coexistence}$(a)$).
			\label{Gardner_along_lines}
		}
	\end{figure}

	\section{Highlights of the main results}\label{sec:main_results}
	
	\begin{figure}
		\centering
		\includegraphics[scale=.9]{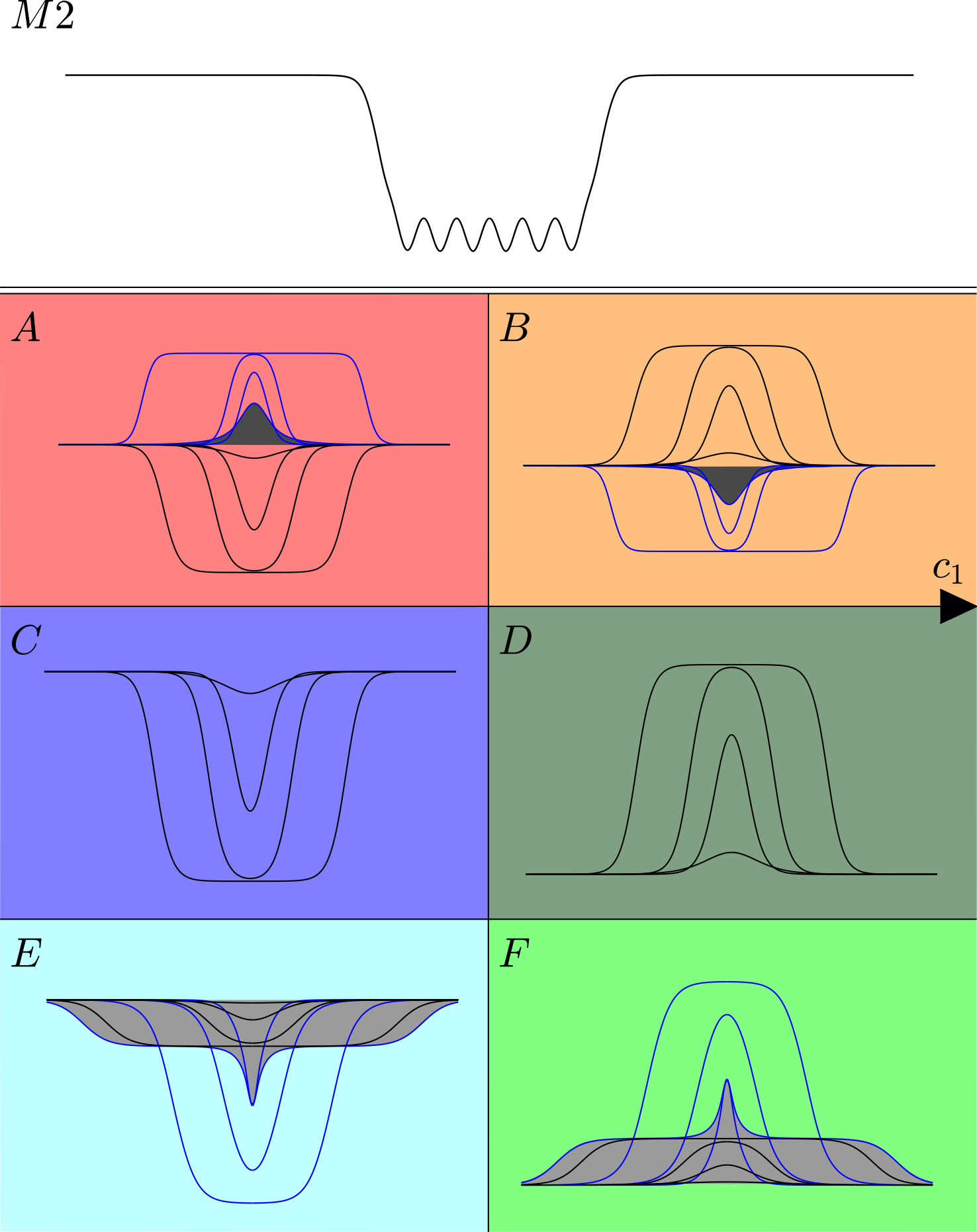}
		\caption{Possible solution spaces for solitary waves with speeds $c>c_0^+$. Here, all solution branches are limited in amplitude. For any set of parameters, the bifurcation of mode-1 waves for the Euler and MMCC3 systems is described by one the of the panels $A-F$. Furthermore, for $H_2$ sufficiently large, one finds mode-2 solitary waves, with a bifurcation structure described in Doak \textit{et al.} (2022), represented here by panel $M2$. When applicable, shaded regions represent a minimum amplitude along a solution branch. Panels A and B show two branches of solitary waves of opposite polarity, both evolving into tabletop solitary waves. Panel $B$ has a finite amplitude bifurcation for the depression branch, while the elevation branch bifurcates from zero amplitude. The opposite is true for panel $A$. Panels $C$ and $D$ show a single solitary wave branch of depression and elevation respectively, developing into a tabletop solitary wave. Panels $E$ and $F$ are new bifurcation structures. There exist two branches, both of depression (panel $E$, $c_1<0$) or elevation (panel $F$, $c_1>0$). One branch bifurcates from zero amplitude and limits to a tabletop solitary wave. The second branch exists with speeds strictly greater than the first branch. The branch starts at the same limiting tabletop solitary wave as the first branch but with an additional bump along the broadened section. This branch then evolves into 
			a different tabletop solitary wave. These new branches are discussed in more detail in \S \ref{sec:numerics}. 
			\label{fig:bif_space}}
	\end{figure}
	
	Having presented the 3-layer models of KdV, mKdV and Gardner in their generality, we outline here the key findings towards a comprehensive classification of mode-1 ISWs in a three-layer fluid with the Euler and MMCC3 theories. Figure~\ref{fig:bif_space} provides a schematic representation of the full range of solutions for this physical configuration. Symmetry about the wave crest is assumed in all our numerical investigations, and the 
	qualitative description applies to all the solutions with speeds $c>c_0^+$ found within both the strongly and fully nonlinear theories.
	We show only the upper interface $\zeta_1$, as for each solution the lower interface has a comparable shape, with the exception of panel $M2$. For any set of parameters the solution space for mode-1 waves will behave like one of the panels $A-F$ (excluding the case when there are no mode-1 solitary waves). The panels $C$ and $D$ are the simplest, each featuring a single branch, which can be of elevation (panel $D$) or depression (panel $C$). Both branches emerge from zero amplitude at the linear long-wave speed $c = c_0^+$ and evolve into tabletop solitary waves as the speed increases, ultimately approaching a front.
	Such behaviour is observed for mode-1 solitary waves in a two-layer fluid (see \citeauthor{choi_camassa_99} \citeyear{choi_camassa_99} and \citeauthor{evans_ford} \citeyear{evans_ford}, for the strongly and fully nonlinear theories, respectively) and the polarity is the same as the KdV polarity, determined by the sign of the quadratic nonlinearity coefficient. In panels $A$ and $B$, we see two branches, both bifurcating from the linear long wave speed. However, in panel $A$, the elevation branch bifurcates from finite amplitude (the smallest amplitude wave being shown by the shaded region, and the solutions along this branch are shown in blue), while the depression branch bifurcates from zero amplitude. The situation is similar in panel $B$, but with the roles of the elevation and depression branch reversed. Similarly to the Gardner theory, the branches of solutions bifurcating from zero amplitude are those with the KdV polarity.

	The behaviour shown in panels $A-D$ share several characteristics with the mode-1 solutions of the Gardner equation, as illustrated in figure~\ref{Gardner_predictions}$(b)$. Although the sign for the quadratic nonlinearity coefficient $c_1$ remains crucial in distinguishing the different solution properties within the strongly and fully nonlinear theories, the sign for the cubic nonlinearity coefficient $c_3$ does not. Furthermore, all branches of solutions have tabletop solitary waves ending in fronts, in contrast with the Gardner theory, where multiple solitary branches - when present - do not have a limiting amplitude. 
	We also note that the behaviour observed in panels $A$--$D$ was also reported in a continuous double-pycnocline stratification by \cite{lamb_2023}.
	
	Panels $E$ and $F$ exhibit novel solution space behaviour with no direct counterpart in Gardner theory. In these panels, two branches of the same polarity are present, both of depression in panel $E$ and both of elevation in panel $F$. However, there is no true multiplicity of solutions: one branch emerges from zero amplitude and gradually approaches a limiting wavefront, while the other bifurcates at a finite amplitude with higher propagation speeds than the first branch. A more detailed discussion of these solution branches is provided in section \ref{sec:numerics}.
	
	Finally, the panel $M2$ shows mode-2 solitary waves with speeds exceeding the linear long wave speed of mode-1 $c_0^+$. Such solutions were found to exist in the strongly nonlinear theory by \cite{barros_choi_milewski}, and were explored in more detail in both the strongly nonlinear and full Euler theories by \cite{doak_et_al}. For sufficiently large $H_2$, these solutions also exist, in conjunction with mode-1 solutions with behaviour from panels $A$ or $B$.

	The results in this paper primarily stem from a conjugate state analysis for the Euler equations. According to \cite{benjamin_66}, two horizontally uniform flows are considered {\it conjugate} if all three basic physical conservation laws of mass, momentum and energy are satisfied. Remarkably, the conjugate states of the Euler equations remain fully preserved within the MMCC3 theory, for which the study of steady solutions is significantly simpler.
	Within this reduced framework, it will be seen how the critical point analysis can be used to explain different and new bifurcation structures in the solutions, which are then validated using the Euler equations. A glimpse of the results is given in figures \ref{Parameter_regions}, \ref{Parameter_regions2}. All technical details are left to subsequent sections \S 4, \S 5, \S 6.
	
	\subsection{Solution space and conjugate states}\label{sec:sols_and_conjugate_sates}
	
\begin{figure}
		\begin{minipage}{.47\textwidth}
			\centering
			\begin{overpic}[width=.9\textwidth]{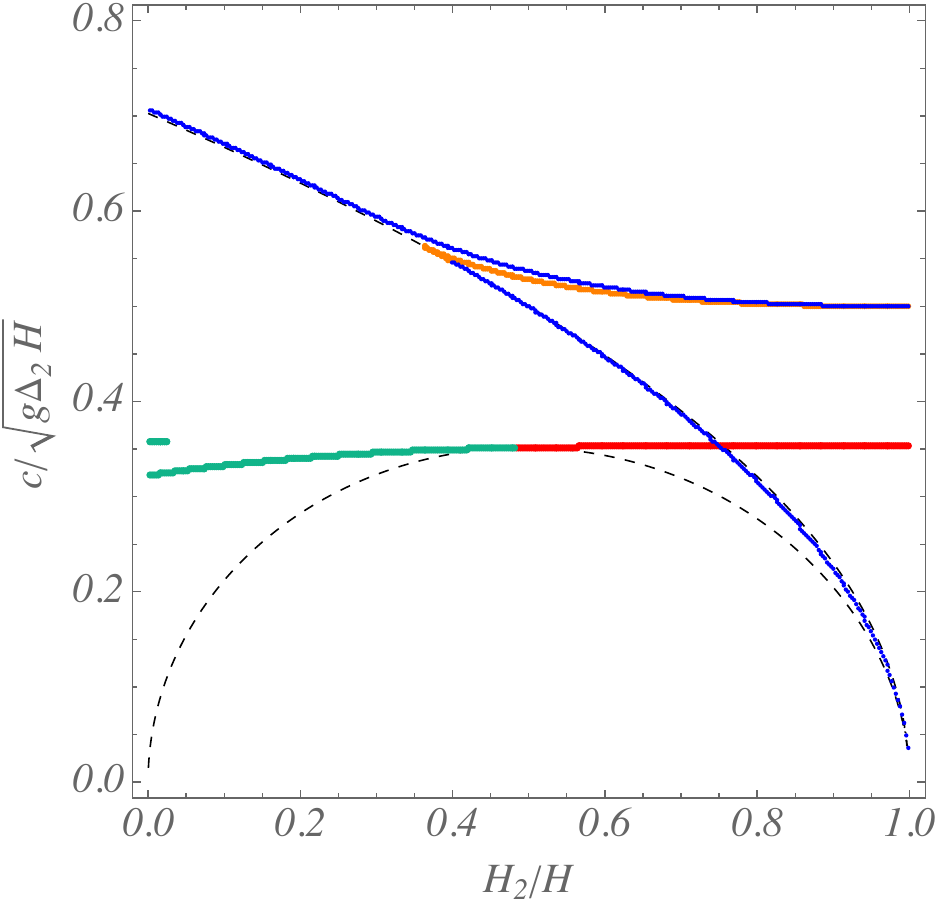}
				\put(-3,92){$(a)$}
			\end{overpic}
		\end{minipage}
		\begin{minipage}{.05\textwidth}
		\end{minipage}
		\begin{minipage}{.47\textwidth}
			\centering
			\begin{overpic}[width=0.9\textwidth]{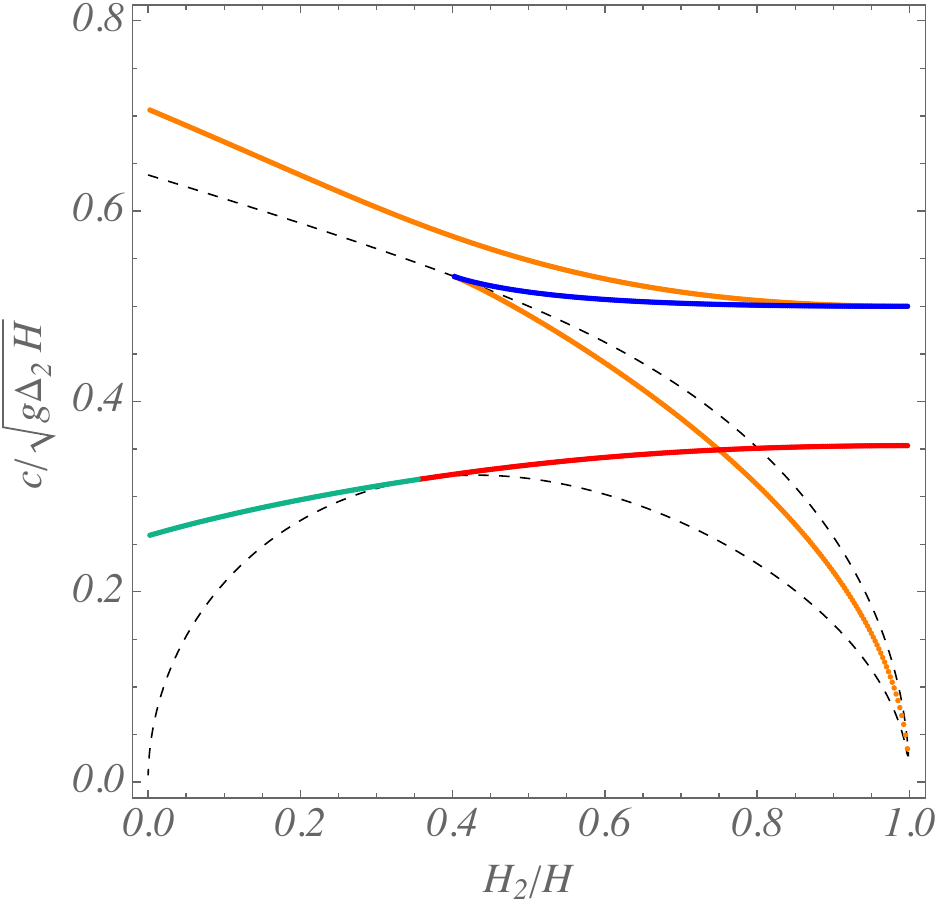}
				
				\put(-5,92){$(b)$}
			\end{overpic}
		\end{minipage}

		\begin{minipage}{.47\textwidth}
			\centering
			\begin{overpic}[width=.9\textwidth]{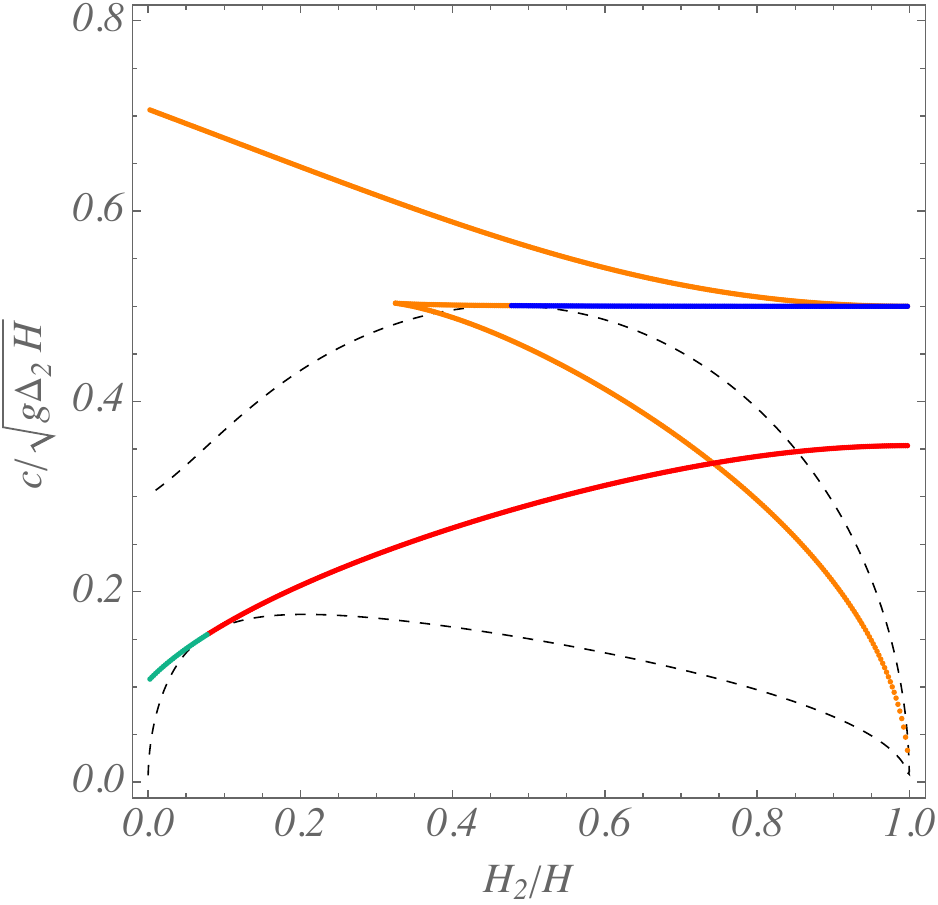}
				\put(-3,92){$(c)$}
			\end{overpic}
		\end{minipage}
		\begin{minipage}{.05\textwidth}
		\end{minipage}
		\begin{minipage}{.47\textwidth}
			\centering
			\begin{overpic}[width=0.9\textwidth]{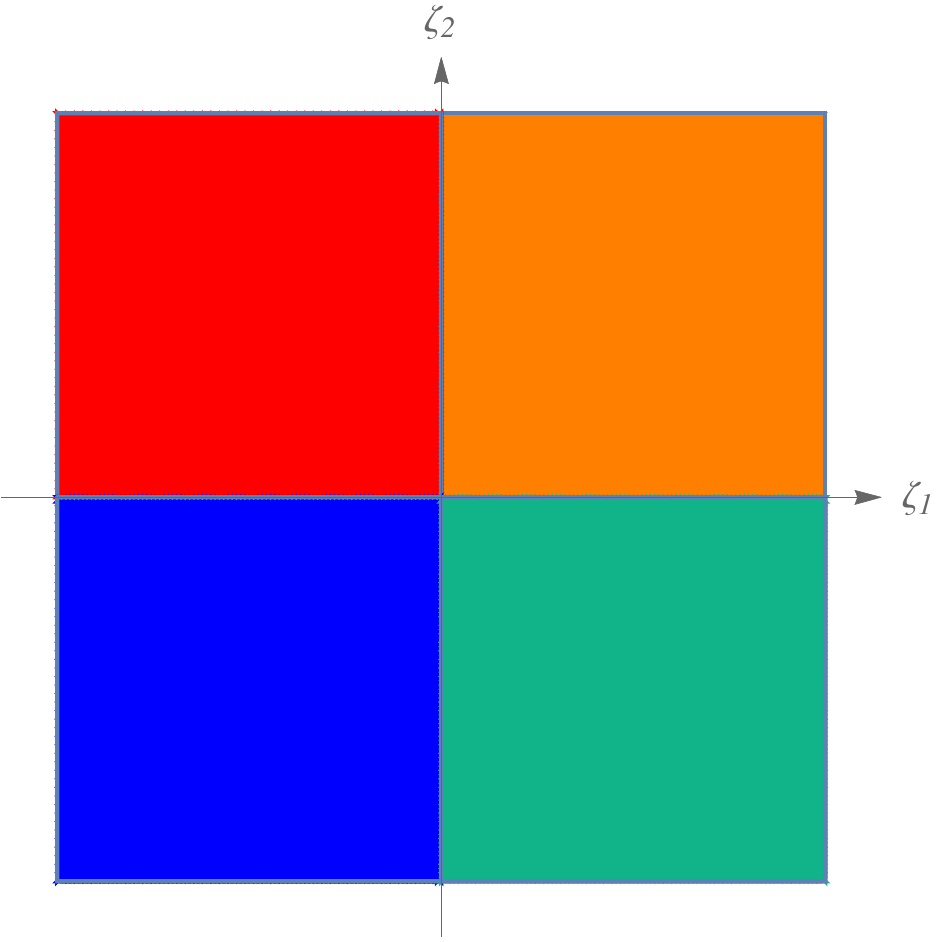}
				\put(-5,92){$(d)$}
			\end{overpic}
		\end{minipage}
		\caption{Solutions to the conjugate state equations \eqref{conjugate_states}, under Boussinesq approximation. In each panel, we plot the branches of conjugate flow speeds that exist for a fixed ratio $H_1/H_3$ and varying $H_2/H$. The physical parameters here correspond to those in figure~\ref{Gardner_along_lines}. Panels $(a)$–$(c)$ show results for increasing values of $H_1/H_3$. The colour scheme in panel $(d)$ allows to identify the quadrant of the $(\zeta_1,\zeta_2)$-plane to which the corresponding conjugate state belongs to. The black dashed curves are the linear long-wave speeds.
			\label{Front_curves}
		}
	\end{figure}

	The MMCC3 model 
	has recently been examined by \cite{barros_choi_milewski} and \cite{doak_et_al}. Notably, \cite{barros_choi_milewski} have shown that ISWs are governed by a Hamiltonian system with two degrees of freedom, for which the potential $V=V(\zeta_1,\zeta_2)$ is a rational function (see \eqref{potential_def}). Furthermore, \cite{doak_et_al} established that the solutions of  
	\begin{equation}\label{conjugate_states}
		V=0, \quad \fpar{V}{\zeta_1}=0, \quad \fpar{V}{\zeta_2}=0,
	\end{equation}
	correspond to uniform flows to the Euler equations, which are {\it conjugate}. 
	This problem was first addressed by \cite{lamb_2000}, who arrived to a set of three algebraic equations - $(12a)$, $(12b)$, $(14)$ therein -  equivalent to our system \eqref{conjugate_states}. To solve the system, \cite{lamb_2000} employed a geometric approach. However, a more insightful procedure involves plotting branches of conjugate state solutions of \eqref{conjugate_states} across different parameter regimes (see {\it e.g.} figure 5 in \citeauthor{doak_et_al} \citeyear{doak_et_al}), following the work of \cite{dias_ilichev} and \cite{barros}. This approach is particularly useful under the Boussinesq approximation, as it allows us to analyse how the number and speed of conjugate state solutions change with increasing values of $H_2/H$ for fixed $\delta$ and $H_1/H_3$, as illustrated in figure~\ref{Front_curves}. 
	The black dashed lines represent the curves $c=c_0^\pm$, corresponding to the linear long wave speeds, and for small values of $H_2/H$, a single branch of solutions with $c>c_0^+$ is present. For thicker intermediate layers, two or even three such branches can be perceived. 
	The branches are colour-coded based on the quadrant of the $(\zeta_1,\zeta_2)$-plane to which the corresponding conjugate solutions belong. Orange and blue will typically correspond to solutions of mode-1, and red and green to solutions of mode-2. However, it can be observed from the figure that solutions sought in odd quadrants do exist with speeds less than $c_0^+$, as well as solutions sought in even quadrants for speeds larger than $c_0^+$ (see red branch for large values of $H_2/H$).

	A method has recently been proposed by \cite{lamb_2023} to determine where in the parameter space ISWs of both polarities exist in a continuous double-pycnocline stratification. The same method can equally be applied to our piecewise-constant stratification, and consists of identifying the physical parameters for which a blue and orange branch coexist with $c>c_0^+$. According to the author ``{\it This method is somewhat tedious as it requires the determination of where the conjugate flow speed is equal to the linear long-wave propagation speed}. 
	Our analysis will confirm that the onset of the desired feature is indeed predicted by instances where the conjugate flow speed matches the linear long wave speed $c=c_0^+$. However, it will also become evident that not all such instances lead to the expected feature (see figure~\ref{front_curves_all_details}). Furthermore, it will be shown that the conjugate state analysis extends beyond this point, and previously unexplored aspects of the conjugate flow speed curves, which for brevity we will refer to as the conjugate curves, will enable the discovery of new types of solution branches {\it cf.} figure~\ref{fig:bif_space}).

\begin{figure}
		\begin{minipage}{.47\textwidth}
			\centering
			\begin{overpic}[width=7cm]{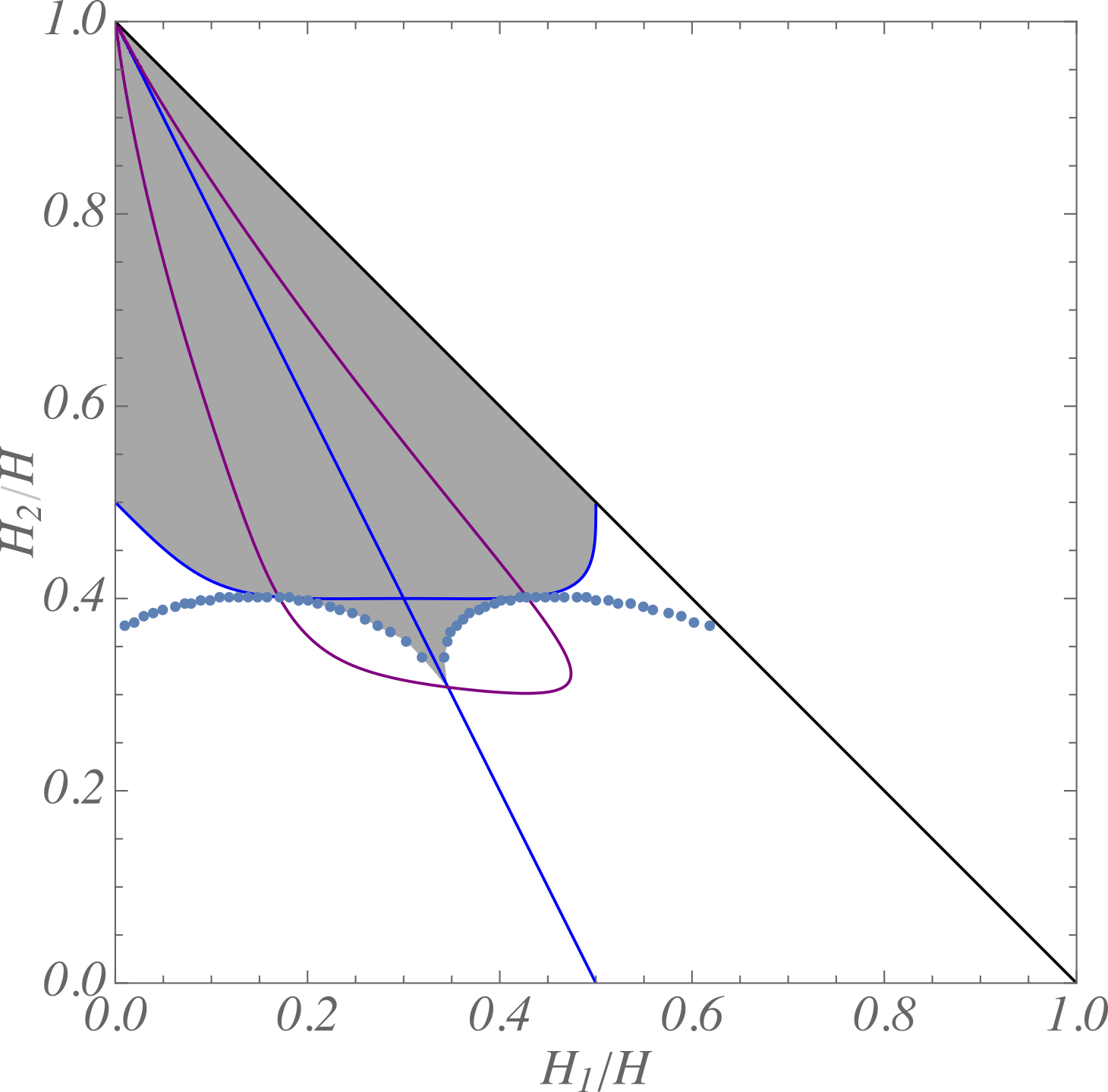}
				\put(45,90){$\delta=1$}
				\put(-5,94){$(a)$}
			\end{overpic}
		\end{minipage}
		\begin{minipage}{.05\textwidth}
		\end{minipage}
		\begin{minipage}{.47\textwidth}
			\begin{overpic}[width=7cm]{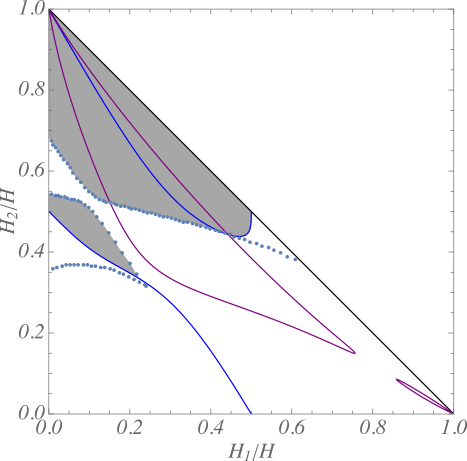}
				\put(45,90){$\delta=0.753$}
				\put(-5,94){$(b)$}
			\end{overpic}
		\end{minipage}
		\caption{Region (shaded) of coexistence of ISWs of opposite polarity for different values of $\delta$, according to the strongly and fully nonlinear theories, under the Boussinesq approximation. Instances where the conjugate flow speed matches the linear long-wave speed $c=c_0^+$ are represented in blue (dots for intersections, and solid line for tangency points). It can be observed that that the set of tangency points coincides with the locus corresponding to the {\it criticality} condition \eqref{criticality_condition} for the KdV equation. The locus in purple corresponds to the vanishing of the cubic nonlinearity coefficient for the Gardner equation and is shown here for comparison. 
			\label{Regions_coexistence}
		}
	\end{figure}

\begin{figure}
		\centering
		\begin{overpic}[scale=1.2]{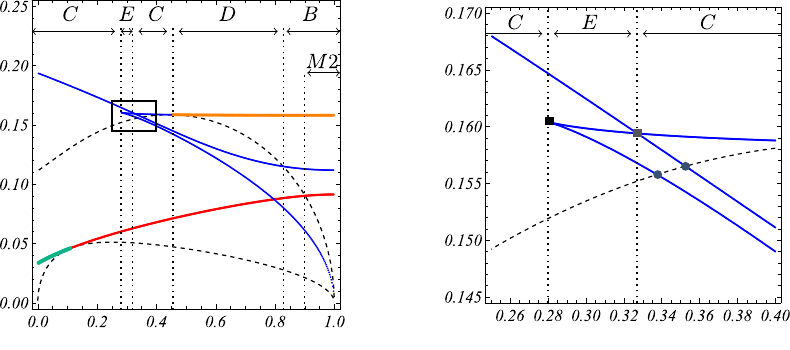}
			\put(-4,41){$(a)$}
			\put(52,41){$(b)$}
			\put(20,0){$H_2/H$}
			\put(77,0){$H_2/H$}
			\put(-4,20){\rotatebox{90}{$c/\sqrt{g \Delta_2 H}$}}
			\put(52,20){\rotatebox{90}{$c/\sqrt{g \Delta_2 H}$}}
		\end{overpic}
		\caption{Solutions to the conjugate state equations \eqref{conjugate_states}, under Boussinesq approximation, for $\delta=0.5$ and $H_1/H_3=0.1$, using the same colour scheme as figure \ref{Front_curves}. The figure also includes vertical dotten curves, which deliminate the various solution space behaviours given by panels $A$-$F$ and $M2$ in figure \ref{fig:bif_space}. Panel $(b)$ shows a blow-up of the region marked by a box in panel $(a)$. In panel $(b)$, we highlight the appearance of two additional conjugate state branches (a maximum and a slower saddle) with $c>c_0^+$ with a black square, the self-intersection of two mode-1 conjugate states (both maxima) with $c>c_0^+$ with a grey square, and mode-1 conjugate states passing through $c=c_0^+$ with a blue circle. \label{front_curves_all_details}}
	\end{figure}

	In figure~\ref{Front_curves}$(c)$, the tangency between the conjugate curve and the curve $c=c_0^+$ is what sets the coexistence of the two branches of opposite polarity (orange and blue) with $c>c_0^+$. In contrast, in figure~\ref{Front_curves}$(a)$, it is the instance when the orange branch intersects the curve $c=c_0^+$ that sets such occurrence. Although not particularly visible in the figure, this intersection precedes a tangency between the orange branch and the curve $c=c_0^+$ (at $H_2/H\approx 0.398$). By preserving the same value $\delta=1$ and varying the ratio $H_1/H_3$, similar figures can be obtained. By doing so, we can cover the entire parameter space as in figure~\ref{Regions_coexistence}$(a)$ and highlight the regions where conjugate curves of opposite polarities exist for $c>c_0^+$, simply by observing the intersections (blue dots) and tangency points (blue solid line) between the conjugate curve and the curve $c=c_0^+$. 
	This naturally raises the question of how this relates to the coexistence of ISWs with opposite polarities. Our extensive numerical tests confirm that when conjugate states of opposite polarities coexist with $c>c_0^+$, wave fronts can be exhibited for each conjugate state (in both the strongly and fully nonlinear theories), ultimately limiting the amplitude of solitary-wave solution branches. This further supports the validity of the criterion proposed by \cite{lamb_2023}. An additional justification for its validity will be presented in the following sections within the framework of strongly nonlinear theory.  
	
	In figure~\ref{Regions_coexistence}, we overlay the locus (in purple) where the cubic nonlinearity of the Gardner equation vanishes, allowing for a comparison with the weakly nonlinear theory. According to the Gardner theory, ISWs of opposite polarity coexist within each closed region enclosed by a purple curve. For
	$\delta=1$, some discrepancies can be observed in panel $(a)$. However, these differences become even more pronounced as 
	$\delta$ varies (see panel $(b)$). 
	
\begin{figure}
		\begin{minipage}{.47\textwidth}
			\centering
			\begin{overpic}[width=7cm]{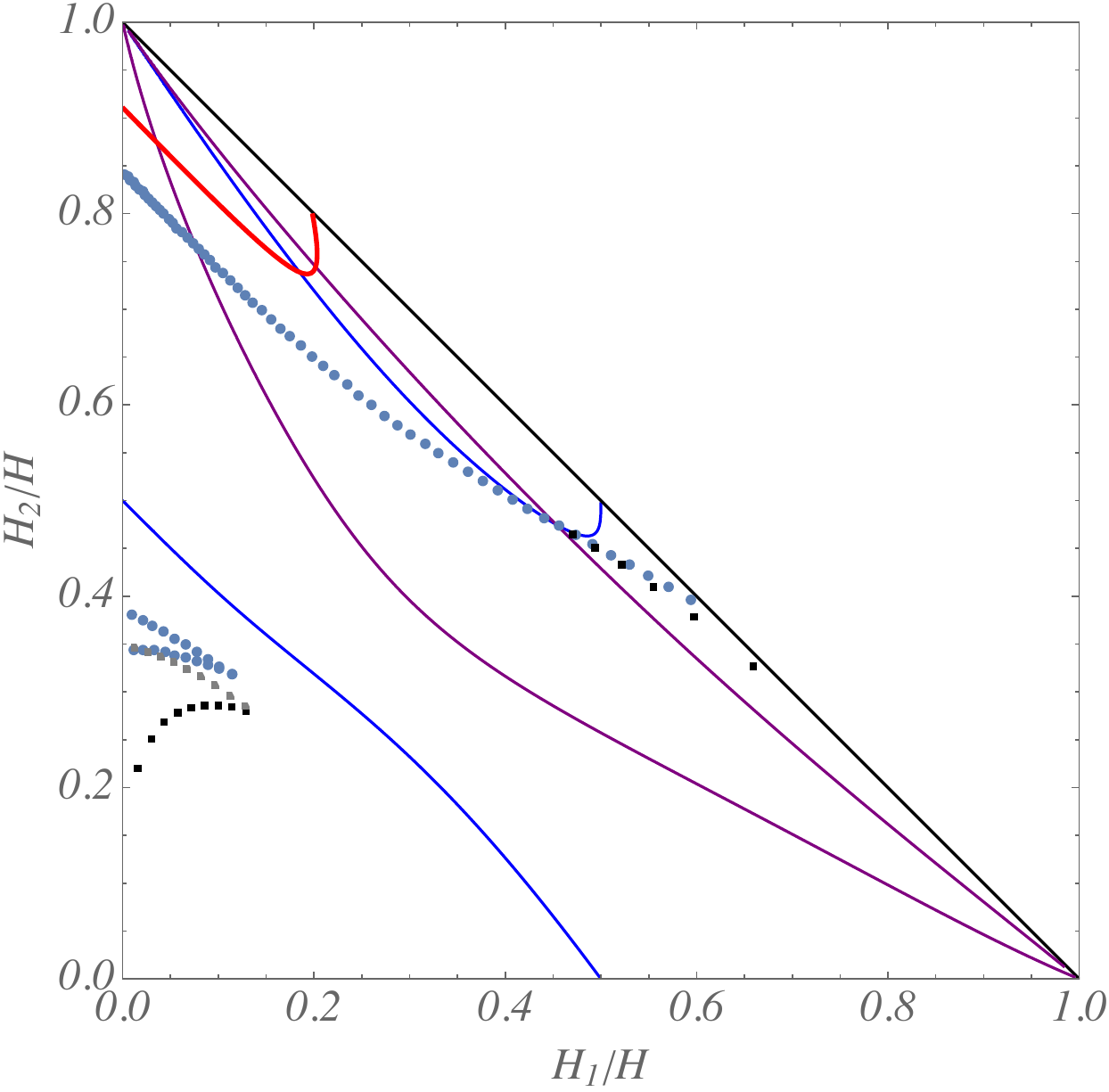}
				\put(45,90){$\delta=0.5$}
				\put(-3,92){$(a)$}
			\end{overpic}
		\end{minipage}
		\begin{minipage}{.47\textwidth}
			\centering
			\begin{overpic}[width=7cm]{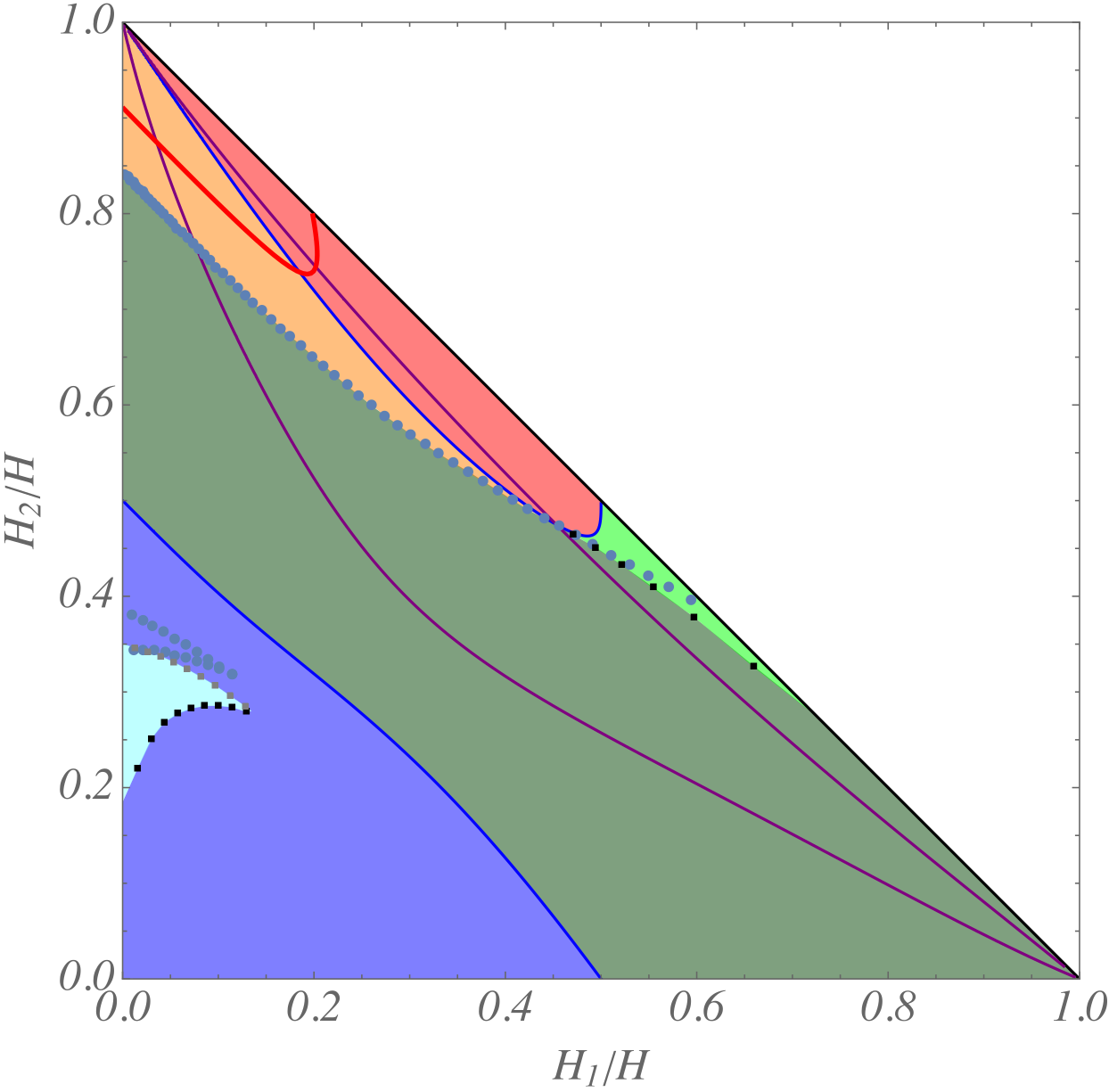}
				\put(45,90){$\delta=0.5$}
				\put(-5,92){$(b)$}
			\end{overpic}
		\end{minipage}
		\caption{All details of our conjugate state analysis encapsulated in one figure. In panel $(a)$ we identify in the parameter region for $\delta=0.5$ all relevant instances for the conjugate curves. The main markers are defined in figures~\ref{Regions_coexistence} and \ref{front_curves_all_details}$(b)$. In addition, the red line denotes the set of all intersections between a red conjugate curve and the linear long wave speed $c=c_0^+$  (see figure~\ref{front_curves_all_details}$(a)$ for large values of $H_2/H$). 
			This information can be used to delimit regions where different properties for mode-1 ISWs can be found, as in panel $(b)$. The colour scheme used here is outlined in figure~\ref{fig:bif_space}. The purple lines represent the condition $c_3=0$, included for comparison with predictions from Gardner theory.  
			\label{skeleton_and_colouring}
		}
	\end{figure}

	Returning to the point made earlier on unexplored aspects of the conjugate curves allowing us to unveil new types of ISWs, we first highlight the intricate behaviour of these curves, as depicted in 
	figure~\ref{front_curves_all_details}. In the figure, with $\delta=0.5$, $H_1/H_3=0.1$, in addition to the intersections and tangencies between the conjugate curves and the curve $c=c_0^+$, we also track the self-intersections (gray squares) of the conjugate curves, and where they first emerge (black square), provided $c>c_0^+$. By varying $H_1/H_3$ while keeping $\delta$ constant, the entire parameter space can be explored. All these markers can then be compiled, as in figure~\ref{skeleton_and_colouring}$(a)$, to delimit regions where distinct ISWs properties arise. This classification is further illustrated in figure~\ref{skeleton_and_colouring}$(b)$, following the colour scheme outlined in  figure~\ref{fig:bif_space} (see also figure~\ref{Parameter_regions} with $\delta=1$ to see how the complete solution space compares with figure~\ref{Regions_coexistence} $(a)$). How these markers separate the bifurcation spaces, and why, is explored in the following sections.

	\section{Strongly nonlinear theory: a critical point analysis}
	The solitary-wave solutions for the MMCC3 model are governed by the following dynamical system (see \citeauthor{barros_choi_milewski} \citeyear{barros_choi_milewski}):
	\begin{multline}\label{dyn_system_eq1}
		c^2 \Bigg\{ \frac{1}{3} \left(\rho_1 \frac{H_1^2}{h_1} + \rho_2 \frac{H_2^2}{h_2} \right) \zeta_1\mydprime + \frac{1}{6} \rho_2 \frac{H_2^2}{h_2} \, \zeta_2\mydprime + \frac{1}{6} \left( \rho_1 \frac{H_1^2}{h_1^2} - \rho_2 \frac{H_2^2}{h_2^2} \right) {\zeta_1^\prime}^2 + \frac{1}{3} \rho_2 \frac{H_2^2}{h_2^2} \,\zeta_1^\prime \zeta_2^\prime + \\
		+\frac{1}{3} \rho_2 \frac{H_2^2}{h_2^2} \,{\zeta_2^\prime}^2 \Bigg\} = \frac{1}{2}c^2 \left[ (\rho_2-\rho_1) + \rho_1 \frac{H_1^2}{h_1^2} - \rho_2 \frac{H_2^2}{h_2^2} \right] - g(\rho_2-\rho_1) \zeta_1, 
	\end{multline}
	\begin{multline}\label{dyn_system_eq2}
		c^2 \Bigg\{ \frac{1}{6} \rho_2 \frac{H_2^2}{h_2} \, \zeta_1\mydprime + \frac{1}{3} \left( \rho_2\frac{H_2^2}{h_2} +\rho_3 \frac{H_3^2}{h_3} \right) \zeta_2\mydprime - \frac{1}{3} \rho_2 \frac{H_2^2}{h_2^2} \, {\zeta_1^\prime}^2 - \frac{1}{3} \rho_2 \frac{H_2^2}{h_2^2} \, \zeta_1^\prime \zeta_2^\prime + \\
		+\frac{1}{6} \left( \rho_2 \frac{H_2^2}{h_2^2} - \rho_3 \frac{H_3^2}{h_3^2} \right) {\zeta_2^\prime}^2 \Bigg\} = \frac{1}{2}c^2 \left[ (\rho_3-\rho_2) + \rho_2 \frac{H_2^2}{h_2^2} - \rho_3 \frac{H_3^2}{h_3^2} \right] - g(\rho_3-\rho_2)\zeta_2, 
	\end{multline}
	where $V(\zeta_1,\zeta_2)$ is defined by
	\begin{multline}\label{potential_def}
		V(\zeta_1,\zeta_2) = -\frac{1}{2} c^2 \left( \rho_2-\rho_1 + \rho_1 \frac{H_1}{h_1} \right) \zeta_1 + \frac{1}{2} (\rho_2-\rho_1)g\,\zeta_1^2 - \\
		- \frac{1}{2} c^2 \left( \rho_3-\rho_2 - \rho_3 \frac{H_3}{h_3} \right) \, \zeta_2 + \frac{1}{2} (\rho_3-\rho_2) g \, \zeta_2^2 - \frac{1}{2} c^2 \rho_2 \frac{H_2}{h_2} (\zeta_2-\zeta_1).
	\end{multline}
	It can be shown that the system is equivalent to a Hamiltonian system with two degrees of freedom, for which $V$ is the potential.
	Furthermore, $V$ has been chosen to satisfy $V(0,0) =0$, {\it i.e.,} the origin lies on the zero energy level of the system.

	The critical points of our dynamical system are precisely the critical points of $V$. These are found as solutions of $\nabla V (\zeta_1,\zeta_2)=(0,0)$, and can be classified by examining the Hessian matrix of $V$, here denoted by ${\cal H}$:
	\begin{equation}\label{Hessian_matrix}
		{\cal H} = \left[ 
		\begin{array}{cc}
			g (\rho_2-\rho_1) -\left( \rho_1\frac{H_1^2}{h_1^3}+\rho_2 \frac{H_2^2}{h_2^3} \right) c^2 & \rho_2 \frac{H_2^2}{h_2^3} c^2\\
			\rho_2 \frac{H_2^2}{h_2^3} c^2 & g (\rho_3-\rho_2) - \left( \rho_2 \frac{H_2^2}{h_2^3}+\rho_3 \frac{H_3^2}{h_3^3} \right) c^2
		\end{array}
		\right].
	\end{equation}
	Clearly, $\nabla V (0,0)=0$, point at which 
	$\det {\cal H}$ is simply a multiple of the left-hand side of \eqref{lin_lw_speed_ast}. As a result, $(0,0)$ is a non-degenerate critical point for the range of speeds $\R^{+} \setminus \{c_0^{\pm}\}$. It becomes a saddle of $V$ when $\det {\cal H}(0,0)<0$, {\it i.e.,} when $c_0^{-}<c<c_0^{+}$, and a local extremum otherwise. More precisely, \eqref{lin_lw_speed_ast} leads to the inequality
	\begin{equation}\label{llwspeeds_ineq}
		\left(c_0^{-}\right)^2 < \frac{g H_1 H_2 (\rho_2-\rho_1)}{\rho_1 H_2 + \rho_2 H_1}<\left(c_0^{+}\right)^2,
	\end{equation}
	from which it follows that the origin is a local minimum (maximum) of $V$ in the range $]0,c_0^{-}[$ ($]c_0^{+}, \infty[$). 
	
	\subsection{Geometric locus of the critical points}
	
	Consider fixed values of $\rho_i$ and $H_i$ ($i=1,2,3$). Then, for a given wave speed $c$, the set of critical points is found by solving $\fpar{V}{\zeta_1}=\fpar{V}{\zeta_2}=0$. The admissible solutions are those within the triangular region $\cal{R}$:
	\begin{equation}\label{admissible_region}
		{\cal R} = \{ (\zeta_1,\zeta_2)\in \R^2: \, h_i>0, \,\, i=1,2,3\}.
	\end{equation}
	Given that the potential $V$ is a rational function, finding its critical points is equivalent to 
	finding the intersection of two plane algebraic curves:
	\begin{equation}\label{curveC1}
		{\cal C}_1 \equiv \frac{1}{2} c^2 \left[ (\rho_2-\rho_1) h_1^2 h_2^2 + \rho_1 H_1^2 h_2^2 - \rho_2 H_2^2 h_1^2 \right] -(\rho_2-\rho_1)g \,\zeta_1 h_1^2 h_2^2 =0,
	\end{equation}
	\begin{equation}\label{curveC2}
		{\cal C}_2 \equiv \frac{1}{2} c^2 \left[ (\rho_3-\rho_2) h_2^2 h_3^2 + \rho_2 H_2^2 h_3^2 - \rho_3 H_3^2 h_2^2 \right] -(\rho_3-\rho_2)g \,\zeta_2 h_2^2 h_3^2 =0.
	\end{equation}
	The fact that the origin remains a critical point for all values of $c$ can be interpreted geometrically by noting that both ${\cal C}_1 = 0$ and ${\cal C}_2 = 0$ are homogeneous curves. Eliminating $c$ from equations \eqref{curveC1} and \eqref{curveC2} yields the following relationship between $\zeta_1$ and $\zeta_2$:
	\begin{equation}\label{geo_locus_critical_points}
		P(\zeta_1,\zeta_2) \equiv (\rho_2-\rho_1) \zeta_1 \,h_1^2 \,G_3(h_2,h_3) - (\rho_3-\rho_2) \zeta_2 \,h_3^2 \,G_1 (h_1,h_2) =0,
	\end{equation}
	where
	\begin{equation}\label{def_G1}
		G_1(h_1,h_2)=(\rho_2-\rho_1)h_1^2 h_2^2 + \rho_1 H_1^2 h_2^2 - \rho_2 H_2^2 h_1^2,
	\end{equation}
	\begin{equation}\label{def_G3}
		G_3(h_2,h_3) = (\rho_3-\rho_2)h_2^2 h_3^2 + \rho_2 H_2^2 h_3^2 - \rho_3 H_3^2 h_2^2.
	\end{equation}
	Within the admissible region $\cal{R}$, the two sets of equations \eqref{curveC1}--\eqref{curveC2} and \eqref{curveC2}--\eqref{geo_locus_critical_points} are equivalent. In particular, any critical point of $V$ lies on the geometric locus defined by the curve $P=0$ in \eqref{geo_locus_critical_points}. Conversely, it can be shown (see Lemma 1 of Appendix B) that {\it for any point $(\zeta_1, \zeta_2)$ on the curve $P=0$ there exists a real value of $c$ for which \eqref{curveC2}--\eqref{geo_locus_critical_points} are satisfied, and hence $(\zeta_1, \zeta_2)$ is a critical point of $V$}.

	\begin{figure}
		\begin{minipage}{.33\textwidth}
			\centering
			\begin{overpic}[width=5cm]{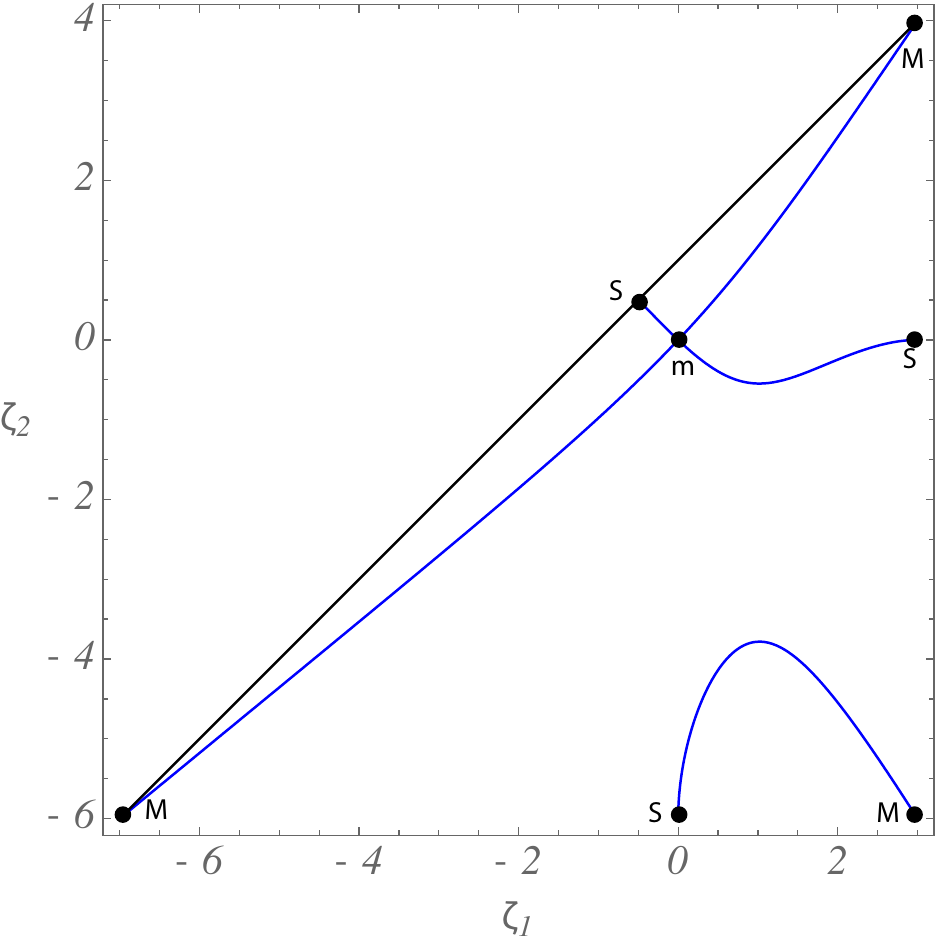}
				\put(-3,94){$(a)$}
			\end{overpic}
		\end{minipage}
		\begin{minipage}{.33\textwidth}
			\centering
			\begin{overpic}[width=5cm]{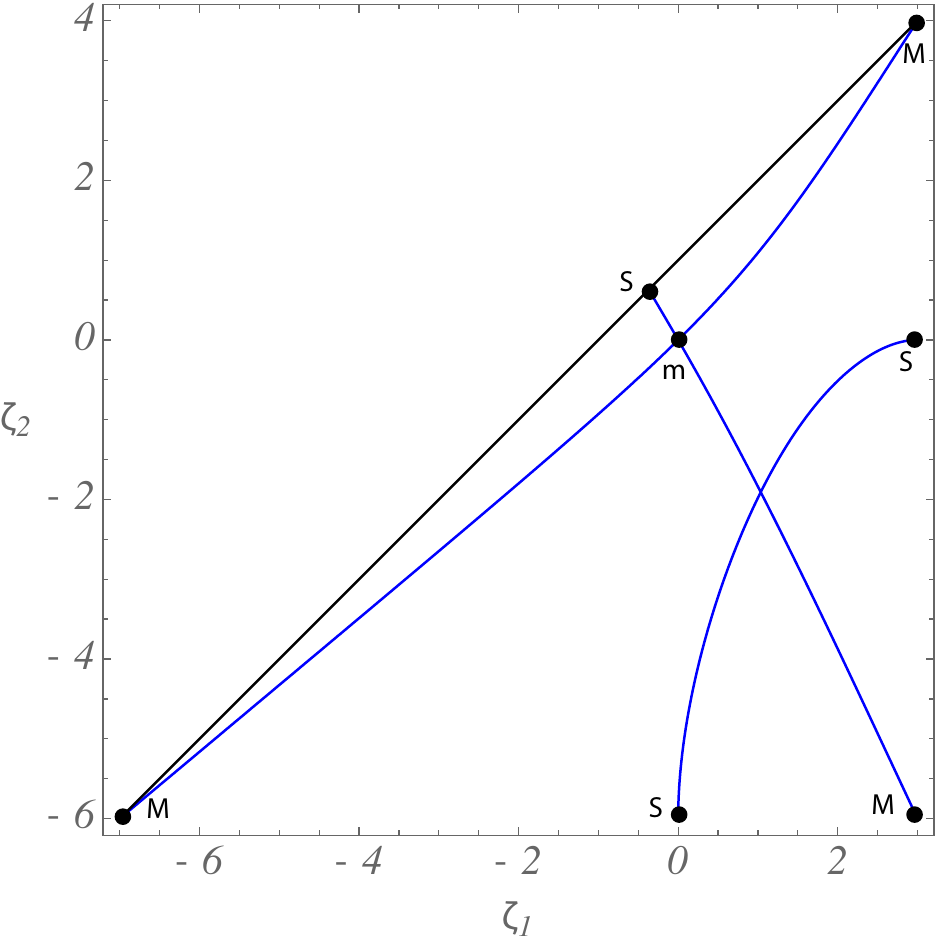}
				\put(-3,94){$(b)$}
			\end{overpic}
		\end{minipage}
		\begin{minipage}{.33\textwidth}
			\centering
			\begin{overpic}[width=5cm]{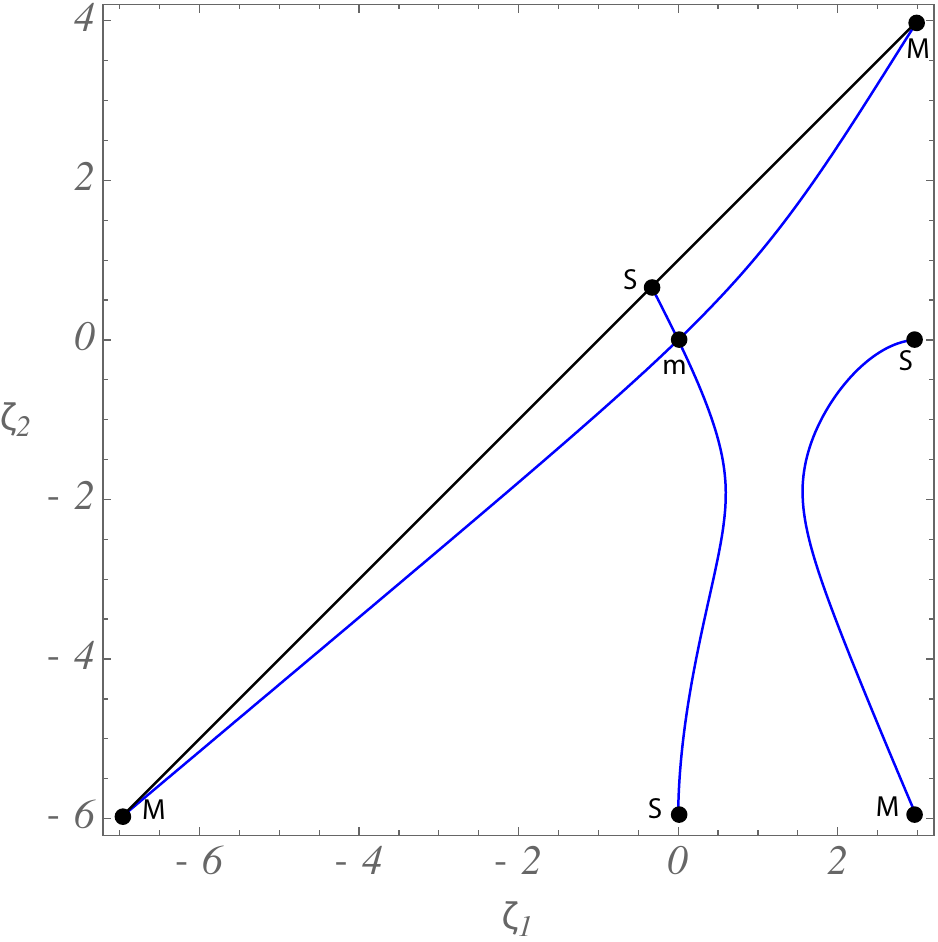}
				\put(-3,94){$(c)$}
			\end{overpic}
		\end{minipage}
		\caption{Plots for the curve $P=0$ in \eqref{geo_locus_critical_points}. We set $(H_1,H_2,H_3)=(3,1,6)$ and adopt the stratification \eqref{stratification} with $\Delta_1=\delta \,\Delta_2$ and $\Delta_2=0.1$. $(a)$ $\delta=1$, $(b)$ $\delta \approx 1.70824$, $(c)$ $\delta =2$. When $c\approx 0$, there are seven critical points of $V$: 3 local maxima (M), 3 saddles (S), 1 minimum (m). 
			\label{geometrical_locus}
		}
	\end{figure}
	
	The curve $P=0$ can be readily visualised in the $(\zeta_1,\zeta_2)$-plane once all physical parameters are specified, revealing two main distinct configurations {\it cf.} panels $(a)$ and $(c)$ of figure~\ref{geometrical_locus}. In either case, the curve $P=0$ consists of three distinct branches: {\em Branch I}, which connects 
	vertices $(-H_2-H_3,-H_3)$ and $(H_1, H_1+H_2)$; {\em Branch II}, which connects the point $\left( -H_2/ (1+\delta), H_2 \delta/(1+\delta) \right)$ to one of the two points: $(0,-H_3)$ or $(H_1,0)$; {\em Branch III}, which connects $(H_1,-H_3)$ to either $(0,-H_3)$ or $(H_1,0)$. All endpoints of these branches 
	correspond to critical points in the limit when $c\rightarrow 0^+$. As $c$ increases, all points (except the origin) move inward. Three maxima (M) depart from the vertices of the triangle ${\cal R}$; three saddles (S) depart from each side of the triangle; the origin remains a minimum (m) until $c$ reaches the value $c_0^-$. 
	These points move according to two basic rules: ($i$) if two points ever meet at any location other than the origin, they cannot pass each other; ($ii$) as $c$ increases, any given point on a branch of $P=0$ must remain on that same branch. The first rule follows from a simple observation: from \eqref{curveC1}, or \eqref{curveC2}, it is evident that each point on $P=0$, apart from the origin, corresponds to a unique value of $c$. The second rule is more subtle and requires a technical justification, which is provided in Appendix D.

	The origin is a somewhat special point. It is the only critical point that will not change its position as the wave speed varies. Moreover, it is not a regular point of $P=0$. In fact, at this {\it double point} (or {\it crunode}), the tangents to the curve $P=0$ can be found as $\zeta_2=\gamma^\pm \,\zeta_1$ as in \eqref{ratio_displacements} (see Lemma 2 in Appendix B).

	For certain exceptional values of the physical parameters, the curve $P=0$ may exhibit an additional singular point. Assuming the stratification given in \eqref{stratification} with $\Delta_1=\delta \Delta_2$, and fixing the parameters $\Delta_2$, $H_i$ ($i=1,2,3$), 
	the specific value of $\delta$ that gives rise to an extra crunode can be determined by solving the system $P=0$, $\frac{\partial P}{\partial \zeta_1}=0$, and $\frac{\partial P}{\partial \zeta_2}=0$, simultaneously. In the particular case depicted in figure~\ref{geometrical_locus}, this critical value is approximately $\delta \approx 1.70824$, as shown in panel $(b)$.
	If, instead, the Boussinesq approximation were applied, this
	transition between the two curve configurations would occur precisely at $\delta=2$ (corresponding to $=H_3/H_1$), as demonstrated in Appendix B.

	\subsection{Collision of critical points}\label{sec:collision}
	Consider fixed values of $\rho_i$ and $H_i$ ($i=1,2,3$). As the wave speed varies, the surface of the potential $V$ is smoothly deformed and the number and nature of critical points remains invariant, unless a collision takes place. Examining when collisions occur is thus a crucial step in the critical point analysis.

	\subsubsection{Trivial collisions}\label{sec:trivial_collisions}
	We begin by examining the collision of critical points at the origin, which we will refer to as “trivial” collisions. It is important to note that the origin is a regular point of both curves ${\cal C}_1 = 0$ and ${\cal C}_2 = 0$. Specifically, the tangents to each curve at the origin are given, respectively, by:
	\begin{gather}\label{tangents_C1_C2}
		\begin{aligned}
			\left[ (\rho_1 H_2 + \rho_2 H_1)c^2 - g(\rho_2-\rho_1)H_1 H_2 \right] \zeta_1 - \rho_2 H_1 c^2 \,\zeta_2 &=0,\\
			\rho_2 H_3 c^2 \,\zeta_1 - \left[(\rho_2 H_3 + \rho_3 H_2)c^2 - g(\rho_3-\rho_2)H_2 H_3 \right] \zeta_2 &=0.
		\end{aligned}
	\end{gather}
	Moreover, it can be shown that $\frac{\partial {\cal C}_k}{\partial \zeta_2} \neq 0$ for any point on the curve ${\cal C}_k$ ($k=1,2$). We can therefore use the following result (\citeauthor{coolidge}, Theorem 4, pp. 16--17): {\it In the vicinity of an ordinary finite point where the tangent is not vertical, $\zeta_2$ may be expressed as a convergent power series in terms of $\zeta_1-\zeta_{1,0}$, where $\zeta_{1,0}$ is the abscissa of the point in question}. For our purposes, we set $\zeta_{1,0}=0$ and seek expansions:
	\begin{equation}\label{parameterization_Ck}
		\zeta_2 = p_k \,\zeta_1 + q_k \,\zeta_1^2 + r_k \,\zeta_1^3 + O(\zeta_1^4), \quad k=1,2.  
	\end{equation}
	The coefficients $p_k$, $q_k$, $r_k$ ($k=1,2$) are all given in \eqref{ps_qs_rs} of Appendix C. 
	A double collision at the origin holds when $p_1=p_2$, which is precisely when \eqref{lin_lw_speed_ast} is satisfied. In other words, this type of collision arises only at the linear long wave speeds $c=c_0^{\pm}$, at which the tangent lines given in \eqref{tangents_C1_C2} become identical. 
	
	For a triple collision at the origin to occur, not only the first- but also the second-order approximation to the curves must coincide, {\it i.e.,}
	\begin{equation}\label{triple_collision}
		p_1=p_2, \quad q_1=q_2.
	\end{equation}
	It is shown in the Appendix C that \eqref{triple_collision} holds precisely at criticality, {\it i.e.,} when \eqref{criticality_condition} is satisfied. 
	
	Lastly, a quadruple collision at the origin occurs when 
	\begin{equation}
		p_1=p_2, \quad q_1=q_2, \quad r_1 = r_2.
	\end{equation}
	As shown in Appendix C, this situation arises precisely when, at criticality, $T_1=0$, with $T_1$ defined by \eqref{def_T1}. In other words, a quadruple collision at the origin occurs exactly when both the quadratic and cubic coefficients, $c_1$ and $c_3$, respectively, of Gardner's equation vanish simultaneously. To understand this, note that the coefficient $c_3$ in \eqref{cubic_non_coef} vanishes when $T_1+T_2+T_3=0$. Since $T_2$ and $T_3$ are both proportional to $c_1$, we deduce that, at criticality ({\it i.e.,} when $c_1=0$), the cubic nonlinearity coefficient can only vanish if $T_1=0$. Another important consequence is that quadruple collisions can only occur at $c=c_0^+$. 
	\begin{figure}
		\begin{center}
			\includegraphics*[width=7cm]{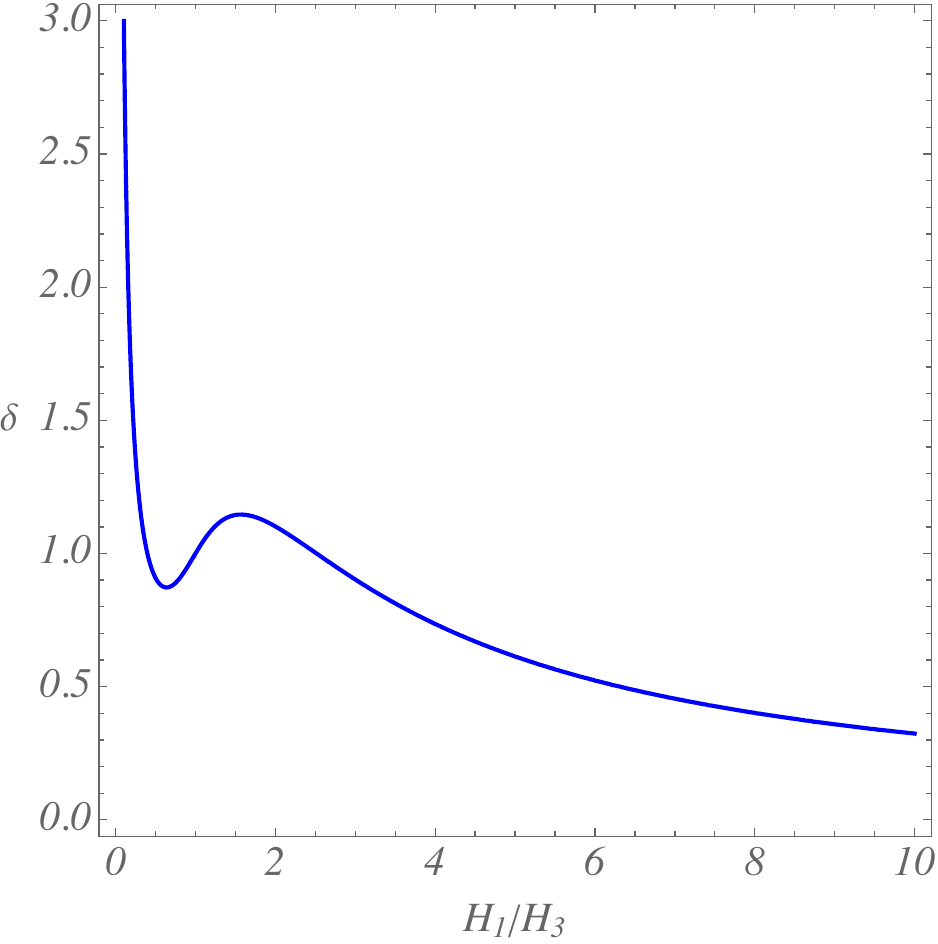}
		\end{center}
		\caption{Instances in the parameter space where a quadruple collision occurs at the origin. Here, the Boussinesq approximation is adopted and, for each value of $\delta=\delta_0$, the number of quadruple collisions may be read from the plot by counting how many times the horizontal line $\delta=\delta_0$ intersects the graph.    
			\label{number_quadruple_collisions}
		}
	\end{figure}

	In figure~\ref{Gardner_predictions}, we can identify in panel $(b)$ three instances in the parameter space where a quadruple collision occurs. In all other panels, only one instance of this type of  collision is possible. To gain a better understanding of how many such collisions can occur in the parameter space, it is helpful to resort to the Boussinesq approximation. 
	Under this reduction, a quadruple collision is equivalent to:
	\begin{equation}\label{quad_col_syst_eqs}
		\tilde{T}_1=0, \quad H_1^2 H_2^2 \,\gamma^3 + H_1^2 H_3^2 \,(1-\gamma)^3 - H_2^2 H_3^2=0, \quad \gamma^2+ \Lambda \gamma - \delta=0,
	\end{equation}
	with $\Lambda=\delta (1+H_2/H_3)-1-H_2/H_1$. The first condition in \eqref{quad_col_syst_eqs} results from viewing $T_1=0$ as the vanishing of a multivariate polynomial, similarly to \eqref{T1_tilde_def}, but under Boussinesq approximation. The second condition is equivalent to the criticality condition. The third results from imposing that the collision occurs at the linear long wave speeds (see (B.4) in \cite{barros_choi_milewski}. Solving this system simultaneously through use of the {\it Gr\"{o}bner basis} (see e.g.~\citeauthor{cox_et_al_book1} \citeyear{cox_et_al_book1}, \citeauthor{cox_et_al_book2} \citeyear{cox_et_al_book2}), we arrive at a condition (more specifically, an algebraic curve) involving only $\delta$ and $H_1/H_3$. This is very convenient, since with one single plot a full description can be provided, as in figure~\ref{number_quadruple_collisions}. For values of $\delta$ close to 1, there will be three instances where a quadruple collision occurs, whereas in general only one quadruple collision is expected. Specifically, for $\delta=1$, quadruple collisions occur for the following values of $H_1/H_3$: $\approx 0.397536, \, 1.0, \, \approx 2.515$. In figure~\ref{Gardner_along_lines}, quadruple collisions correspond to intersections between the blue and purple curves. Quadruple collisions help define regions within the parameter space where distinct ISWs properties arise, as illustrated in figure~\ref{skeleton_and_colouring} and further demonstrated in figures \ref{Parameter_regions}, \ref{Parameter_regions2}.

	\subsubsection{Non-trivial collisions}
	
\begin{figure}
		\begin{minipage}{.47\textwidth}
			\centering
			\begin{overpic}[width=7cm]{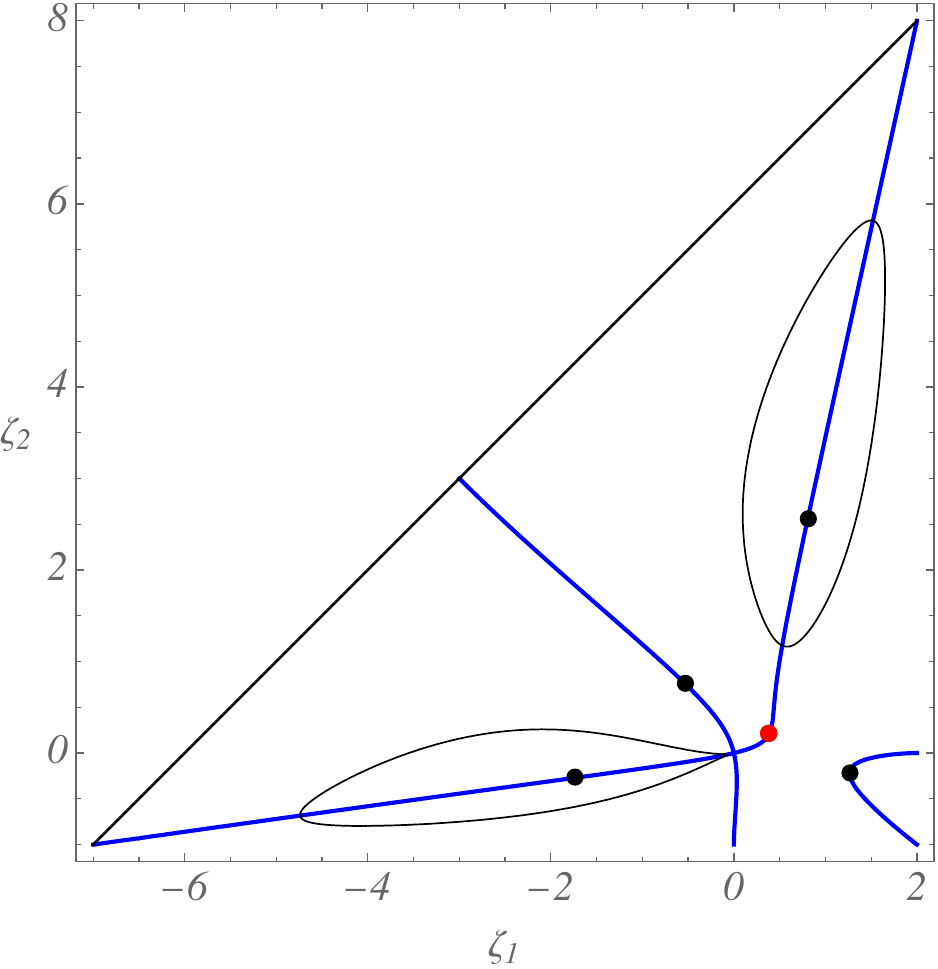}
				\put(-3,95){$(a)$}
			\end{overpic}
		\end{minipage}
		\begin{minipage}{.05\textwidth}
		\end{minipage}
		\begin{minipage}{.47\textwidth}
			\centering
			\begin{overpic}[width=7cm]{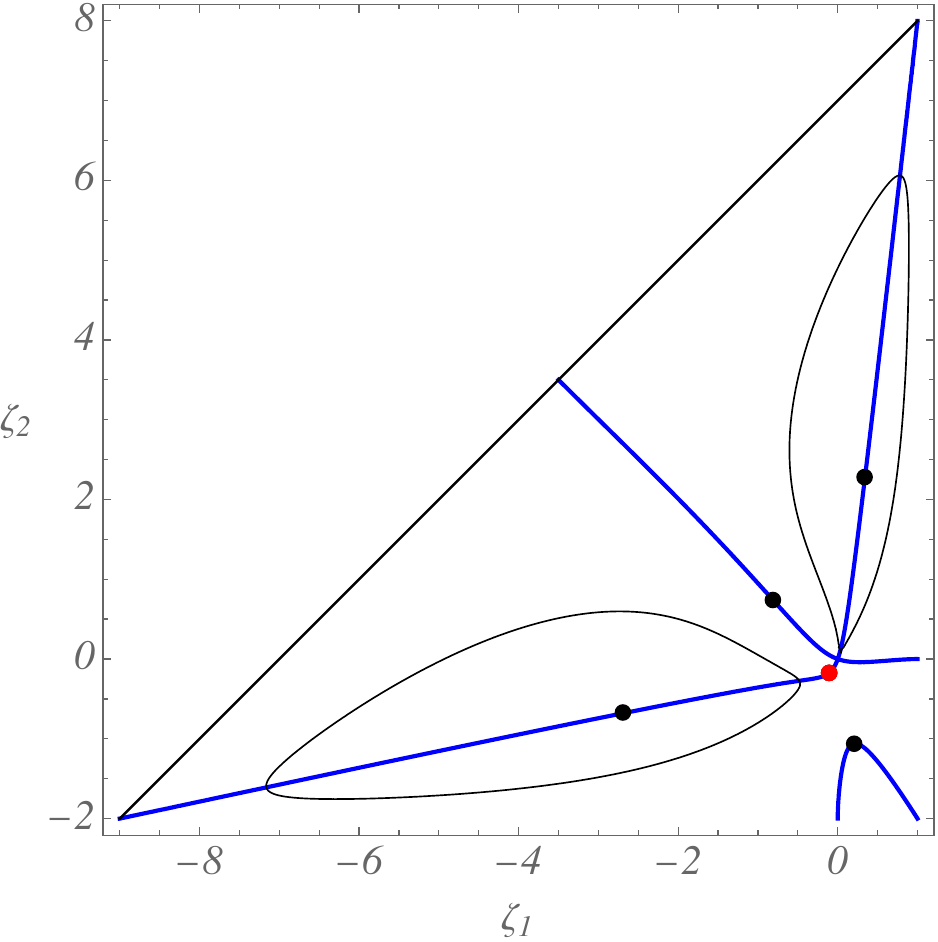}
				\put(-5,95){$(b)$}
			\end{overpic}
		\end{minipage}
		
		\begin{minipage}{.47\textwidth}
			\centering
			\begin{overpic}[width=7cm]{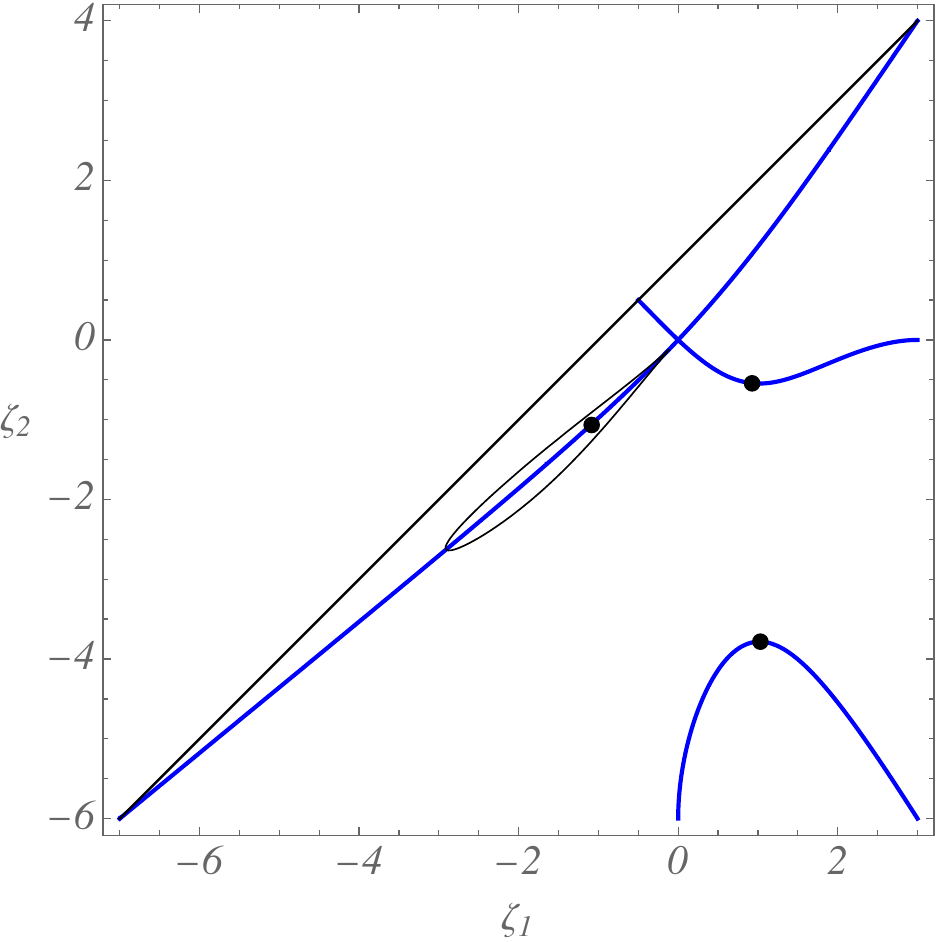}
				\put(-3,95){$(c)$}
			\end{overpic}
		\end{minipage}
		\begin{minipage}{.01\textwidth}
		\end{minipage}
		\begin{minipage}{.47\textwidth}
			\centering
			\begin{overpic}[width=7cm]{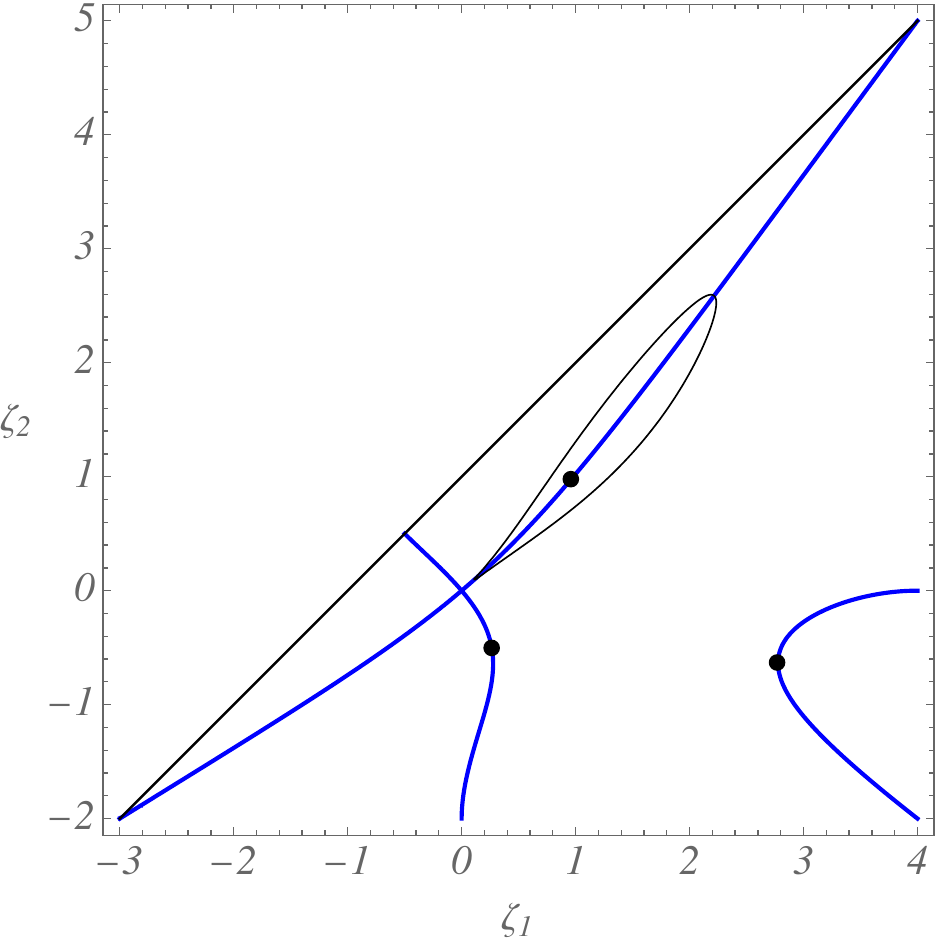}
				\put(-5,95){$(d)$}
			\end{overpic}
		\end{minipage}
		\caption{Plots of the curve $P=0$ in \eqref{geo_locus_critical_points}, and points where non-trivial collisions occur. The stratification is given by \eqref{stratification}, with $\Delta_1= \Delta_2=0.1$, and the undisturbed thicknesses of each layer parameters are set as follows. $(a)$ $(H_1,H_2,H_3)=(2,6,1)$, $(b)$ $(H_1,H_2,H_3)=(1,7,2)$, $(c)$ $(H_1,H_2,H_3)=(3,1,6)$, $(d)$ $(H_1,H_2,H_3)=(4,1,2)$. The points in black (red) correspond to collisions leading to a decrease (increase) of the number of critical points. In each panel, we overlay the zero contour of the potential $V$, obtained for  $(c/c_0^+)^2=1.01$. The mode-1 ISWs for the physical parameters in panels $(a)$--$(d)$ resemble, respectively, those illustrated in panels $A$--$D$ of figure~\ref{fig:bif_space}.
			\label{non_trivial_collisions}
		}
	\end{figure}

\begin{figure}
		\begin{minipage}{.45\textwidth}
			\centering
			\begin{overpic}[width=.9\textwidth]{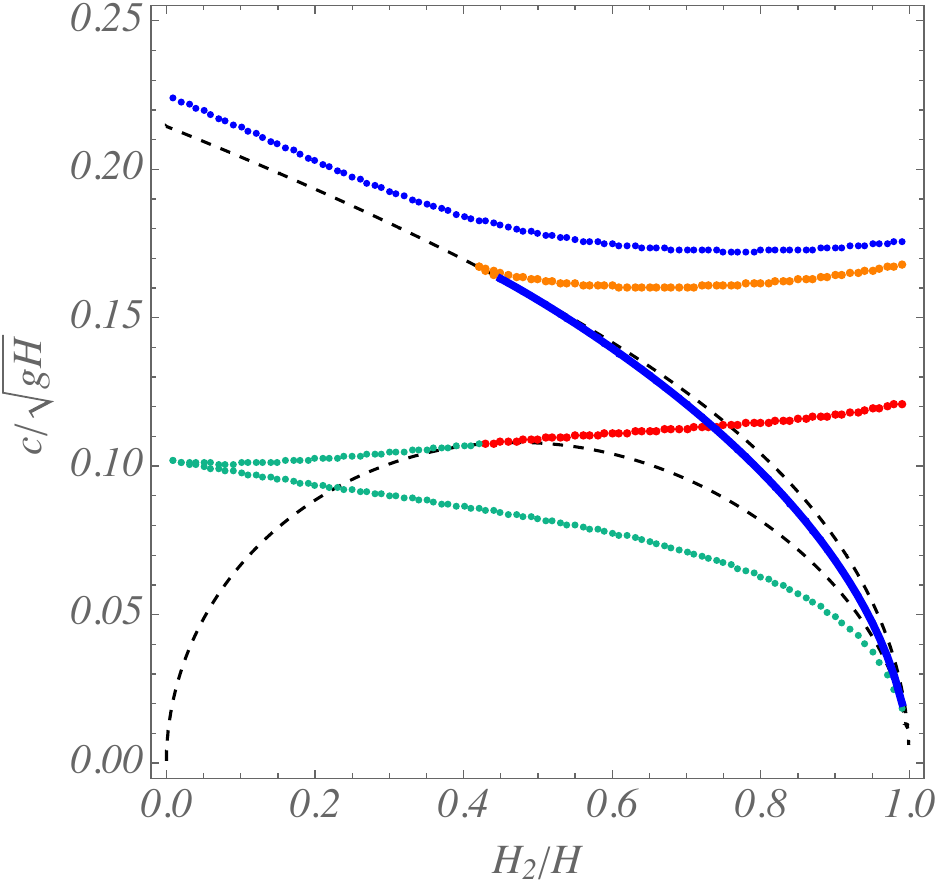}
				\put(-1,90){$(a)$}
			\end{overpic}
		\end{minipage}
		\begin{minipage}{.05\textwidth}
		\end{minipage}
		\begin{minipage}{.45\textwidth}
			\centering
			\begin{overpic}[width=.9\textwidth]{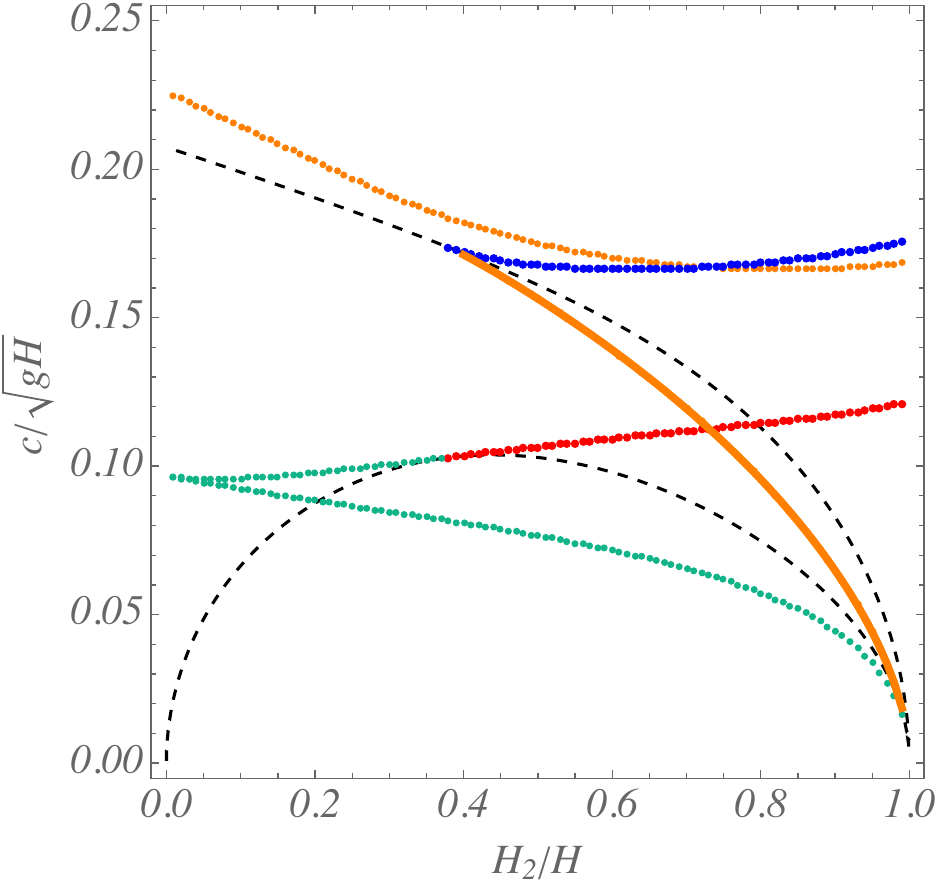}
				\put(-3,90){$(b)$}
			\end{overpic}
		\end{minipage}
		\caption{Collision curves on the $(H_2/H,c/\sqrt{gH})$-plane for the stratification \eqref{stratification}, with $\Delta_1= \Delta_2=0.1$. $(a)$ $H_1/H_3=0.5$, $(b)$ $H_1/H_3=2$. The black dashed curves are the linear long wave speeds, at which trivial collisions occur. Non-trivial collisions are represented by dotted (solid) lines when leading to a decrease (increase) of the number of critical points. We adopt the same colour scheme as in figure~\ref{Front_curves} to reveal the quadrant of the $(\zeta_1,\zeta_2)$-plane where the collisions occur. 
			\label{collisions_vs_Froude}
		}
	\end{figure}

	Whenever two critical points coalesce, they form a degenerate critical point, which can be identified by the vanishing of the determinant of the Hessian matrix in \eqref{Hessian_matrix} at that specific location. Let $(\zeta_1,\zeta_2)\neq(0,0)$ be the coordinates of such a point. Then, 
	\begin{equation}\label{segway_to_curve}
		\det {\cal H} (\zeta_1,\zeta_2)=0
	\end{equation}
	is a condition that also depends on the value of speed $c$. However, given that $(\zeta_1,\zeta_2)$ is a critical point, the corresponding value of $c$ can be easily determined from \eqref{curveC2} and substituted back into \eqref{segway_to_curve}. Eliminating denominators and simplifying the expression leads to a family of plane algebraic curves given by $F(\zeta_1,\zeta_2)=0$, which depends on the parameters $\rho_i$, $H_i$ ($i=1,2,3$). Once these parameters are fixed, the coordinates $(\zeta_1,\zeta_2)$ corresponding to non-trivial collisions can be found as the intersection points of the two plane algebraic curves $P=0$ and $F=0$. Figure~\ref{non_trivial_collisions} shows some demonstrative examples of the locations of non-trivial collisions. This figure will be revisited in section \S 5 to explain the bifurcation of panels $A$-$D$ in figure \ref{fig:bif_space}. Note that
	while the explicit expression for $F$ is unfortunately too cumbersome to be presented here, the results in figure~\ref{non_trivial_collisions} can be reproduced using any symbolic program, such as {\sc Mathematica}.
	In order to enable a more thorough investigation of collisions, we may employ the same methodology used for the conjugate states (see figure~\ref{Front_curves}) to generate figure~\ref{collisions_vs_Froude}. This approach reveals not only the quadrants of the $(\zeta_1,\zeta_2)$-plane in which collisions occur, but also the corresponding wave speeds at which these events take place. As illustrated, for fixed values of the density and undisturbed thickness of each layer, there exist either three or five non-trivial collisions.
	
	In general, a non-trivial collision leads to a decrease in the number of critical points. This is an immediate consequence of the first rule how critical points evolve: since two points cannot move past one another, they must annihilate upon collision. This also poses a constraint on the nature of the critical points that collide. As a result of a Theorem by Kronecker (see Theorem 2.4.6 in \citeauthor{emelyanov_et_al} \citeyear{emelyanov_et_al}), the total sum of topological indexes of critical points is 1 and must remain invariant as the value of $c$ is increased (see \citeauthor{barros} \citeyear{barros} for more details). Thus, no local maximum (or minimum) of the potential $V$ can collide with another point other than a saddle. In addition, after all collisions have taken place, only a single critical point remains: the origin, as a local maximum of $V$. Let $c=c_f$ ($>c_0^+$) denote the speed at which the final collision takes place. Beyond this speed, no solitary-wave solutions can exist. This critical speed thus serves as an upper bound for the propagation speed of mode-1 ISWs, as illustrated by the blue dotted curve in figure~\ref{collisions_vs_Froude}$(a)$, and it inherently limits their amplitude.

\begin{figure}
		\begin{minipage}{.33\textwidth}
			\centering
			\begin{overpic}[width=5cm]{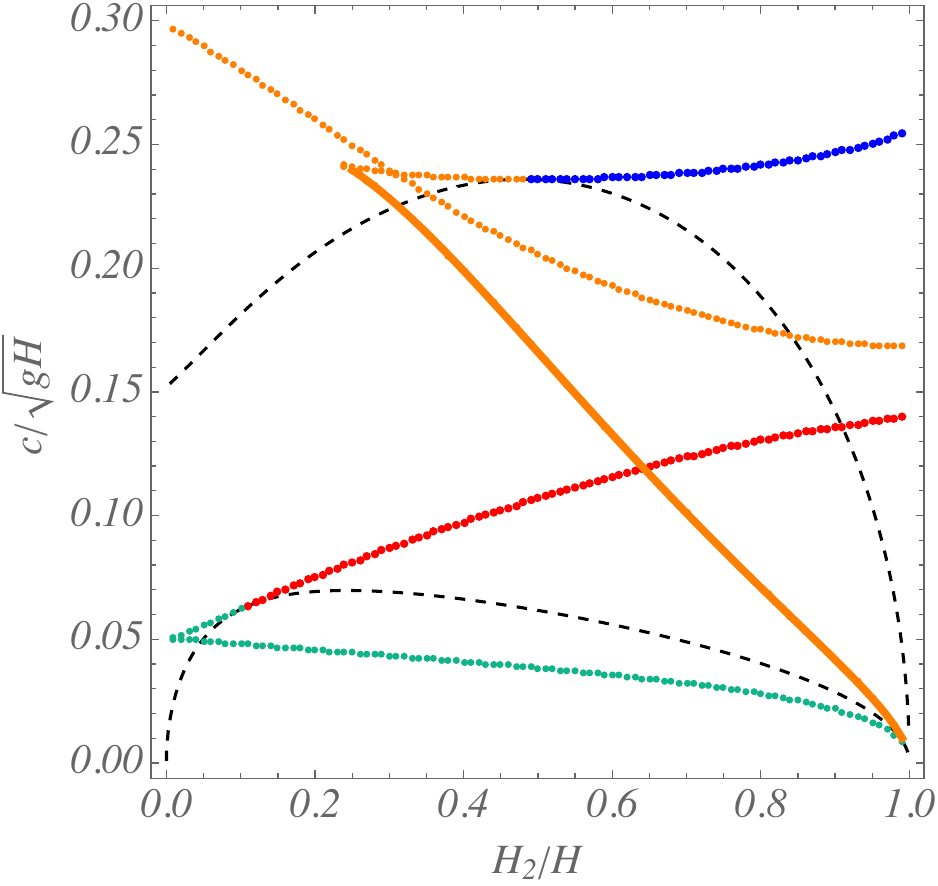}
				\put(-6,90){$(a)$}
			\end{overpic}
		\end{minipage}
		\begin{minipage}{.33\textwidth}
			\centering
			\begin{overpic}[width=5cm]{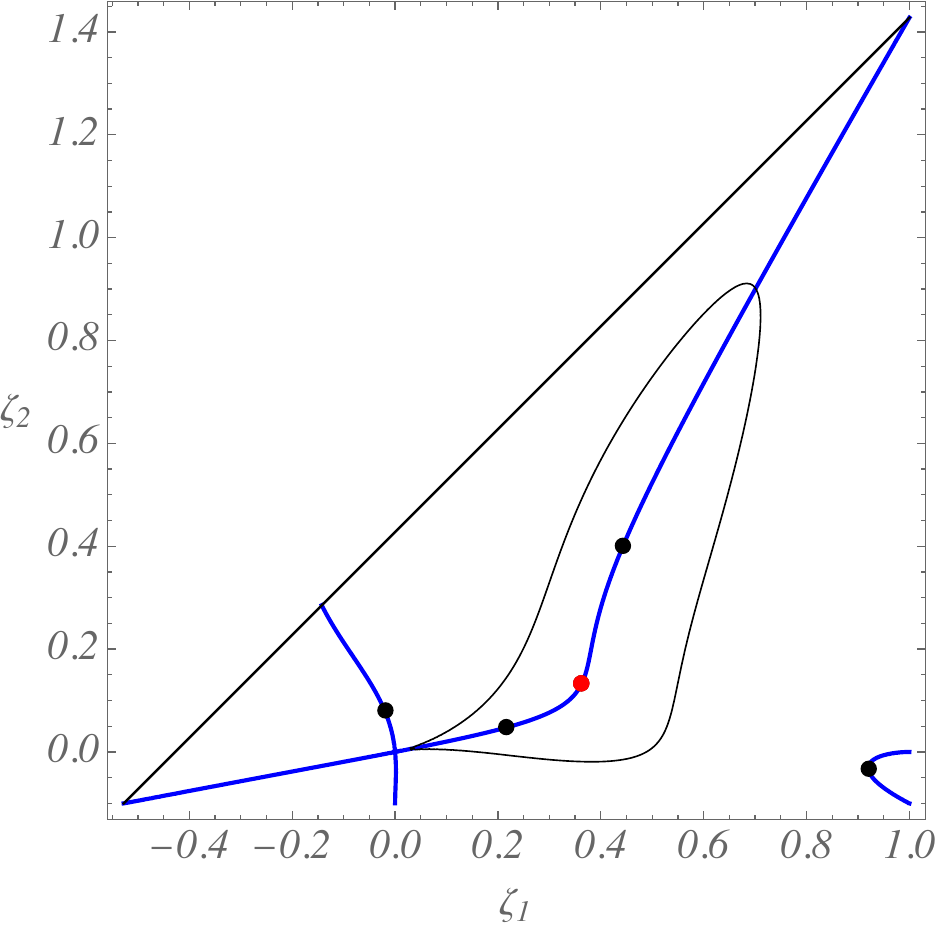}
				\put(-6,93){$(b)$}
			\end{overpic}
		\end{minipage}
		\begin{minipage}{.33\textwidth}
			\centering
			\begin{overpic}[width=5cm]{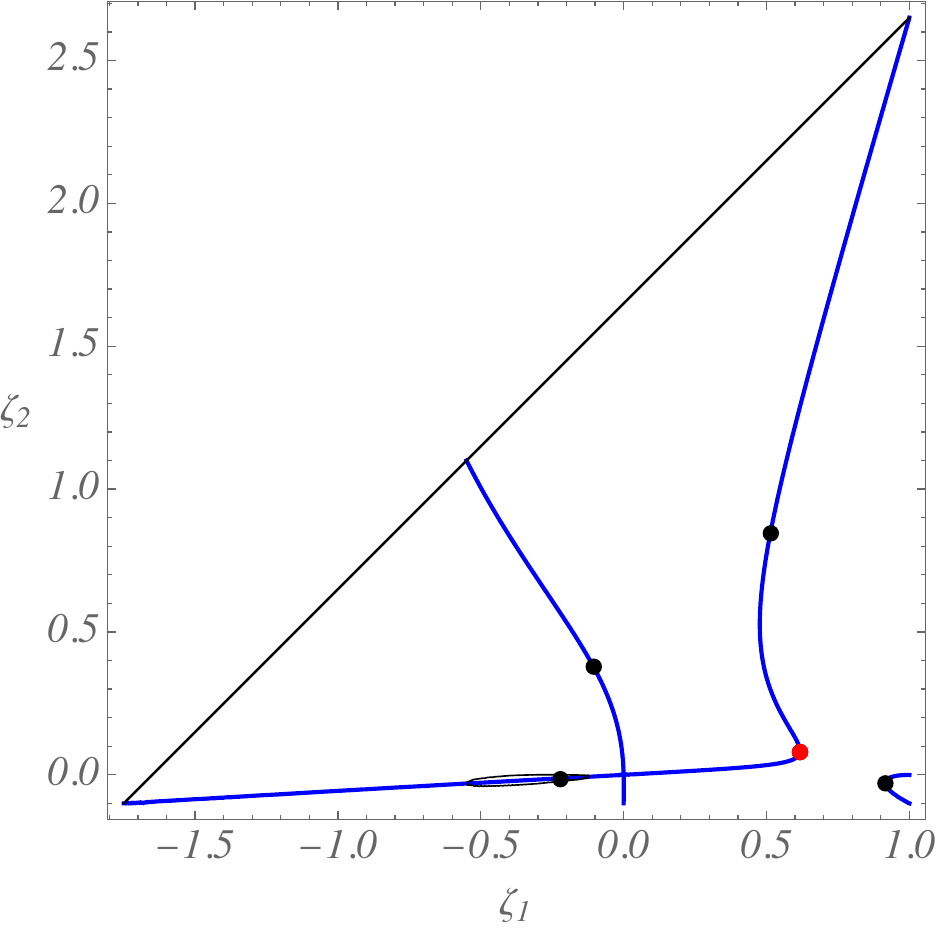}
				\put(-6,93){$(c)$}
			\end{overpic}
		\end{minipage}
		\caption{In panel $(a)$, similar to figure~\ref{collisions_vs_Froude}, collisions curves are displayed on the $(H_2/H,c/\sqrt{gH})$-plane for the parameters $\Delta_1=0.2$, 
			$\Delta_2=0.1$, and $H_1/H_3=10$. The two other panels illustrate distinct situations in panel $(a)$, obtained with different values of $H_2/H$. 
			$(b)$ $H_2/H=0.28$, $(c)$ $H_2/H=0.6$. In both plots, we set $H_1=1$ and overlay the zero contour of the potential $V$, obtained  for $(c/c_0^+)^2=1.01$.
			\label{collisions_vs_Froude_complicated}
		}
	\end{figure}

	We now turn to the exceptional case in which, unexpectedly, two critical points emerge spontaneously. These occurrences are indicated by red markers in figure~\ref{non_trivial_collisions} and by solid lines in figure~\ref{collisions_vs_Froude}. The panels $(b)$, $(c)$ of figure~\ref{non_trivial_collisions} are related to figure~\ref{collisions_vs_Froude}$(a)$, since in both cases $H_1/H_3=0.5$. Similarly, the panels $(a)$, $(d)$ of figure~\ref{non_trivial_collisions} are related to figure~\ref{collisions_vs_Froude}$(b)$, since in both cases $H_1/H_3=2$. From figure~\ref{collisions_vs_Froude}, emergent collisions are to be expected for large values of $H_2/H$, which explains the contrast between the bottom and top panels of figure~\ref{non_trivial_collisions}.
	While ``typical'' collisions -- where critical points annihilate -- occur along all branches of the curve $P=0$, our extensive numerical investigations (see figures~\ref{collisions_vs_Froude}, \ref{collisions_vs_Froude_complicated}) indicate that such emergent collisions occur exclusively along {\em Branch I}. From this, we conclude that these events correspond to the simultaneous emergence of a saddle point and a local maximum.
	In the two cases considered in figure~\ref{collisions_vs_Froude}, such occurrences are observed only for speeds below $c_0^+$. However, as illustrated in figure~\ref{collisions_vs_Froude_complicated}, emergent collisions can also arise at speeds exceeding $c_0^+$.

	\section{The strongly nonlinear solution space: distinguishing the cases $A$--$F$}\label{sec:coexistence_ISWs}
	
	In this section, we present a way to distinguish the characteristics of the solution space for mode-1 solitary waves for the MMCC3 system by tracking the behaviour of the critical points of the potential. 
	From a dynamical systems perspective, mode-1 ISWs for the MMCC3 model arise as  trajectories departing from the local maximum at the origin towards a saddle located on one or both sides of {\em Branch I} of $P=0$. The return point of such trajectories is a point at which $V=0$. 
	
	Let us revisit figure~\ref{non_trivial_collisions}. The top panels show five non-trivial collisions of critical points, with one highlighted in red to indicate an emergent collision, whereas the lower panels display only three such collisions. Also depicted are the zero contours of $V$ at $c=c_0^+(1+\epsilon)$ ($\epsilon \ll1$), 
	which appear as simple closed loops superimposed on the plots of $P=0$. Each loop contains a special point that serves as a return point for the trajectory of a mode-1 ISW. 
	
	Assuming mode-1 ISWs exist for given physical parameters and no emergent collisions occur as c increases, the scenario corresponds to that shown in the lower panels of figure~\ref{non_trivial_collisions}, where only a single loop appears, indicating that solutions of opposite polarity do not coexist. The figure further reveals that mode-1 solutions may be of depression (panel $(c)$), or elevation (panel $(d)$), depending on the quadrant in which the loop resides.
	The polarity predictions from the strongly nonlinear theory are in full agreement with those of the KdV theory, as shown in Appendix D. According to figures~\ref{collisions_vs_Froude}, \ref{collisions_vs_Froude_complicated}, this behaviour is expected when the ratio $H_2/H$ is sufficiently small. In this regime, the solutions resemble those illustrated in panels $C$ and $D$ of figure~\ref{fig:bif_space}. It is important to note, however, that the behaviour in panels $C$ and $D$ can also be observed in other regimes, as shown below.
	
	In what follows, we establish the existence of conjugate states of opposite polarity with $c>c_0^+$ as the criteria for the coexistence of solutions of opposite polarity (panels $A-B$), as found by \cite{lamb_2023} for a smooth double-pycnocline stratification. We then distinguish between the panels which do not exhibit multiplicity of solutions (panels $C-F$). Finally, we sweep the parameter space in the Boussinesq case, presenting our findings for varying $\delta$ on the $(H_1/H,H_2/H)$-space.
	
	\subsection{Criteria for the coexistence of ISWs of opposite polarity (panels $A$--$B$)}

	Based on the discussion above, we conclude that an emergent collision is a necessary condition for the coexistence of mode-1 waves with opposite polarities. This is exemplified by the physical parameters corresponding to the top panels of figure~\ref{non_trivial_collisions}. However, as illustrated in figures~\ref{collisions_vs_Froude_complicated}$(b)$ and $(c)$, the presence of an emergent collision alone is not sufficient: solutions of opposite polarity fail to materialise in both cases.
	To guarantee the coexistence of solutions of opposite polarity, we require that, at $c=c_0^+$, two local maxima of $V$ exist on either side of {\em Branch I}, with both satisfying $V>0$. In light of how critical points evolve, this implies that: $(i)$ an emergent collision (yielding a saddle and a maximum) must occur at $c<c_0^+$; $(ii)$ the resulting maximum must reach the origin as $c$ approaches $c_0^+$. Furthermore, the polarity of the solution branch bifurcating from zero amplitude at the linear long wave speed $c=c_0^+$ is determined by the quadrant of the $(\zeta_1,\zeta_2)$-plane where the emergent collision takes place. Specifically, in figure~\ref{non_trivial_collisions}$(a)$, the emergent collision occurs in the first quadrant, resulting in a depression branch bifurcating from zero amplitude, while the elevation branch bifurcates from finite amplitude. A similar situation, but with the roles of the elevation and depression branch reversed, can be observed in figure~\ref{non_trivial_collisions}$(b)$, where the emergent collision occurs in the third quadrant. In the regimes considered, the solutions resemble those illustrated in panels $A$ and $B$ of figure~\ref{fig:bif_space}, respectively.  
	
	Identifying the parameter region for the coexistence of solutions with opposite polarity is quite challenging when relying solely on the analysis of collisions. Based on figure~\ref{collisions_vs_Froude}, one might be inclined to think that the desired feature holds for large values of $H_2/H$, since collision curves in orange and blue exist for $c>c_0^+$ (following an emergent collision at $c<c_0^+$). This is a clear indication that, at $c=c_0^+$, $V$ has two maxima on either side of {\em Branch I}, which eventually collide with saddle points as $c$ increases. However, this approach does not {\it a priori} provide a guarantee of the desired outcome, as we cannot confirm if, originally at $c=c_0^+$, $V>0$ at the two maxima.
	This can be overcome by observing that, for that to happen, the two loops on either side of {\em Branch I} should both shrink to a point, as $c$ increases from the value $c=c_0^+$. When that happens (for each loop), a maximum becomes an equilibrium at the same energy level as the origin, {\it i.e.}, $V=\fpar{V}{\zeta_1}= \fpar{V}{\zeta_2}=0$, and thus a conjugate state is reached and \eqref{conjugate_states} is recovered. 
	Therefore, if conjugate states of opposite polarity coexist for $c > c_0^+$, the presence of ISWs with opposite polarity is assured. 
	This finding supports the validity of the criterion proposed by \cite{lamb_2023} (see \S\ref{sec:sols_and_conjugate_sates}), at least within the scope of the strongly nonlinear theory.  
	
	\begin{figure}
		\begin{center}
			\includegraphics[scale=1]{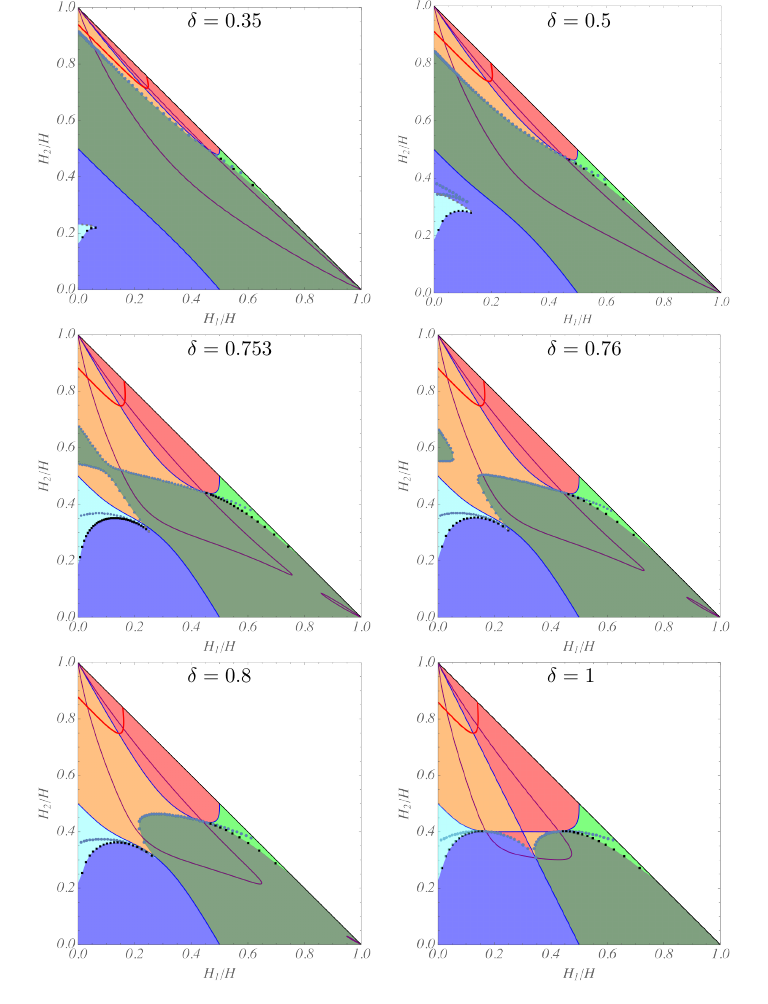} 
		\end{center}
		
		\caption{Partition of the (Boussinesq) parameter space into regions associated with distinct solution types.
			The dotted curves in each panel are the instances in the parameter space at which the front curve crosses the curve $c=c_0^+$. The blue curve corresponds to criticality, at which the front curve and the curve $c=c_0^+$ are tangent. The other markers in the panels (when applicable) are the black and grey squares discussed in figure~\ref{skeleton_and_colouring} $(b)$. 
			The red curve represents the instance when the front curve bearing characteristics of mode-2 crosses the curve $c=c_0^+$. In the region above this red line, not only coexistence of ISWs of opposite polarity exist, but also large amplitude solutions with mode-2 characteristics. The panels show results for different values of $\delta$. The colour scheme used here is outlined in figure~\ref{fig:bif_space}. The purple lines represent the condition $c_3=0$, included for comparison with predictions from Gardner theory.   
			\label{Parameter_regions}
		}
	\end{figure}
	\begin{figure}
		\begin{center}
			\begin{overpic}[width=200pt]{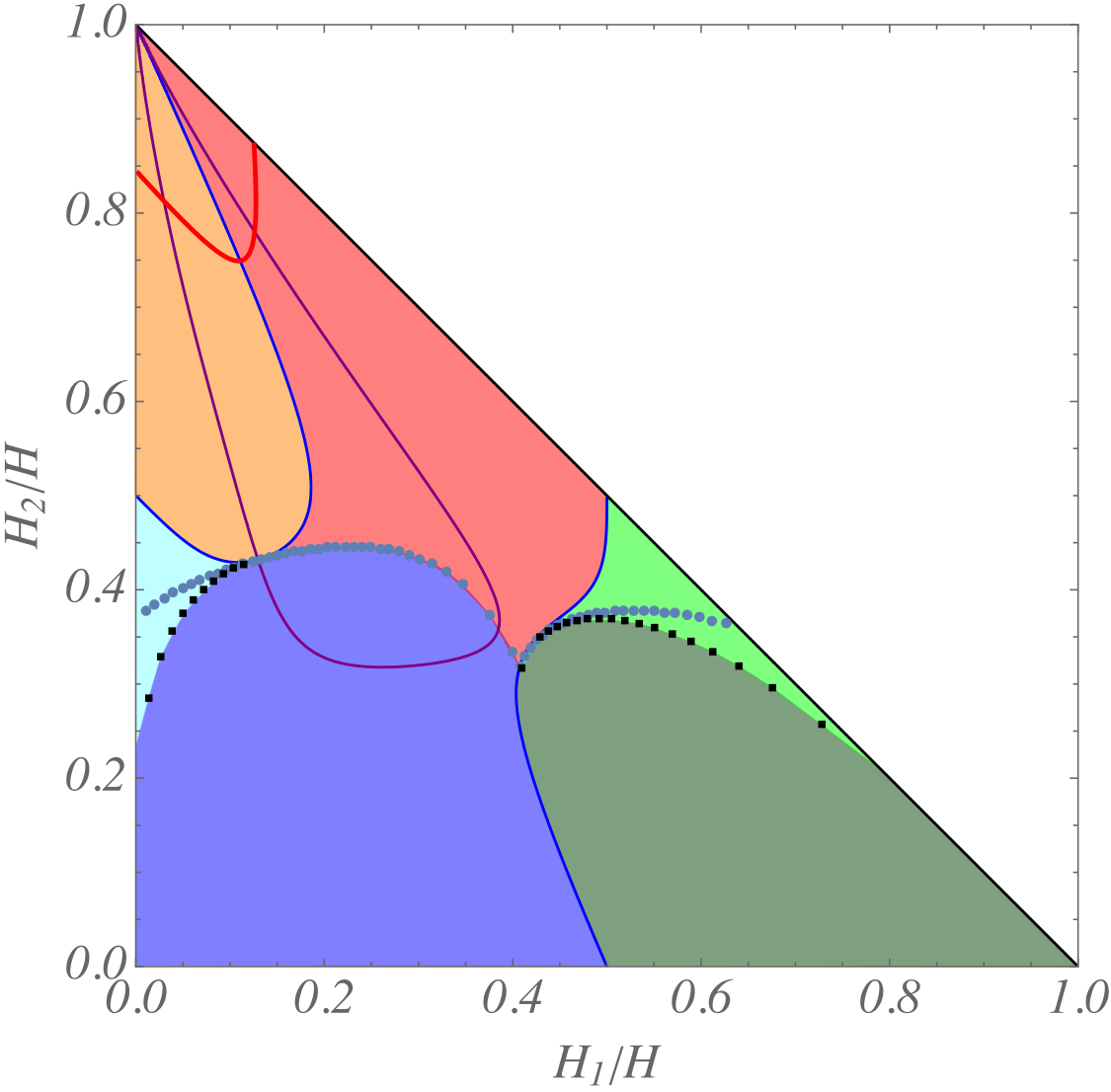}
				\put(45,90){$\delta=1.2$}
			\end{overpic}
			\quad
			\begin{overpic}[width=200pt]{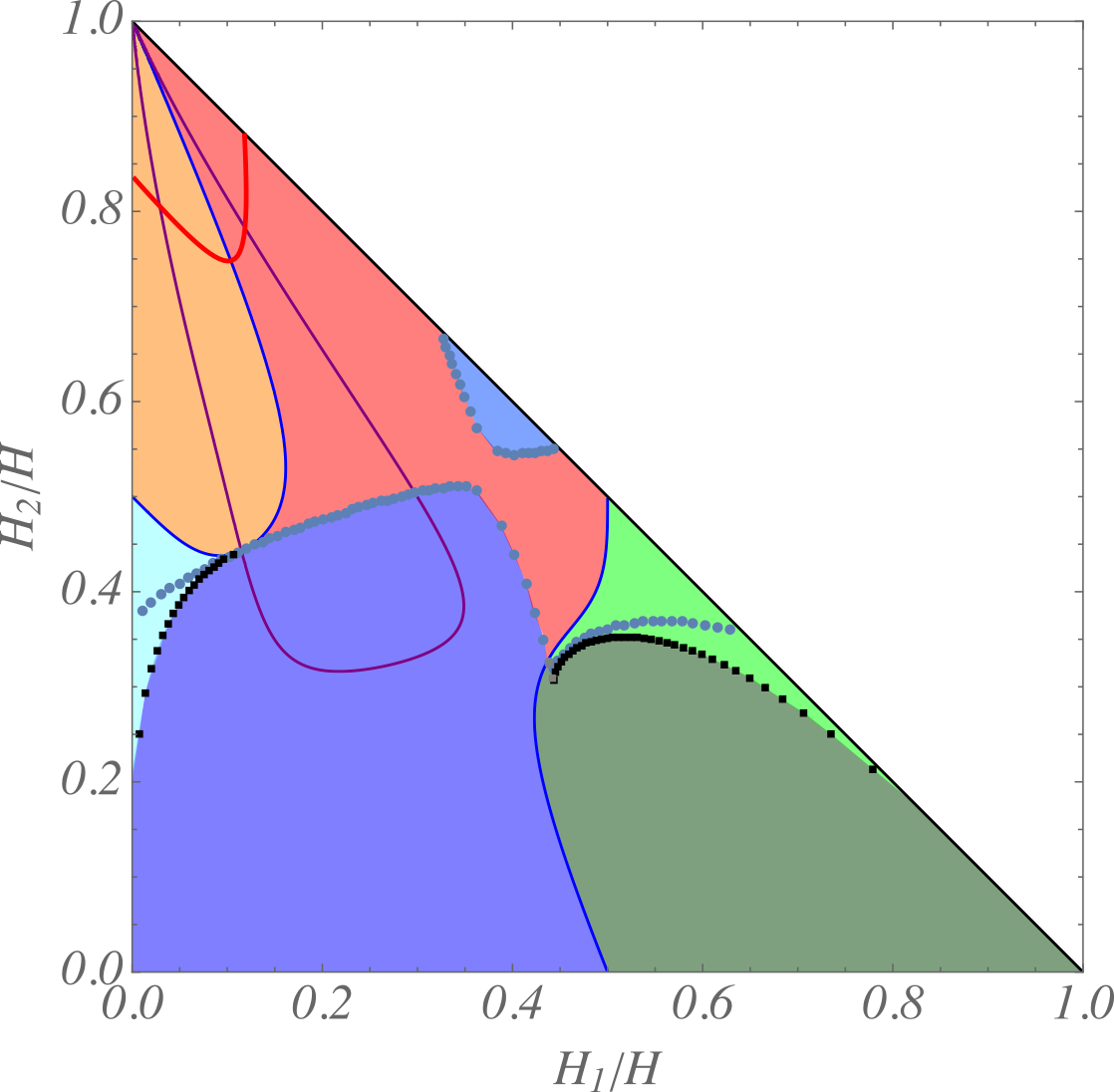}
				\put(45,90){$\delta=1.32$}
			\end{overpic}
			\begin{overpic}[width=200pt]{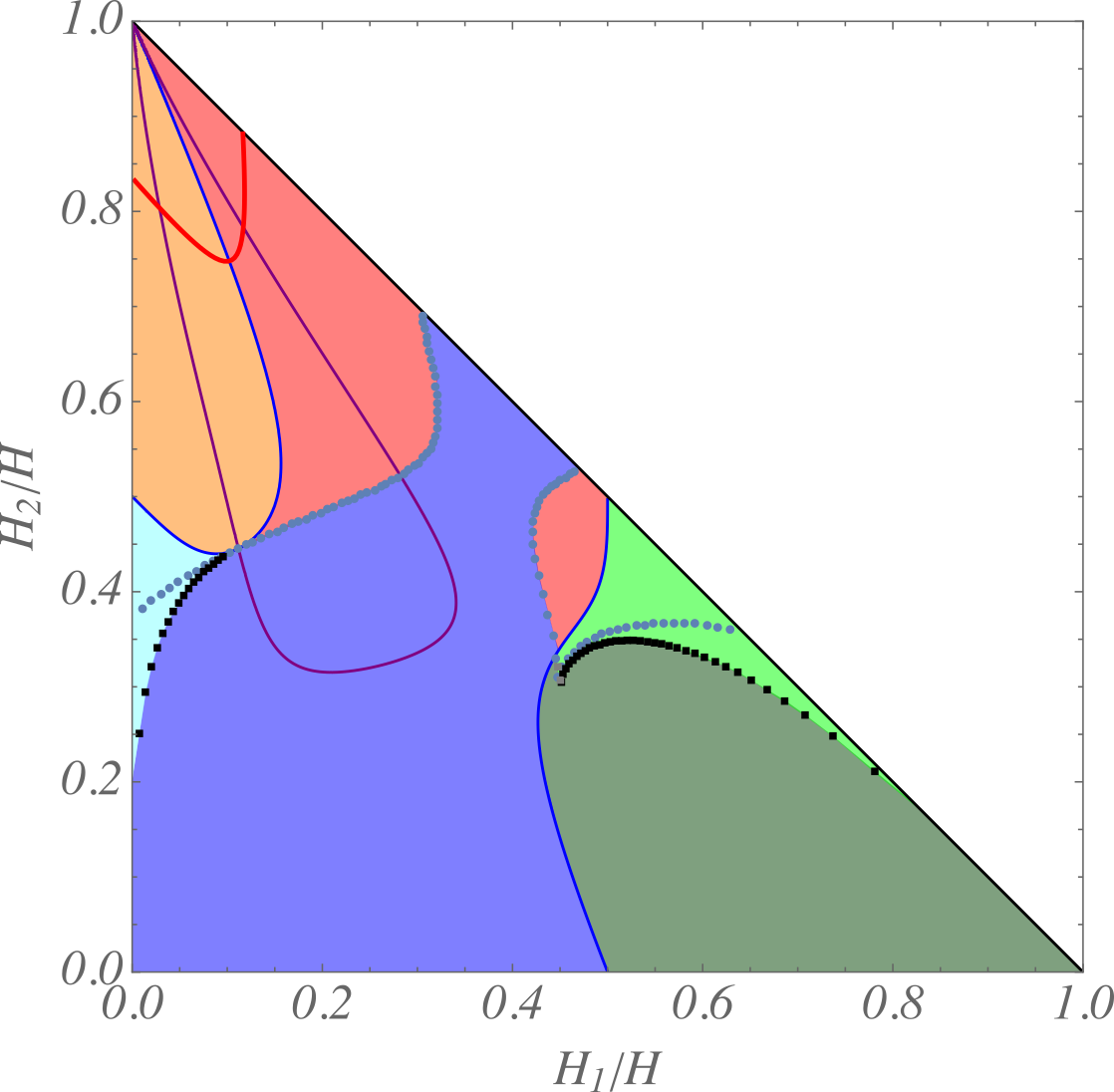}
				\put(45,90){$\delta=1.35$}
			\end{overpic}
			\quad
			\begin{overpic}[width=200pt]{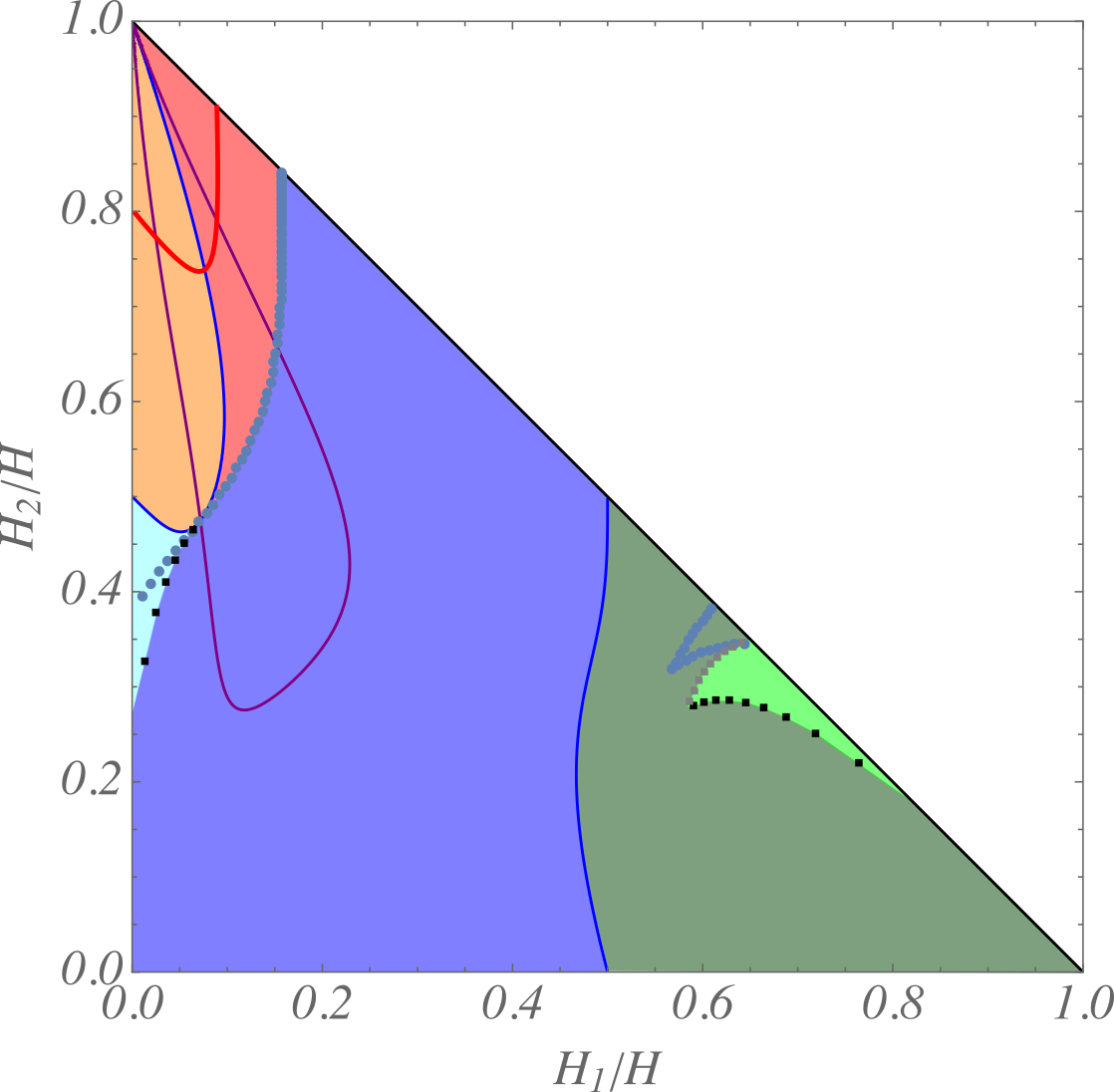}
				\put(45,90){$\delta=2$}
			\end{overpic}
		\end{center}
		
		\caption{Same as in figure \ref{Parameter_regions}, but for $\delta>1$.
			\label{Parameter_regions2}
		}
	\end{figure}
	
	\subsection{Criteria to distinguish panels $C$--$F$}
	The above highlights the benefits of using a conjugate state analysis to identify specific behaviours of solutions. Accordingly, the criteria presented hereafter are based on the properties of the conjugate states for this physical system. 
	
	Assume, for a given set of parameters, that conjugate states of opposite polarity, both with $c>c_0^+$, do not exist. In this case, extensive numerical tests reveal that for all $c>c_0^+$, all critical points of the potential $V$ along \emph{Branch I} lie in the first (third) quadrant if $c_1>0$ ($c_1<0$). 
	Hence, the KdV theory correctly predicts the polarity of the solitary waves in panels $C$--$F$. We note that it is also possible, for the chosen parameters, that a conjugate state exists on \emph{Branch II}. As a result, in addition to the waves discussed above, there are other solutions exhibiting mode-2 characteristics, which will be addressed separately in \S 5.3.
	
	In terms of conjugate states for the system, a few cases must be considered. 
	The simplest case is that there is only one mode-1 conjugate state with $c>c_0^+$. In such cases, there exists a single solution branch of mode-1 solitary waves (panel $C$ or $D$), starting from zero amplitude and limiting to a tabletop solitary wave. 
	
	The case when there are two or three mode-1 conjugate states with $c>c_0^+$ in the first or third quadrant requires more delicate consideration. Without loss of generality, consider the case where two or three mode-1 conjugate states with $c>c_0^+$ in the first quadrant. For this to occur, we must have that $c_1>0$, and the emergent collision of critical points must of occurred in the first quadrant along \emph{Branch I}, such as observed in figure \ref{collisions_vs_Froude_complicated}(b). Note this emergent collision occurs in between the origin and the maximum departing from $(H_1,H_1+H_2)$ in the limit when $c\rightarrow 0^+$, since each point along the curves $P=0$ corresponds to a unique value of $c$. 
	The criteria for distinguishing whether we have one branch of solutions (panel $D$) or two branches (panel $F$) is as follows. As long as the speed at which the emergent maximum becomes a conjugate state is less than the speed at which the other maximum does, we have panel $F$. Otherwise, we have panel $D$. Analogous consideration applies to the case when two or three mode-1 conjugate states with $c>c_0^+$ exist in the third quadrant.
	This is shown, for example, in figure \ref{front_curves_all_details}.  At the black square, where the emergent collision occurs at $V=0$ with $c>c_0^+$, the speed of the emergent maximum conjugate state (given by the black square) is less than the other maximum (the blue curve above the black square).  However, as $H_2$ increases, the two front curves cross at the grey square. Hence, past the grey square, the speed of the emergent maximum as a conjugate state is faster than than of the other maximum. We see here that the black square and grey square signal changes in behaviour, while the crossing of the front curves through $c=c_0^+$ (blue circle) in this case do not. Greater detail on the solution space for panels $E$-$F$ is given in section \ref{sec:numerics}. Next, we use the above criteria to present the Boussinesq solution space for all parameters.

	\subsection{Parameter sweep of the Boussinesq solution space}
	The above provides a mechanism for determining the solution space from the behaviour of the conjugate states. Figures \ref{Parameter_regions} and \ref{Parameter_regions2} demonstrate how the solution space evolves as one varies $\delta$ within the Boussinesq regime. The colour scheme corresponds to panels $A-F$ in the figure \ref{fig:bif_space}. 
	
	Plots such as figure \ref{Front_curves} show conjugate states as a function of $H_2$ for given $\delta$ and varying $H_1/H_3$, where each value of $H_1/H_3$ represents a ray in the parameter space, as shown in figure~\ref{Gardner_along_lines}. 
	To create figures \ref{Parameter_regions} and \ref{Parameter_regions2}, a representative set of rays spanning the parameter space is selected. From each ray, key markers are extracted as illustrated in figure~\ref{front_curves_all_details}, including: blue circles (mode-1 conjugate state with $c=c_0^+$), black squares (emergent collision coinciding with being a conjugate state), grey squares (two conjugate states in the same quadrant having the same speed), and the red curve (mode-2 conjugate state with $c=c_0^+$). All of these points are recovered numerically. The plots also display blue and purple lines, which are obtained analytically. As can be seen from figures~\ref{Front_curves} and \ref{front_curves_all_details}, each point of tangency between a conjugate curves and $c=c_0^+$ is marked by a colour change (from blue to orange or vice-versa). Within the MMCC3 theory, this means that tangencies correspond to a triple collision at the origin, which holds at criticality -- specifically when $c_1=0$ in the KdV equation (see Appendix C) -- and is depicted as a blue line. The purple lines represent the condition $c_3=0$, included for comparison with predictions from Gardner theory.   
	
	For small values of $H_2$ small, the solution space always corresponds to either panel $C$ or $D$, depending on the sign of $c_1$. As $H_2$ increases, various events -- such as the emergence of critical points, the crossing of front curves, their intersection with the curve $c=c_0^+$, and criticality -- can significantly affect the structure of the solution space. Figures \ref{Parameter_regions}-\ref{Parameter_regions2} capture all the key topological transitions observed in the solution space. More precisely, between two consecutive panels ({\it i.e.,} successive values of $\delta$) the structure of the solution space remains qualitatively similar to that corresponding to the lower value of $\delta$. 
	It is important to note that, although not indicated in the colour scheme, for any point in parameter space lying above the red curve, along with the solutions described by panel $A$ or $B$ we also have solutions with characteristics of mode-2 corresponding to panel $M2$ in figure \ref{fig:bif_space}  (see \citeauthor{doak_et_al} \citeyear{doak_et_al} for further details).
	
	Finally, it is worth noting that any coloured panel corresponding to $\delta>1$ can be transformed into one with with $1/\delta$ via the mapping $(x,y)\mapsto(1-x-y,y)$, with the caveat that the colour scheme is reversed. This reflects  the upside-down symmetry inherent in both the Euler and MMCC3 systems, and also highlights the fact that any behaviour exhibited by one interface is mirrored by the other.

	\section{Numerical solutions for the MMCC3 and Euler systems under the Boussinesq approximation}\label{sec:numerics}
	In this section, we adopt the Boussinesq approximation and compare the numerical solutions of the corresponding MMCC3 and Euler systems. Throughout, we choose mass, length, and time-scales such that $\rho_0=1$, $H=1$, and $g=1$. In addition, when referring to the potential associated with the MMCC3 system under the Boussinesq approximation, we mean the potential $V_B$ defined in \eqref{potential_Boussinesq}; for brevity, however, we denote it simply as $V$.
	The MMCC3 system is solved using a second order central difference finite-difference scheme. The three-layer Euler solutions are recovered using the numerical method presented in \cite{guan}, based on a non-local representation of the equations which arise from conformal mappings and boundary integral formulae. All solutions are assumed to be symmetric about $x=0$, halving the domain over which one has to the numerically solve the system.

	We first describe in detail the bifurcation of the solutions corresponding to parameters that gives panels $E-F$. We choose the parameters $\delta=0.5$, $\Delta_2=0.01$, $H_1/H_3=0.1$, and $H_2=0.3$. For such parameters, there are three mode-1 conjugate states in the third quadrant of the $(\zeta_1,\zeta_2)$-plane (see figure \ref{front_curves_all_details}). The fastest two of these are maxima of the potential $V$, and have speeds $c_2^*\approx1.059c_0^+$ and $c_1^*\approx 1.042c_0^+$. The third (slowest) conjugate state corresponds to a saddle of the potential, for which there does not exist any heteroclinic orbits connecting this critical point to the origin. At what speed this critical point is a conjugate state tells us nothing about the solution space.
	\begin{figure}
		\centering
		\includegraphics[scale=1.2]{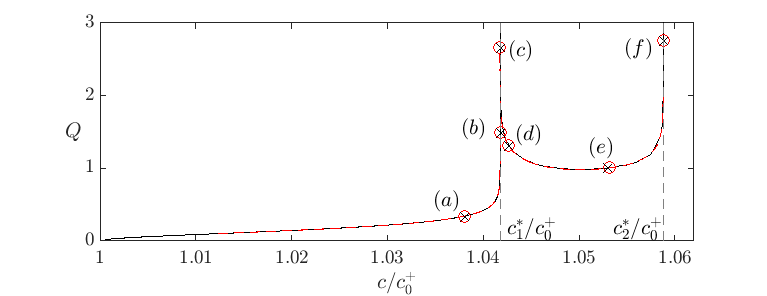}
		\caption{\label{fig:EF_branch} Mode-1 solitary wave solution branches for the MMCC3 (black solid) and Euler (red dashed) systems with $\delta=0.5$, $\Delta_2=0.01$, $H_2=0.3$, and $H_1/H_3=0.1$. There are two solution branches, the first bifurcating from  $c=c_0^+$, and gradually evolving into a tabletop solitary wave with a limiting speed $c=c_1^*\approx1.042c_0^+$. The second branch has speeds $c>c_1^*$. The solutions with speeds close to $c_1^*$ have a tabletop solitary wave-like structure, with an additional hump about $x=0$. As the speeds increase, the volume of the wave decreases, before again increasing to another tabletop solitary wave with a limiting speed $c=c_2^*\approx 1.059c_0^+$. The profiles of the solutions $(a)$--$(f)$ marked by black crosses (MMCC3) and red circles (Euler) are illustrated in figure \ref{fig:EF_solns}.  The black dashed vertical lines show the values $c=c_1^*$ and $c=c_2^*$, to which the solution branches asymptote.}
	\end{figure}
	\begin{figure}
		\centering
		\includegraphics[scale=1.2]{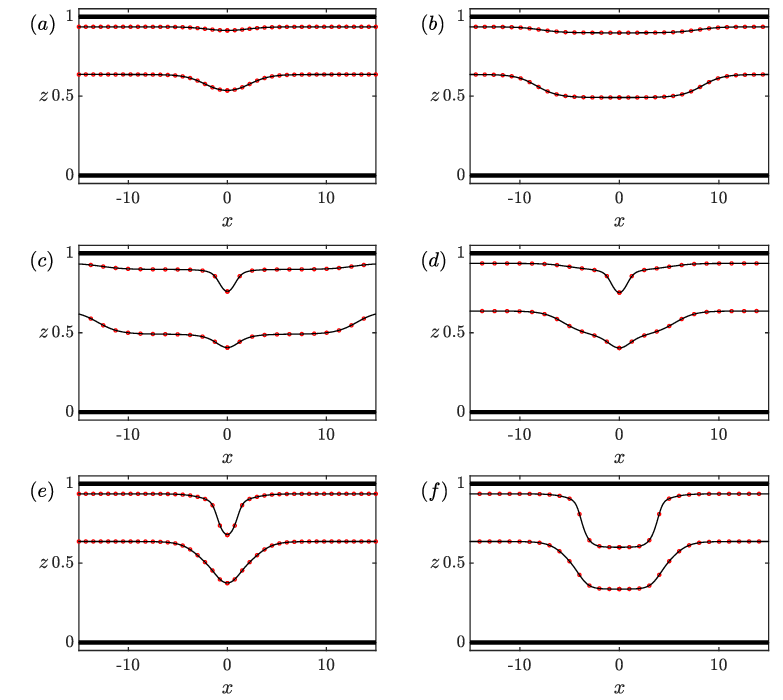}
		\caption{\label{fig:EF_solns} Solutions $(a)-(f)$ from figure \ref{fig:EF_branch} for the MMCC3 (black lines) and Euler (red circles, every fifteenth mesh point plotted). When comparing the profiles, we chose solutions such that they have the same volume $Q$. Solutions $(a)$-$(b)$ are on the first solution branch, with $(c)-(f)$ being on the second. The speeds of the MMCC3 solutions $(a)-(f)$ are (to four decimal places) 
			$c/c_0^+=1.0380,1.0418,1.0418,1.0427,1.0529,1.0588$ while for Euler they are the same to that accuracy except solution $(d)$ which has $c/c_0^+=1.0427$, and $(e)$ with $c/c_0^+=1.0532$. }
	\end{figure}
	\begin{figure}
		\centering
		\includegraphics[scale=1.2]{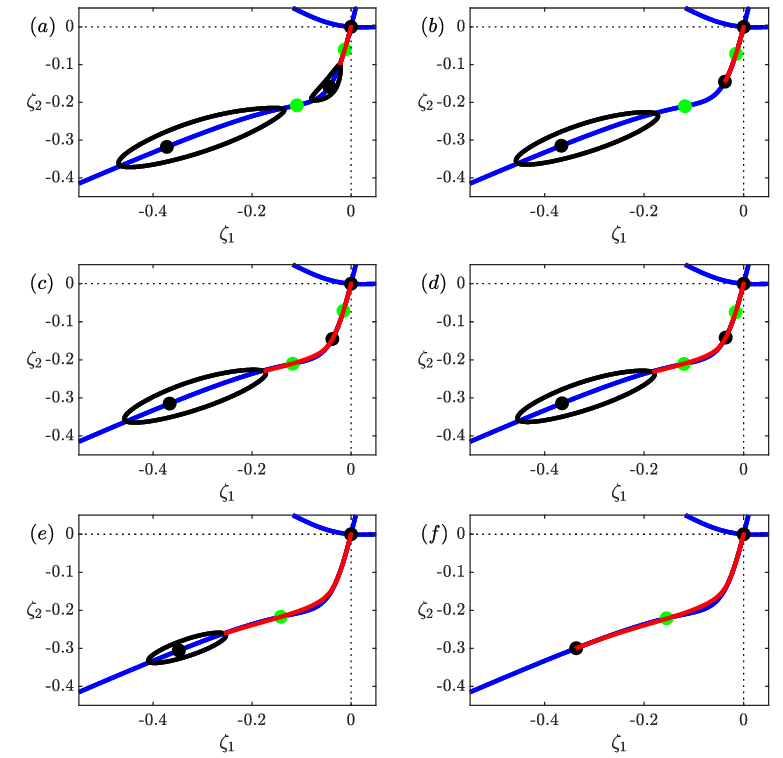}
		\caption{\label{fig:EF_solns_V} Trajectories (in red) of solutions $(a)-(f)$ from figure \ref{fig:EF_branch} for the MMCC3 model, when projected onto the $(\zeta_1,\zeta_2)$-plane. The straight solid black curve corresponds to $\zeta_2-\zeta_1=H_2$, the dashed curves show the origin. The blue curve is $P=0$, the black curves are the equipotential $V=0$, and the green (black) dots are saddles (maxima) of $V$. }
	\end{figure}
	
	\begin{figure}
		\centering
		\begin{overpic}[scale=1.2]{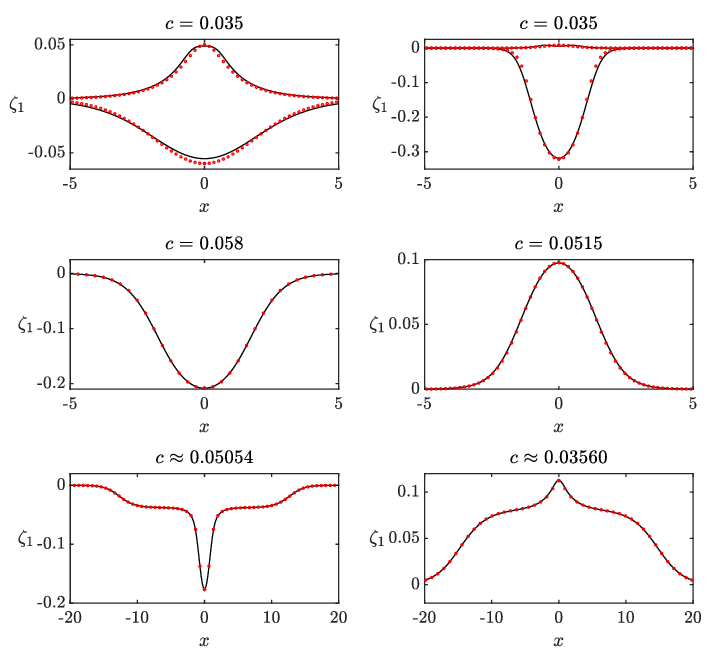}
			\put(0,85){$(a)$}
			\put(0,54){$(c)$}
			\put(0,23){$(e)$}
			\put(51,85){$(b)$}
			\put(51,54){$(d)$}
			\put(51,23){$(f)$}
		\end{overpic}
		\caption{Each panel shows the upper interface $\zeta_1(x)$ of a solution of the MMCC3 (black curves) and full Euler (red circles) system.
			All solutions have $\delta=0.5$ and $\Delta_2=0.01$. The total depth is normalised to $H_1+H_2+H_3=1$, and the undisturbed heights  $(H_1,H_2)$ for each panel are 
			$(a)$ $(0.35,0.6)$, 
			$(b)$ $(0.1,0.85)$,
			$(c)$ $(0.2,0.1)$,
			$(d)$ $(0.3,0.5)$,
			$(e)$ $(0.0636,0.3)$,
			$(f)$ $(0.5714,0.4)$.
			The parameters for $H_i$ were chosen such that each panel $(a)-(f)$ corresponds to the panels $A-F$ from figure \ref{fig:bif_space}.
			For a given speed, shown above the panel, there exists either one solution (as in panels $(c)-(f)$, or two solutions, as in panels $(a)-(b)$. 
			\label{fig:Euler_MCC}}
	\end{figure}

	Numerical results reveal there are two solution branches, one with $c\in[c_0^+,c_1^*]$, and the other with $c\in(c_1^*, c_2^*]$. These are shown in figure \ref{fig:EF_branch}, where the horizontal axis is the speed and the vertical axis is a measure of the volume of the wave $Q$, given by
	\begin{align}
		Q = \int_0^L |\zeta_1| + |\zeta_2| \, \mathrm{d}x.
	\end{align}
	We integrate over only half the domain in $x$ since symmetry is assumed.
	The profiles corresponding to solutions $(a)-(f)$ are shown in figure \ref{fig:EF_solns}. The first branch bifurcates from zero amplitude and limits to a heteroclinic orbit between the origin and the conjugate state with speed $c=c_1^*$. Solutions $(a)$ and $(b)$ show this behaviour. To describe the second solution branch, we first highlight the solution in panel $(c)$. This solution has a speed slightly greater than $c_1^*$. The solution is similar to the tabletop solitary wave from the first branch (panel $(b)$), but is characterised by an additional localised hump along the broadened section. As with tabletop solitary waves, one can compute solutions with arbitrarily large broadened sections as $c\rightarrow c_1^*$, but for the second branch the limit is from above. Due to the hump along the broadened section, this solution cannot limit to a heteroclinic orbit, and hence must exist for speeds strictly greater than $c_1^*$, unlike the first branch, where the limiting solution is a heteroclinic orbit with $c=c_1^*$. Following the branch the other way, for a small increase in speed the broadened section decreases in volume, resulting in a more typical solitary wave profile (solution $(d)$). The solution branch then increases in amplitude (solution $(e)$), before limiting to a heteroclinic orbit to the second faster conjugate state given by $c=c_2^*$ (solution $(f)$).

	To better understand this previously unseen limiting behaviour of the second branch near $c=c_1^*$, it is helpful to consider the critical points of the potential $V$. In figure \ref{fig:EF_solns_V}, we plot the trajectories of the 
	MMCC3 solutions $(a)-(f)$ projected onto the $(\zeta_1,\zeta_2)$-plane, shown in this figure as a red curve. We also show for these solutions the critical points of $V$, with a black and blue circle representing maxima and saddles respectively. Finally, we show the curve $P=0$ (blue, see equation \eqref{geo_locus_critical_points}) and the contour $V=0$ (black). Solution $(a)$ is a moderate amplitude solution from the first solution branch. As can be seen, the solution is a homoclinic orbit originating at the origin, and reaching the $V=0$ contour, before returning. The solution $(b)$ is a tabletop solitary along the first branch. Following the branch further, the solution limits to a heteroclinic orbit, connecting the origin to the non-trivial maximum closer to the origin as $c\rightarrow c_1^*$ from below. For speeds just greater than $c_1^*$, we have solutions on the second branch, such as solution $(c)$. This solution has a broadened shelf region, where the amplitude of the shelf is predicted by the conjugate state with speed $c=c_1^*$. However, since the speed is just greater than $c_1^*$, the maximum corresponding to that conjugate state has $V<0$. It can be seen that the solution also has a hump at $x=0$. As $c\rightarrow c_1^*$ from above, the broadened section becomes wider, and the limiting solution is a near-heteroclinic orbit connecting the origin to the conjugate state, and then a near-homoclinic orbit originating from that same conjugate state. 
	Heuristically, the non-trivial maximum closer to the origin, which is a conjugate state at $c=c_1^*$, \emph{blocks} the trajectory from the origin in to the second valley where the additional bump lies for speeds $c\leq c_1^*$. This explains why the two solution branches exist for different ranges of speeds. Following the branch further, as $c$ increases the broadened shelf shrinks (solution $(d)$), and as the speed increases further the solution limits to a heteroclinic orbit connecting the origin with the second maximum (solution $(f)$).
	
	As mentioned earlier, in terms of partitioning the parameter space into regions associated with distinct types of solutions, the procedure should remain valid within both the MMCC3 and Euler theories. The reason being that all criteria proposed  
	to distinguish the different regions rely on the shared conjugate state structure between the MMCC3 model and the full Euler equations. What remains to be confirmed is that the solution behaviour predicted by the critical point analysis for the MMCC3 model, for each one of those regions, is well reproduced by the solutions to the Euler equations. The numerical results presented above for solutions of the type shown in panels $E-F$ are certainly encouraging. Keeping the same values $\delta=0.5$, $\Delta_2=0.01$, we now choose the values of $H_i$ such that the six regions in figure~\ref{skeleton_and_colouring} are covered. Figure \ref{fig:Euler_MCC} shows representative solutions from each panel $A-F$ for the Euler (red circles) and MMCC3 (black curves).  
	We show only the upper interface displacement $\zeta_1$, as for each solution the lower interface has a comparable shape. For each panel, we fix the speed, and show the corresponding solutions. Note that panels $(a)$ and $(b)$ show two solutions, which is only possible since they are in the parameter regime for which there is true multiplicity of solutions. Even in cases where the quantitative agreement is not especially strong, such as in panel $(a)$, the qualitative agreement between the MMCC3 and Euler system is excellent.
	
\begin{figure}
		\centering
		\includegraphics[width=16cm]{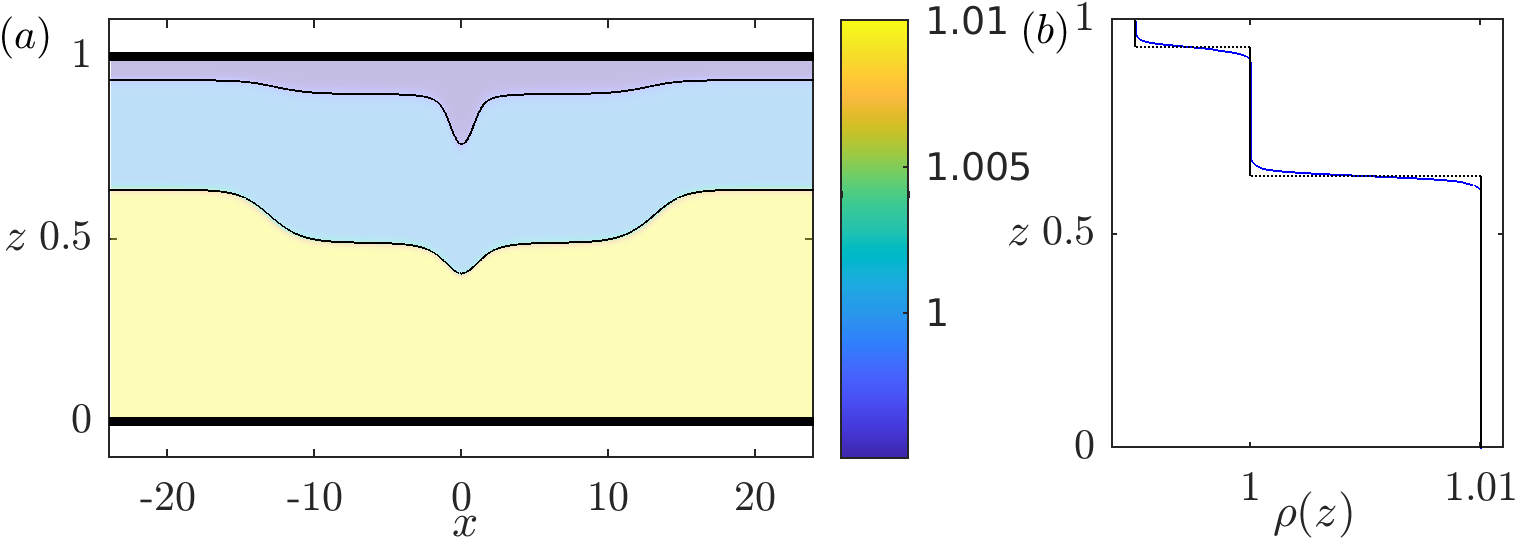}
		\caption{\label{fig:DJL} In panel $(a)$, the black curves shows solution $(c)$ from figure \ref{fig:EF_branch} for the MMCC3 model. The colourmap shows the density field of a comparable DJL solution, where the undisturbed density in the far-field is given by \ref{eq:DJL_rho}. The parameters are set as $d=1/80$, $\Delta_1=0.005$, $\Delta_2=0.01$, $H_2=0.3$ and $H_3=7/11$ for which
			the DJL solution obtained has speed $c=0.0500$, while the MMCC3 solution has speed $c=0.0505$. Panel $(b)$ shows the undisturbed density in the far-field for DJL (blue) model, approximated by the piecewise-constant stratification for the the MMCC3 (black) model.}
	\end{figure}

	We conclude the section by pointing out that the unusual solution behaviour shown in panels $E-F$ is not exclusive to a layered model. As an example, we consider the 
	double-pycnocline stratification defined by 
	\begin{align}\label{eq:DJL_rho}
		\rho(z) = 1+\frac{\Delta_2}{2} - \frac{\Delta_1}{2} - \frac{\Delta_2}{2}\tanh\lrr{(z-H_3)/d} - \frac{\Delta_1}{2}\tanh\lrr{(z-H_3-H_2)/d},
	\end{align}
	as in \cite{lamb_2023}, for which the steady-state solutions of the Euler equations must satisfy the DJL equation (see \citeauthor{marek_book} \citeyear{marek_book}) 
	The parameter $d$ in \eqref{eq:DJL_rho} determines the thicknesses of the two pycnoclines; smaller values of $d$ correspond to thinner pycnoclines. Here, we consider small values of $d$ and small density increments ($d=1/80=0.0125$, $\Delta_1=0.005$, $\Delta_2=0.01$) to enable a meaningful comparison with a three-layer fluid system, as illustrated in figure~\ref{fig:DJL}$(b)$, under the Boussinesq approximation. We set $H_2=0.3$ and $H_3=7/11 \approx 0.63636$, or equivalently $H_1/H_3=0.1$, and compare the solution $(c)$ from figure \ref{fig:EF_branch} for the MMCC3 model with an akin solution for the DJL equation. The DJL solution has speed $c=0.0500$, while the MMCC3 solution has speed $c=0.0505$.
	As can be seen from figure~\ref{fig:DJL}$(a)$, the agreement between the two interfaces of the layered model and the pycnoclines in the continuous stratification is excellent. The DJL solution was computed using a finite difference method, similar to the one used by \cite{tung_et_al}.
	
	\section{Concluding remarks} 
	In this paper we present a classification of mode-1 ISWs in a three-layer fluid. The results contribute to a deeper understanding of the various types of solitary waves that can arise in this physical system. More significantly, criteria are proposed for delineating regions within the parameter space associated with the distinct solution types. This is illustrated in figures~\ref{Parameter_regions}-- \ref{Parameter_regions2}, which provide a complete treatment of the problem under the Boussinesq approximation. The figures show how the structure of the solution space changes with different values of $\delta$. It is evident that, irrespective of the specific value of $\delta$, when the intermediate layer is sufficiently thin (approximately $H_2/H < 0.2$), the solitary waves resemble those shown in panels $C$ or $D$, depending on the values of $H_1$ and $H_3$. In contrast, for thicker intermediate layers, ISWs exhibits a much richer and complex behaviour. Nevertheless, it seems that a significant part of the parameter region yields solutions similar to those shown in panels $A-D$, and often the regions associated with panels $E$ and $F$ are rather limited. This may help explain why these types of solutions have remained unnoticed in earlier investigations.
	
	The criteria proposed to partition the parameter space into regions associated with distinct types of solutions relies on the shared conjugate state structure between the MMCC3 and Euler equations. As such, we expect the approach to be valid under both theoretical frameworks. What is particularly striking, however, is that the behaviour of solutions based on the critical point analysis for the MMCC3 model is also exhibited by solutions of the Euler equations, \textit{e.g.} the behaviour illustrated in figure \ref{fig:EF_solns_V} for a solution of the type shown in panel $E$. In that case, we have demonstrated that the maximum closer to the origin effectively \emph{blocks} the trajectory from entering orbits that would otherwise approach the loop surrounding the other maximum. This scenario changes when the maximum closer to the origin drops below the threshold where the potential $V < 0$. Remarkably, the same dynamics is observed in both systems.
	While the correspondence between their critical points is formally established only when they represent conjugate states, these findings suggest that there may be deeper connections between the critical points of the MMCC3 and Euler equations.

	Despite the relatively small size of the region associated with panel $E$ in figure~\ref{Parameter_regions}, we were able to find comparable solution behaviour in the corresponding continuous double-pycnocline stratification \eqref{eq:DJL_rho}. It would be interesting to explore how variations in the parameter $d$ for the pycnocline thickness affect the structure and extent of the different regions in the parameter space.

	\subsection*{Acknowledgements}{The authors would like to thank K. Khusnutdinova for helpful discussions and for bringing the paper by \cite{maltseva_1989} to their attention.}
	
	\subsection*{Funding}{R.B and P.A.M. would like to acknowledge funding from Leverhulme Trust (RPG-2023-264). A.D. would like to acknowledge funding from EPSRC NFFDy Fellowships (EPSRC grants EP/X028607/1). W.C. was supported by the US National Science Foundation (Grant No. DMS-2108524). The authors would like to thank the Isaac Newton Institute for Mathematical Sciences, Cambridge, for support and hospitality during the programme "Emergent phenomena in nonlinear dispersive waves", where part of the work on this paper was undertaken. This work was supported by EPSRC grant EP/R014604/1.}
	
	\subsection*{Declaration of interests}{The authors report no conflict of interest.}
	
	\subsection*{Author ORCIDs}
	{R. Barros https://orcid.org/0000-0003-2977-1485;\newline
		A. Doak https://orcid.org/0000-0003-4205-2688;\newline
		W. Choi https://orcid.org/0000-0002-4433-3013;\\
		P.A. Milewski https://orcid.org/0000-0003-4454-8441.}
	
	\secta{Appendix A. Derivation of the Gardner equation}
	The Gardner equation \req{Gardner_3layer_dim} can be obtained using the method presented, {\it e.g.},  by \cite{choi_camassa_99}, but the derivation is lengthy and tedious.
	Here we sketch the derivation of the nonlinear terms in the Gardner equation without the linear dispersive term, $\zeta_{1,xxx}$, which was previously obtained for the KdV equation \citep{barros_choi_milewski}. 
	
	The three-layer (non-dispersive) hydrostatic model that neglects any dispersive effects can be found by disregarding the right-hand sides of the MMCC equations, equations (2.1)--(2.2)) in \citet{barros_choi_milewski}. In a frame of reference moving with the linear long wave speed $c_0$, the hydrostatic model can be written, for the local layer thicknesses $h_i(x,t)$, thickness-averaged velocities $\overline{u}_i$, and the hydrostatic pressures $P_i$, as
	\beq
	h_{i,t}-c_0\,h_{i,x}+ (h_i\, \overline{u}_i)_x=0\,,\quad \overline{u}_{i,t}-c_0\,\overline{u}_{i,x}+\overline{u}_i \,\overline{u}_{i,x}+g\, \eta_{i,x}+{P_{i,x}\over \rho_i}
	=0\quad \mbox{for $i=1,2,3$}\,,
	\labeleq{nh3}
	\eeq
	where $h_i$ and $\eta_i$ are defined in \eqref{thickness_def} and figure~\ref{setup}. 
	In \req{nh3}, $P_1$ is the unknown pressure on the top rigid surface while $P_2$ and $P_3$, the pressures at the upper and lower interfaces, respectively, are given by $P_{i+1}=P_i+\rho_i gh_i$ for $i=1,2$.
	
	After non-dimensionalizing all variables with respect to $g$ and $H$ (or equivalently, setting $g=H=1$),
	we expand $w= (\zeta_i, \, u_i,\, P_1)$, for weakly nonlinear wave, as
	\beq
	w= \alpha w^{(1)}+\alpha^2\, w^{(2)}+\alpha^3\, w^{(3)}+\cdots\,,
	\labeleq{expzu}
	\eeq
	where $\alpha$ is a small nonlinear parameter given by $\alpha=a/H\ll 1$ with $a$ being the wave amplitude.
	After substituting \req{expzu} into \req{nh3} with introducing slow space and time variables given by $\xi=\alpha x$ and $\tau_n=\alpha^n\, t$ $(n=1,2,\cdots)$, respectively,  one can find equations, {\it e.g.},  for $\zeta_1^{(n)}$ valid at $O(\alpha^n)$.
	
	At $O(\alpha)$, after expressing $u_i$ and $P_1$  in terms of $\zeta_i^{(1)}$ as
	\beq
	u_1^{(1)}={c_0\over H_1}\,\zeta_1^{(1)}\,,\ \ u_2^{(1)}={c_0\over H_2}\,(\zeta_1^{(1)}-\zeta_2^{(1)})\,,\ \ 
	u_3^{(1)}={c_0\over H_3}\,\zeta_2^{(1)}\,,\ \ P_1^{(1)}=-{\rho_1c_0^2\over H_1}\,\zeta_1^{(1)}\,,
	\eeq
	a linear system for 
	${\bm\zeta^{(1)}}= \left (\zeta_1^{(1)}, \zeta_2^{(1)} \right )^T$ can be obtained as
	\beq
	\bf A\, {\bm \zeta}^{(n)} =0\,, 
	\qquad
	\bf A=
	\left ( 
	\begin{array}{c c}
		a_{11} & a_{12} \\[2pt]
		a_{21}  &  a_{22} \\[2pt]
	\end{array} \right )
	\,,
	\labeleq{ntheq}
	\eeq 
	where  and $a_{jl}$ ($j,l=1,2$) are given, in dimensional form, by
	\beqa
	&a_{11}= - ( {\rho_1}{H_2} + {\rho_2} {H_1}  )\,c_0^2  + g\,( {\rho_2}- {\rho_1})   {H_1}  {H_2} \,,\qquad
	a_{12}= {\rho_2}\, {H_1}c_0^2\,,
	\\
	&a_{21}=-\rho_1 H_3\,c_0^2 +g\, ({\rho_2}- {\rho_1})   {H_1} {H_3} \,,\qquad
	a_{22}= - \rho_3 {H_1}\, c_0^2+ g \,( {\rho_3}- {\rho_2})    {H_1} {H_3} \,.
	\eeqa
	The fact that the determinant of $\bf A$  given by $a_{11}a_{22}-a_{12}a_{21}$ must vanish for a non-trivial solution for ${\bm\zeta^{(1)}}$ yields the linear dispersion relation for $c_0$. 
	To find the form in \req{lin_lw_speed_ast}, it should be noted that
	\beq
	a_{21}={H_3\over H_2}a_{11}+ {\rho_2H_1H_3\over H_2}\,c_0^2\,,\ \ 
	a_{22}={H_1\over H_2} \Big [ -( {\rho_2}{H_3} + {\rho_3} {H_2}  )\,c_0^2  
	+ g\,( {\rho_3}- {\rho_2})   {H_2}  {H_3} \Big]+\rho_2{H_1H_3\over H_2}c_0^2\,.
	\eeq
	
	At $O(\alpha^n)$ ($n=2,3$), we have $\bf A\, {\bm \zeta}^{(n)} =\bm f^{(n)}$ with $\bm f^{(n)}=\left (f_1^{(n)}, f_2^{(n)} \right )^T$, where
	$f_i^{(2)}$ and $f_i^{(3)}$ ($i=1,2$) are given by
	\beq
	f_i^{(2)} = \beta_{i1}^{(2)} \ \zeta_{1,\tau_1}^{(1)}+ \beta_{i2}^{(2)} \ \zeta_{1}^{(1)}\,\zeta_{1,\xi}^{(1)}\,,
	\eeq
	\beq
	f_i^{(3)} = \beta_{i1}^{(3)} \ \zeta_{1,\tau_2}^{(1)}+ \beta_{i2}^{(3)} \ \zeta_{1,\tau_1}^{(2)}+ \beta_{i3}^{(3)} \ \zeta_{1}^{(1)}\,\zeta_{1,\xi}^{(2)}
	+ \beta_{i4}^{(3)}\ \zeta_{1}^{(2)}\,\zeta_{1,\xi}^{(1)}+ \beta_{i5}^{(3)} \ {\zeta_{1}^{(1)}}^2\,\zeta_{1,\xi}^{(1)}\,.
	\eeq
	Here the expressions for $\beta_{ij}^{(n)}$ are too long and are omitted. Then, by imposing the solvability condition for the nonhomogenous system $\bf A\, {\bm \zeta}^{(n)} =\bm f^{(n)}$ given by
	\beq
	a_{22}\,f_1^{(n)}-a_{12}\,f_2^{(n)} = 0\ \ \  \mbox{or} \ \ \  a_{21}\,f_1^{(n)}-a_{11}\,f_2^{(n)} = 0\,.
	\eeq
	one can obtain the following evolution equations for $\zeta_1^{(n)}$ for $n=1$ and 2 at $O(\alpha^2)$ and $O(\alpha^3)$, respectively
	\beqa
	&\zeta_{1,\tau_1}^{(1)}+ c_1\,\zeta_{1}^{(1)}\,\zeta_{1,\xi}^{(1)}
	=0\,,
	\labeleq{zteq2}
	\\
	&\zeta_{1,\tau_2}^{(1)}+ \zeta_{1,\tau_1}^{(2)}+ c_1\, \zeta_{1}^{(1)}\,\zeta_{1,\xi}^{(2)}
	+ c_1\, \zeta_1^{(2)}\,\zeta_{1,\xi}^{(1)}+ c_3 \left ({\zeta_1^{(1)}}^3\right )_\xi=0\,.
	\labeleq{zteq3}
	\eeqa
	where $c_1$ and $c_3$ are given, in dimensional form, by \req{coef_c1} and \req{cubic_non_coef}, respectively. 
	From $\zeta_1=\alpha\, \zeta_1^{(1)}+\alpha^2\,\zeta_1^{(2)}+\cdots$, equation \req{zteq3} can be written, after recovering the original variables,  as
	\beq
	\zeta_{1,t}+c_0\,\zeta_{1,x}+c_1\,\zeta_1\zeta_{1,x}+c_3 \left({\zeta_1}^3\right)_x = 0\,.
	\eeq
	When this is combined with the linear dispersive term given by $c_2\,\zeta_{1,xxx}$, we obtain the Gardner equation given by \req{Gardner_3layer_dim}.
	
	\sectb{Appendix B. Some properties of the curve $P=0$}
	
	Consider fixed values for the densities $\rho_i$ and depths $H_i$ ($i=1,2,3$). Clearly, any critical point of $V$ obtained for a given value of $c$ will belong to the curve $P=0$. We prove here the ``converse'' statement. 
	\begin{lemma}
		Let $(\zeta_1,\zeta_2)$ be point of the admissible region ${\cal R}$ satisfying $P(\zeta_1,\zeta_2)=0$. Then, it must be a critical point of $V$ for some real value $c$.
	\end{lemma}
	\textbf{Proof} \,
	Clearly, the origin is a trivial critical point of $V$ for all $c>0$. Let $(\zeta_1, \zeta_2)\in{\cal R}\setminus\{(0,0)\}$, verifying $P=0$. Then, a value of $c$ can be found such that \eqref{curveC2} holds, by setting
	\begin{equation}\label{admissible_condition}
		c^2= 2 g(\rho_3-\rho_2) h_2^2 h_3^2 \,\frac{\zeta_2}{G_3(h_2,h_3)},
	\end{equation}
	with $G_3$ defined by \eqref{def_G3}. For $c$ to be real, it remains to be proven that the right-hand side of \eqref{admissible_condition} is positive. 
	Notice that the level set $G_3=0$ can be viewed as the graph $\zeta_1 = f(\zeta_2)$, for a certain monotonic function $f$. As a consequence, the curve $G_3=0$ splits ${\cal R}$ into two separate connected components. Moreover, 
	$G_3(\zeta_1,\zeta_2)>0$ ($<0$) when $(\zeta_1,\zeta_2)$ lies above (below) the level set $G_3=0$. 
	To prove our claim it suffices then to show that $P=0$ is above (below) $G_3=0$ for $\zeta_2>0$ ($\zeta_2<0$). 
	
\begin{figure}
		\begin{center}
			\includegraphics*[width=190pt]{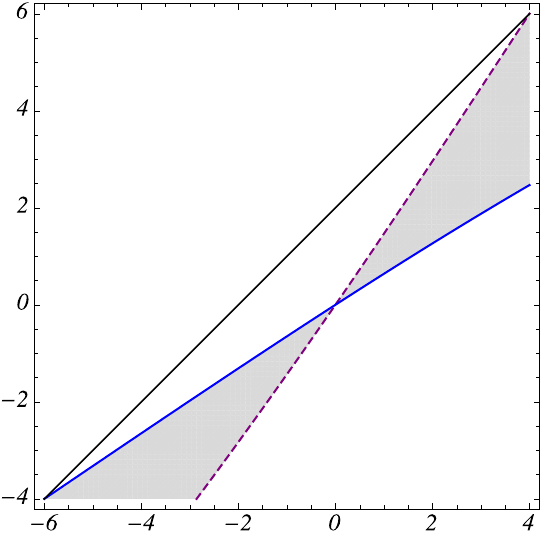}
		\end{center}
		\caption{A typical region on the $(\zeta_1,\zeta_2)$-plane where $G_1$ and $G_3$ defined by \eqref{def_G1}, \eqref{def_G3}, respectively, have the same sign (shaded region). The level sets $G_1=0$ and $G_3=0$ are shown by a dashed (purple) line and a solid (blue) line, respectively. 
			\label{region_G3_G1_same_sign}
		}
	\end{figure}

	From \eqref{geo_locus_critical_points}, it follows that if $(\zeta_1,\zeta_2)$ belongs to  $P=0$, then $\sgn(\zeta_1/\zeta_2) = \sgn (G_1/G_3)$. Since $G_3=0$ is included in the first and third quadrants, only these regions need to be considered. 
	We find that $G_1$ and $G_3$ have the same sign on the region between the level sets $G_1=0$ and $G_3=0$ (see figure~\ref{region_G3_G1_same_sign}). Given that $G_1=0$ is above (below) the level set $G_3=0$ for $\zeta_2>0$ ($\zeta_2<0$), the result follows.
	\qed
	
	\vspace{2mm}
	We consider now one more important feature for the curve $P=0$.
	\begin{lemma}
		The origin is a crunode of the curve $P=0$. Moreover, the tangents to the curve at the origin are given by $\zeta_2=\gamma^\pm \zeta_1$, as in \eqref{ratio_displacements}.   
	\end{lemma}
	\textbf{Proof} \,
	It is easy to check that the origin satisfies $P=0$, $\frac{\partial P}{\partial \zeta_1}=0$, $\frac{\partial P}{\partial \zeta_2}=0$, so it is indeed a finite singular point. To show that it is a crunode, we will show that, at the origin, two branches of the curve have distinct tangent lines. 
	Locally at the origin, the curve $P=0$ is described by
	\begin{multline}
		\rho_2(\rho_1-\rho_2) H_1 H_3 \,\zeta_1^2 + \Big[ \rho_1 (\rho_2-\rho_3) H_2 H_3 - \rho_2 (\rho_1-2\rho_2+\rho_3) H_1 H_3 - \rho_3 (\rho_1-\rho_2) H_1 H_2 \Big] \zeta_1 \zeta_2 - \\
		- \rho_2 (\rho_2-\rho_3)H_1 H_3 \,\zeta_2^2 = 0, 
	\end{multline}
	which splits up into two lines
	\begin{equation}
		\zeta_1 - v_0^{\pm} \zeta_2 = 0,
	\end{equation}
	where $v_0^+$ and $v_0^-$ have opposite signs and are defined by
	\begin{equation}\label{P_slopes_origin}
		v_0^{\pm} = \frac{\rho_1 \rho_2 H_3 + \rho_1 \rho_3 H2 + \rho_2 \rho_3 H_1}{g\rho_2 (\rho_1-\rho_2)H_1 H_3} \left(c_0^{\pm}\right)^2 + \frac{\rho_2 H_3+\rho_3 H_2}{\rho_2 H_3}.
	\end{equation}
	By expanding \eqref{lin_lw_speed_ast} as a quadratic equation for $c^2$, we see that
	\begin{equation}
		(c_0^-)^2 (c_0^+)^2 = \frac{g^2 H_1 H_2 H_3 (\rho_1-\rho_2)(\rho_2-\rho_3)}{\rho_1 \rho_2 H_3 + \rho_1 \rho_3 H_2 + \rho_2 \rho_3 H_1},
	\end{equation}
	which together with \eqref{def_gamma} and \eqref{P_slopes_origin} implies that $v_0^\pm = (\gamma^\mp)^{-1}$. In other words the tangent lines to the curve $P=0$ at the origin are given by $\zeta_2=\gamma^\pm \zeta_1$. 
	\qed
	
	\bigskip

	\noindent {\bf Boussinesq approximation}
	The Hamiltonian structure of the dynamical system governing the ISWs of the MCC3 model is preserved under the Boussinesq approximation. Under this approximation, once all physical parameters are set, the critical points can be found as the solution of $\nabla V_B=(0,0)$, for the corresponding potential $V_B$, specifically:
	\begin{equation}\label{potential_Boussinesq}
		V_B (\zeta_1,\zeta_2)=\frac{1}{2} \left\{ -c^2 \left( \frac{\zeta_1^2}{h_1}+\frac{\zeta_2^2}{h_3} + \frac{(\zeta_2-\zeta_1)^2}{h_2} \right) + g_1^\prime \,\zeta_1^2 + g_2^\prime \,\zeta_2^2 \right\},  
	\end{equation}
	or, equivalently, as the intersection of the following plane algebraic curves: 
	\begin{equation}\label{curveC1B}
		{\cal C}_{1_B} \equiv \frac{1}{2} c^2 \left( H_1^2 h_2^2 - H_2^2 h_1^2 \right) -g_1^\prime \,\zeta_1 h_1^2 h_2^2 =0,
	\end{equation}
	\begin{equation}\label{curveC2B}
		{\cal C}_{2_B} \equiv \frac{1}{2} c^2 \left( H_2^2 h_3^2 - H_3^2 h_2^2 \right) -g_2^\prime \,\zeta_2 h_2^2 h_3^2 =0.
	\end{equation}
	Once again, the geometrical locus of all critical points can be found by eliminating the wave speed $c$ from these two equations. As a result: 
	\begin{equation}\label{geo_locus_critical_points_Bouss}
		P_B (\zeta_1,\zeta_2) \equiv \delta \,\zeta_1 \,h_1^2 \left(H_2^2 h_3^2 - H_3^2 h_2^2\right) - \zeta_2 \,h_3^2 \left(H_1^2 h_2^2 - H_2^2 h_1^2 \right) =0,
	\end{equation}
	with $\delta=\Delta_1/\Delta_2$ in \eqref{stratification}.
	Similarly to what is observed for $P=0$, there are also two distinct configurations for the curve $P_B=0$. The transition between the two configurations can be more easily studied here. It transpires that for $\delta=H_3/H_1$ the curve $P_B=0$ becomes degenerate, and splits up into two components, one of which is the line of equation 
	\begin{equation}\label{component_curve_PB}
		\zeta_2=-(H_3/H_1) \,\zeta_1.
	\end{equation} 
	The coordinates $(\zeta_1,\zeta_2)$ of the extra finite singular point can be determined as follows. First, we solve
	\begin{multline}
		2H_1^2 H_2 (H_1+H_3) + H_1 \left( H_1^2 - 3H_2^2 - 6H_2 H_3 + H_3^2 + 2 H_1 (-3H_2 + H_3)\right) \zeta_1 + \\
		+\left( H_2^2 - 3(H_1+H_3)^2 \right) \zeta_1^2=0,
	\end{multline}
	to find $\zeta_1$, and then use \eqref{component_curve_PB} to find $\zeta_2$.

	\sectc{Appendix C. Power series expansion of the curves ${\cal C}_1$ and ${\cal C}_2$ and critical point collision at the origin}
	As we have mentioned in the text, in the vicinity of the origin, each curve ${\cal C}_k=0$ ($k=1,2$) can be expressed as a convergent power series in terms of $\zeta_1$: 
	\begin{equation}
		\zeta_2 = p_k \,\zeta_1 + q_k \,\zeta_1^2 + r_k \,\zeta_1^3 + \ldots, \quad k=1,2.  
	\end{equation}
	Such is the case because at no point of the curves ${\cal C}_k$ we have $\frac{\partial {\cal C}_k}{\partial \zeta_2}=0$. Indeed, simple calculations lead to 
	\begin{equation}
		\frac{\partial {\cal C}_1}{\partial \zeta_2}=-\frac{2}{h_2} \left[ \frac{1}{2}c^2 G_1 - g(\rho_2-\rho_1)\zeta_1 h_1^2 h_2^2 + \frac{1}{2} c^2 \rho_2 H_2^2 h_1^2 \right], 
	\end{equation}
	with $G_1$ defined in \eqref{def_G1}. If $(\zeta_1,\zeta_2) \in {\cal C}_1=0$, then it follows from \eqref{curveC1} that 
	\begin{equation}
		\frac{\partial {\cal C}_1}{\partial \zeta_2} = -\frac{2}{h_2} \left( \frac{1}{2} c^2 \rho_2 H_2^2 h_1^2 \right) <0.
	\end{equation}
	Similar steps can be taken to show that $\frac{\partial {\cal C}_2}{\partial \zeta_2}\neq 0$ at any point on the curve ${\cal C}_2$.
	
	Now that the conditions are fulfilled to legitimate the use of power series to expand the curves in the vicinity of the origin, it remains to find the coefficients in each expansion. To that purpose, we use Newton's method (see e.g.~\citeauthor{christensen} \citeyear{christensen}) and find: 
	\begin{multline}\label{ps_qs_rs}
		p_1=\frac{R_2}{\rho_2 H_1 \,c^2}, \quad \quad p_2 = \frac{\rho_2 H_3 \,c^2}{S_2}, \quad q_1=\frac{3 (\rho_2-\rho_1) H_2 \,R_4}{2 \rho_2^2 H_1^2 \,c^4}, \\
		q_2 = \frac{3\rho_2 (\rho_2-\rho_3) H_2 H_3 \,c^2 \,S_4}{2 \,S_2^3}, \quad r_1= \frac{(\rho_1-\rho_2) H_2 \,R_6}{2 \rho_2^3 H_1^3 \,c^6 }, \quad r_2 = \frac{\rho_2 (\rho_2-\rho_3)H_2 H_3 \,c^2 \,S_8}{2 \,S_2^5},
	\end{multline}
	where polynomials $R_j$, $S_j$ ($j=2,4$) are defined as:
	\begin{align}
		R_2 (c) &= (\rho_1H_2 + \rho_2 H_1)\, c^2-g(\rho_2-\rho_1)H_1 H_2, & S_2 (c) &=  (\rho_2 H_3 + \rho_3 H_2) \,c^2-g(\rho_3-\rho_2)H_2 H_3, \\
		R_4 (c) &=  \rho_1 c^2 (c^2 + 2gH_1)+g^2H_1^2 (\rho_1-\rho_2), & S_4 (c) &= \rho_3 c^2 (c^2-2gH_3) + g^2 H_3^2 (\rho_3-\rho_2),
	\end{align}
	and the polynomials $R_6(c)$ and $S_8(c)$ are given by:
	\begin{equation}
		R_6(c)= 5 \rho_1^2 (c^2+ g H_1)^3 - \rho_1 \rho_2 (c^2 + g H_1) \left( 4 c^4 + 5 c^2 g H_1 + 10 g^2 H_1^2\right) + 5 g^3 H_1^3 \rho_2^2,
	\end{equation}
	\begin{multline}
		S_8(c) = 4 g^4 H_2 H_3^4 (\rho_2-\rho_3)^3 - g^3 H_3^3 (\rho_2-\rho_3)^2 (5 \rho_2 H_3 - 16 \rho_3 H_2)\,c^2-\\
		-3 g^2 H_3^2 \rho_3 (\rho_2-\rho_3) (3\rho_2 H_2 + 5 \rho_2 H_3 - 8 \rho_3 H_2) \,c^4+ \rho_3 \left( -4 \rho_2^2 H_3 + 5 \rho_2 \rho_3 (H_2+H_3) - 4 \rho_3^2 H_2 \right) c^8 + \\
		+ g H_3 \rho_3 \Big( \rho_2^2 (-4 H_2 + 9 H_3) - 3 \rho_2 \rho_3 (6 H_2+5 H_3)+16 \rho_3^2 H_2 \Big) \,c^6.
	\end{multline}
	For a double collision of critical points to occur at the origin, we must impose $p_1=p_2$, resulting in: 
	\begin{equation}\label{disp_rel_alt}
		Q_4(c) \equiv R_2(c) \, S_2(c) - \rho_2^2 \,H_1 H_3 \,c^4=0,
	\end{equation}
	which is precisely the condition \eqref{lin_lw_speed_ast} for the linear long wave speeds. For a triple collision at the origin we must have 
	\begin{equation}
		p_1=p_2, \quad q_1=q_2,
	\end{equation}
	which amounts to imposing:
	\begin{equation}\label{triple_collision_again}
		Q_{10}(c) \equiv (\rho_2-\rho_1) \,R_4 (c) \,S_2^3(c) +\rho_2^3 (\rho_3-\rho_2) H_1^2 H_3 \,c^6 \,S_4(c)=0 \quad \text{at} \,\,c=c_0.
	\end{equation}
	The polynomials $Q_4$ and $Q_{10}$ introduced in \eqref{disp_rel_alt} and \eqref{triple_collision_again} play an important role in predicting the polarity of ISWs (see Appendix D). 
	
	To simplify the expression of $Q_{10}(c_0)$ we first observe that from \eqref{disp_rel_alt} it follows that $S_2(c_0) = \rho_2 H_3 c_0^2 \,\gamma^{-1}$, with $\gamma$ defined by \eqref{def_gamma}. Simple calculations lead to 
	\begin{equation}
		(\rho_2-\rho_1) R_4(c_0)=\rho_2^2 c_0^4  \left[ \frac{\rho_1}{\rho_2} - \frac{H_1^2}{H_2^2} (1-\gamma)^2 \right],
	\end{equation}
	\begin{equation}
		(\rho_3-\rho_2) S_4(c_0)=\rho_2^2 c_0^4  \left[ -\frac{\rho_3}{\rho_2} + \frac{H_3^2}{H_2^2} (1-\gamma^{-1})^2 \right].
	\end{equation}
	When these are put together, simple manipulations of \eqref{triple_collision_again} lead to
	\begin{equation}\label{criticality_once_again}
		\frac{\rho_3}{H_3^2} \gamma^3 + \frac{\rho_2}{H_2^2} (1-\gamma)^3 - \frac{\rho_1}{H_1^2} =0,
	\end{equation}
	which is precisely the criticality condition obtained in \eqref{criticality_condition}.
	
	Lastly, for a quadruple collision, we must impose:
	\begin{equation}
		p_1=p_2, \quad q_1=q_2, \quad r_1 = r_2.
	\end{equation}
	At the linear long wave speeds, $r_1=r_2$ amounts to consider
	\begin{equation}\label{quadruple_collision}
		(\rho_2-\rho_1) R_6(c_0) \,S_2^5(c_0) - \rho_2^4 H_1^3 H_3 \,c_0^8 \,(\rho_3-\rho_2) S_8(c_0)=0.
	\end{equation}
	We use the same idea employed above to simplify the expression, obtaining:
	\begin{multline}\label{pathway_quadruple_collision}
		-4 \rho_1 \rho_2 H2^3 H3^4 - 9 \rho_1 \rho_2 H_1 H_2^2 H_3^4 (1-\gamma) + \\
		+H_1^3 \Big[ \rho_2^2 H_3^4 (1-\gamma)^4 (5+9\gamma) - \rho_2 \rho_3 H_2^2 H_3 \gamma^3 [ 4 H_2 \gamma + 9 H_3(1-\gamma)(1-2\gamma)] + 9 \rho_3^2 H_2^4 \gamma^5\Big]=0.
	\end{multline}
	Since a quadruple collision at the origin is obtained by requiring that both \eqref{criticality_once_again} and \eqref{pathway_quadruple_collision} are satisfied simultaneously, we may cast \eqref{pathway_quadruple_collision} into the form:
	\begin{multline}
		-4 \rho_1 \rho_2 H2^3 H3^4 - 9 \rho_1 \rho_2 H_1 H_2^2 H_3^4 (1-\gamma) + \\
		+H_1^3 \Big[ \rho_2^2 H_3^4 (1-\gamma)(1-\gamma)^3 (5+9\gamma) - \rho_2 \rho_3 H_2^2 H_3 \gamma^3 [ 4 H_2 \gamma + 9 H_3(1-\gamma)(1-2\gamma)] + 9 \rho_3^2 H_2^4 \gamma^2 \gamma^3 \Big]=0
	\end{multline}
	If we now use \eqref{criticality_once_again} and manipulate the expression as follows: 
	\begin{multline}
		-4 \rho_1 \rho_2 H2^3 H3^4 - 9 \rho_1 \rho_2 H_1 H_2^2 H_3^4 (1-\gamma) + \\
		+H_1^3 \Bigg\{ \rho_2^2 H_3^4 (1-\gamma)\frac{H_2^2}{\rho_2} \left( \frac{\rho_1}{H_1^2}-\frac{\rho_3}{H_3^2} \gamma^3 \right) (5+9\gamma) - \\
		- \rho_2 \rho_3 H_2^2 H_3 \gamma^3 [ 4 H_2 \gamma + 9 H_3(1-\gamma)(1-2\gamma)] + 9 \rho_3^2 H_2^4 \gamma^2 \frac{H_3^2}{\rho_3} \left( \frac{\rho_1}{H_1^2}-\frac{\rho_2}{H_2^2} (1-\gamma)^3 \right)  \Bigg\}=0,
	\end{multline}
	we conclude that a quadruple collision occurs when the following multivariate polynomial vanishes:
	\begin{multline}
		W \equiv -4\rho_2\rho_3 H_1^3 H2^2 H_3 (H_2+H_3)\, \gamma^4 + 13 \rho_2 \rho_3 H_1^3 H_2^2 H_3^2 \,\gamma^3  - \\
		-9H_1H_2^2 H_3^2 \left( \rho_1 \rho_2 H_3^2 - \rho_1 \rho_3 H_2^2 + \rho_2 \rho_3 H_1^2 \right) \gamma^2 + 13 \rho_1 \rho_2 H_1 H_2^2 H_3^4 \, \gamma -4\rho_1\rho_2 H_2^2 H_3^4 (H_1 + H_2)=0.
	\end{multline}
	Let us introduce the multivariate polynomials $\tilde{T}_1$, resulting from $T_1$ in \eqref{def_T1} according to:
	\begin{equation}\label{T1_tilde_def}
		T_1 = \frac{c_0}{4 \rho_2 H_1^3 H_2^3 H_3^3} \,\tilde{T}_1,
	\end{equation}
	and $Z$, resulting from the criticality condition \eqref{criticality_once_again}:  
	\begin{equation}\label{Z_def}
		Z = \rho_3 H_1^2 H_2^2 \,\gamma^3 + \rho_2 H_1^2 H_3^2 \,(1-\gamma)^3 - \rho_1 H_2^2 H_3^2.
	\end{equation}
	Then, it is easy to check that 
	\begin{equation}
		W - H_3 \tilde{T}_1 = - 5 H_1 H_3^2 (1-\gamma) \,Z.
	\end{equation}
	Hence, {\bf at criticality}, we have $Z=0$, and so $W=0$ if and only if $\tilde{T}_1=0$. In other words, the quadruple collision at the origin can be linked to the simultaneous vanishing of the quadratic and cubic nonlinearity coefficients ($c_1,c_3$, respectively) of the Gardner equation.

	\sectd{Appendix D. Polarity of interfacial waves}

	Appendix C shows that, as long as we remain away from criticality, a double collision of critical points will occur at the origin when $c=c_0^\pm$. One of the points involved in the collision is the origin itself. The other point can be found (approximately) as long as the wave speed $c$ is close enough to $c_0^\pm$. Indeed, from \eqref{parameterization_Ck}, it follows that in a neighborhood of the origin, $(\zeta_1,\zeta_2) \in {\cal C}_1 \cap {\cal C}_2$ provided 
	\begin{equation}
		\zeta_1 \left[ (p_1-p_2) + (q_1 - q_2)\zeta_1 + O(\zeta_1^2) \right]=0.
	\end{equation}
	Therefore, at leading order two points of ${\cal C}_1 \cap {\cal C}_2$ can be found, with ordinates prescribed by 
	\begin{equation}\label{ordinates_col_points}
		\zeta_1 =0, \quad (q_1-q_2)\zeta_1 = p_2-p_1. 
	\end{equation}
	Then, from \eqref{ps_qs_rs}:
	\begin{equation}\label{aux_step}
		p_2-p_1 = \frac{- Q_4 (c)}{\rho_2 H_1 \,c^2 \,S_2 (c)}, \quad q_1-q_2 = \frac{3H_2 \,Q_{10} (c)}{2\rho_2^2 H_1^2 \,c^4 \,S_2^3(c)}.
	\end{equation}
	Here, $Q_4$ and $Q_{10}$ are defined in \eqref{disp_rel_alt} and \eqref{triple_collision_again}. 
	Also, it can be easily checked that 
	\begin{align*}
		&Q_4(c) <0  \quad \text{if and only if} \quad (c_0^-)^2 <c^2< (c_0^+)^2, \\
		&S_2(c)<0 \quad \text{if and only if} \quad c^2< g H_2 H_3 (\rho_3-\rho_2)/(\rho_2 H_3 + \rho_3 H_2).
	\end{align*}
	
	We examine now the collision at the linear long wave speed $c_0^-$. Just before collision, we can assume that $c^2/(c_0^-)^2 = 1-\epsilon$ ($\epsilon \ll 1$), check the sign of $p_2-p_1$ and $q_1-q_2$, and use \eqref{ordinates_col_points} to find the way the non-trivial critical point approaches the origin. From \eqref{aux_step} it follows:
	\begin{equation}
		\sgn (p_2-p_1) =\sgn (Q_4) =1, \quad \sgn(q_1-q_2) = - \sgn (Q_{10}), 
	\end{equation}
	and from \eqref{ordinates_col_points}: 
	\begin{equation}
		\sgn(\zeta_1) = \sgn \left( \frac{p_2-p_1}{q_1-q_2}\right) = \sgn(q_1-q_2) = - \sgn (Q_{10}).
	\end{equation}
	Therefore, just before collision, the critical point (a saddle of $V$) approaches the origin from the left (right) when $Q_{10} >0 \,(<0)$. Similarly, just after collision, we can assume that $c^2/(c_0^-)^2 = 1+\epsilon$ ($\epsilon \ll 1$) to conclude:
	\begin{equation}
		\sgn (p_2-p_1) =\sgn (Q_4) =-1, \quad \sgn(q_1-q_2) = - \sgn (Q_{10}), 
	\end{equation}
	and hence:
	\begin{equation}
		\sgn(\zeta_1) = \sgn \left( \frac{p_2-p_1}{q_1-q_2}\right) = -\sgn(q_1-q_2) = \sgn (Q_{10}).
	\end{equation}
	This means that after collision, the critical point (now a minimum of $V$) leaves the origin towards the right (left) when $Q_{10} >0 \,(<0)$. More can be said about the location of this critical point, before and after the collision at the origin. It suffices to notice that \eqref{parameterization_Ck} implies that $\sgn(\zeta_2) = \sgn(p_1 \zeta_1) = -\sgn(\zeta_1)$ and so, during the collision process such point can only belong to the second and fourth quadrants. This implies, in particular, that any critical point on {\em Branch II} of $P=0$ will remain on that branch. 
	
	Let $\rho_i$ and $H_i$, $i=1,2,3$ be physical parameters such that $Q_{10} (c_0^-)\neq0$. Then, there exist $\epsilon>0$ ($\ll 1$) such $Q_{10}\neq 0$ within the range $1-\epsilon < c^2/(c_0^-)^2 < 1+\epsilon$. It can be deduced that mode-2 homoclinic orbits are obtained by leaving the origin in the direction of the unique minimum on {\em Branch II}. Consequently, $Q_{10}(c_0^-) >0$ ($<0$) corresponds to a wave of elevation (depression) for the upper interface. We also observe that for mode-2 waves, $\sgn(c_1)=\sgn(Q_{10}(c_0^-))$, indicating that the KdV equation correctly predicts the solution polarity.

	The double collision at the origin when $c=c_0^+$ can be examined in an analogous way. As before, it can be established that any critical point on {\em Branch I} of $P=0$ will remain on that branch. The key difference is that, depending on the parameters considered, two distinct branches of mode-1 solutions may exist. The branch that bifurcates from zero amplitude is the one corresponding to homoclinic orbits departing from the origin in the direction of the saddle on {\em Branch I} closest to the origin, {\it i.e.,} the point involved in the collision at the origin at $c=c_0^+$. This assertion implies that if a branch of solutions bifurcates from zero amplitude, it must exhibit KdV polarity. This results from $-Q_{10} (c_0^+)>0$ ($<0$) corresponding to a wave of elevation (depression) of the upper interface, together with the fact that for mode-1 waves, $\sgn(c_1)=-\sgn(Q_{10}(c_0^+))$.

\end{document}